\pdfoutput=1
\documentclass[11pt,twoside,a4paper]{article}

\usepackage[utf8]{inputenc}
\usepackage[hang,normal,footnotesize]{caption}
\usepackage{subcaption}
\usepackage{wrapfig}
\usepackage[sort&compress,numbers,colon,merge]{natbib}

\usepackage{a4wide}
\usepackage{fancyhdr}
\usepackage{graphicx}
\usepackage{url}
\usepackage{amsfonts}
\usepackage{amsmath,amssymb,amsthm}

\usepackage{mathrsfs}

\usepackage{multirow}
\usepackage{color}
\definecolor{orange}{rgb}{1,0.5,0}
\definecolor{darkred}{rgb}{0.5,0,0}
\definecolor{darkgreen}{rgb}{0,0.5,0}
\definecolor{darkblue}{rgb}{0,0,0.5}

\usepackage{hyperref}
\hypersetup{ colorlinks,
linkcolor=darkgreen,
urlcolor=darkred,
citecolor=darkblue }

\usepackage{cancel}
\usepackage{bbm}
\usepackage{ulem}

\usepackage{units}

\usepackage[hang,flushmargin]{footmisc} 

\usepackage{booktabs}
\hypersetup{
pdfauthor={Julian Heeck, Martin Holthausen, Werner Rodejohann, Yusuke Shimizu},
pdftitle={Higgs \texorpdfstring{$\to \mu \tau$}{-> Mu Tau} in Abelian and Non-Abelian Flavor Symmetry Models}
}

\makeatletter
\renewcommand{\fnum@table}{\textbf{\tablename~\thetable}}
\renewcommand{\fnum@figure}{\textbf{\figurename~\thefigure}}
\makeatother

\numberwithin{equation}{section}

\newcommand{\re}{\ensuremath{\mathrm{Re}}}
\newcommand{\im}{\ensuremath{\mathrm{Im}}}

\newcommand{\diag}{\ensuremath{\mathrm{diag}}}
\newcommand{\I}{\ensuremath{\mathrm{i}}}

\newcommand{\GeV}{\ensuremath{\,\mathrm{GeV}}}

\newcommand{\ev}[1]{\ensuremath{\left\langle#1\right\rangle}}
\newcommand{\VEV}[1]{\ensuremath{\left\langle#1\right\rangle}}
\newcommand{\vev}[1]{\ensuremath{\left\langle#1\right\rangle}}
\newcommand{\hc}{\ensuremath{\text{h.c.}}}

\newcommand{\be}{\begin{equation}}
\newcommand{\ee}{\end{equation}}
\newcommand{\ba}{\begin{eqnarray}}
\newcommand{\ea}{\end{eqnarray}}
\newcommand{\SU}[1]{\ensuremath{\mathrm{SU}(#1)}}
\newcommand{\U}[1]{\ensuremath{\mathrm{U}(#1)}}

\newcommand{\Eqref}[1]{Eq.~(\ref{#1})}

\newcommand{\Tabref}[1]{Tab.~\ref{#1}}

\renewcommand{\subsubsection}[1]{\vspace{1ex}\mathversion{bold}{\bf #1:}\mathversion{normal}}

\newcommand{\Rep}[1]{\ensuremath{\underline{\mbox{\textbf{#1}}}}}
\newcommand{\MoreRep}[2]{\ensuremath{\underline{\mbox{\textbf{#1}}} _{\mbox{{#2}}}}}

\newcommand{\abs}[1]{\ensuremath{\left\vert#1\right\vert}}

\newcommand{\ra}[1]{\renewcommand{\arraystretch}{#1}}

\def \L {\mathcal{L}} 

\def \epsilon {\varepsilon} 

\def \vec#1{{\boldsymbol{#1}}}
\newcommand{\matrixx}[1]{\begin{pmatrix} #1 \end{pmatrix}} 
\newcommand{\BR}{\mathrm{BR}}
\newcommand{\M}{\mathcal{M}}

\begin{document}
\allowdisplaybreaks[1]


\begin{titlepage}

\ \vspace*{-15mm}
\begin{flushright}
ULB-TH/14-17
\end{flushright}
\vspace*{5mm}

\begin{center}
 {\huge\sffamily\bfseries\mathversion{bold} 
 Higgs $\to \mu \tau$ in Abelian and Non-Abelian Flavor Symmetry Models
 \mathversion{normal}
 }
\\[10mm]
{\large
Julian Heeck\footnote{\texttt{julian.heeck@ulb.ac.be}}$^{(a,b)}$,~
Martin Holthausen\footnote{\texttt{martin.holthausen@mpi-hd.mpg.de}}$^{(a)}$,\\
Werner Rodejohann\footnote{\texttt{werner.rodejohann@mpi-hd.mpg.de}}$^{(a)}$,
Yusuke Shimizu\footnote{\texttt{yusuke.shimizu@mpi-hd.mpg.de}}$^{(a)}$}
\\[5mm]
{\small\textit{$^{(a)}$
Max-Planck-Institut f\"ur Kernphysik, Saupfercheckweg 1, 69117
Heidelberg, Germany
}}\\[.5cm]
{\small\textit{$^{(b)}$
Service de Physique Th\'{e}orique,
Universit\'{e} Libre de Bruxelles,\\
Boulevard du Triomphe, CP225, 1050 Brussels, Belgium
}}

\end{center}
\vspace*{1.0cm}

\begin{abstract}
\noindent
We study lepton flavor violating Higgs decays in two models, with the 
recently found hint for Higgs $\to \mu \tau$ at CMS as a benchmark value 
for the branching ratio. The first model uses the discrete 
flavor symmetry group $A_4$, broken at the electroweak scale, while the 
second is renormalizable and based on the Abelian gauge group $L_\mu - L_\tau$. 
Within the models we find characteristic predictions for other
non-standard Higgs decay modes, charged 
lepton flavor violating decays and correlations 
of the branching ratios with neutrino oscillation parameters. 
 
\end{abstract}

\end{titlepage}

\setcounter{footnote}{0}

\section{Introduction}

After the discovery of the Higgs boson in 
2012~\cite{Aad:2012tfa,Chatrchyan:2012ufa}, the obvious next step is to check 
whether the new particle behaves exactly as predicted by the Standard
Model (SM). 
Expectations for departure from SM behavior are based on
the fact that a variety of new physics scenarios can cause
deviations. In particular in light of flavor symmetries, which seem necessary to explain 
the peculiar structure of lepton mixing, 
one expects non-trivial Higgs decays, be it unusual decays in 
SM particles or in new particles, see 
e.g.~Refs.~\cite{Bhattacharyya:2010hp,Frampton:2010uw,Cao:2011df,Ho:2013hia}. 
A particularly interesting possible departure from the Higgs standard properties 
is flavor violation in its decays~\cite{Blankenburg:2012ex,Harnik:2012pb}. 

Indeed, in the first direct search for lepton flavor violating (LFV) Higgs decays,  
the CMS collaboration has recently reported on an interesting hint for a 
non-zero branching ratio~\cite{CMS:2014hha}, namely 
\begin{equation} \label{eq:yeah}
\BR (h\to\mu\tau) = \left( 0.89_{-0.37}^{+0.40} \right)\%\,.
\end{equation}
Translated into Yukawa couplings defined by the Lagrangian
\begin{align}
-\L_Y = y_{\mu\tau} \overline{\mu}_L \tau_R h +y_{\tau\mu} \overline{\tau}_L \mu_R h +\hc,
\end{align}
with decay rate $\Gamma (h \to \mu \tau) = (|y_{\mu\tau}|^2 + |y_{\tau\mu}|^2)
m_h/8\pi\, $, one needs to explain values around
\begin{align}
\sqrt{|y_{\mu\tau}|^2 + |y_{\tau\mu}|^2} \simeq 0.0027 \pm 0.0006 \,.
\label{eq:cmsyukawas}
\end{align} 
Though~\eqref{eq:yeah}  represents only a $2.5\sigma$ effect, the 
measurement has caused some 
attention~\cite{Dery:2014kxa, Campos:2014zaa, Celis:2014roa, Sierra:2014nqa, Lee:2014rba}. 
While the signal in Eq.~\eqref{eq:yeah} is not unlikely a statistical fluctuation, it is surely tempting to apply flavor symmetry 
models to the branching ratio given above, to study 
the necessary structure of models that can generate it, and to investigate other 
testable consequences of such models. At least it demonstrates again 
that some flavor symmetry models have testable consequences outside the purely 
leptonic sector, and that precision 
studies of the Higgs particle can put constraints on such models. 
In this paper we show that the signal in Eq.~\eqref{eq:yeah} can be generated in 
two different approaches based on quite different flavor symmetries: a 
continuous Abelian approach and a more often studied non-Abelian discrete Ansatz.

It is clear that in 
order to enforce non-standard Higgs phenomenology one needs to introduce new physics 
around the electroweak scale. The Higgs could also be 
the member of a larger multiplet of states. These aspects occur frequently in 
flavor symmetry or other models, and in particular in one of 
the approaches that we follow. 
Our first model applies the frequently used 
non-Abelian discrete flavor symmetry group $A_4$, broken at the 
electroweak scale,\footnote{As usual, the discrete symmetry group is broken in different directions at different scales. The ``visible'' breaking takes place at the electroweak scale. For colliders, the neutrino masses are irrelevant and the other breaking is therefore ``invisible''.}
and features the Higgs particle as a member of a scalar 
$A_4$ triplet. The second approach gauges the difference between muon and 
tau flavor, $L_\mu - L_\tau$, and is therefore an anomaly-free 
Abelian gauge symmetry. 
Both models have in common that there are additional Higgs doublets with 
non-trivial and specific Yukawa coupling structure. They are 
distinguishable and falsifiable.
We demonstrate that charged lepton flavor violation bounds are
fulfilled: the model based on gauged $L_\mu - L_\tau$ is broken in such a way that only the 
$\mu\tau$ sector is affected, where constraints are in general weaker than in decays 
involving electrons. The $A_4$ model benefits essentially 
from a residual $Z_3$ symmetry that 
survives the $A_4$ breaking, sometimes known as
triality~\cite{Ma:2010gs}. However, its breaking causes in particular the 
decay $\mu \to e \gamma$, inducing constraints on the model.  
Anomalous Higgs decays other than 
$h\to\mu\tau$ are predicted, most noteworthy $h\to e\tau$, whose testable correlations with $h\to\mu\tau$ 
are governed by the model parameters. As the breaking of the respective 
flavor symmetry also generates lepton mixing, we investigate the impact of the 
Higgs branching ratios on observables in the neutrino sector. For
example, the Abelian 
model links the chiral nature of the leptons in the $h\to\mu\tau$ decays with 
the octant of $\theta_{23}$ and the neutrino mass ordering. 

In what follows we first deal with the 
non-Abelian model based on $A_4$ (Sec.~\ref{sec:A4}), before turning to the Abelian model in Sec.~\ref{sec:LmuLtau}. We summarize our results in Sec.~\ref{sec:conclusion}.

\section{Non-Abelian case: An \texorpdfstring{$A_4$}{A4} example}
\label{sec:A4}

Non-Abelian discrete flavor symmetries have been used to account for
the large mixing angles measured in the lepton sector
\cite{Ishimori:2010au,King:2013eh}. 
The symmetry $A_4$ 
is the smallest discrete group with a 3-dimensional 
representation~\cite{Ma:2001fk,Babu:2002dz,*Ma:2004zv,Altarelli:2005yp,Babu:2005se,Altarelli:2005yx,He:2006dk} and is therefore an economic and popular choice given the three generations of leptons in the SM. 
In typical models the discrete symmetry is broken to non-commuting subgroups, which form remnant symmetries of the charged lepton and neutrino mass matrices~\cite{Lam:2007qc,Lam:2008rs,Lam:2008sh}.
In the vast majority of models the breaking of the flavor symmetry happens at very high 
and untestable scales. 

Here we aim to employ non-Abelian discrete symmetries with a slightly different point of view, namely we want to emphasize the possibility of additional phenomenology 
of non-Abelian flavor symmetries at the electroweak scale~\cite{Ma:2001fk,Ma:2006km,Kubo:2006kx,Hirsch:2009mx,Ibanez:2009du,Ma:2010gs,Adelhart-Toorop:2010fk, Adelhart-Toorop:2010uq,Bhattacharyya:2010hp,Cao:2011df,Cao:2010mp,Bhattacharyya:2012ze,*Bhattacharyya:2012pi,Holthausen:2012wz, Holthausen:2011vd}. Thus, instead of only concentrating on predicting mixing angles, we have additional tests of models at our disposal, e.g.\ 
lepton flavor violation in the Higgs sector.  

Related to this topic there are two aspects of non-Abelian discrete symmetries that are worth pointing out: 
first, embedding the SM Higgs in a multiplet of Higgs fields allows one to predict the Yukawa couplings of the additional Higgs fields. 
We will put electroweak scalar doublets into an $A_4$ triplet, which then 
automatically induces LFV Higgs phenomenology. 
Second, the often occurring possibility that breaking of $A_4$ results in a remaining $Z_3$ subgroup -- which helps obeying charged lepton flavor violating bounds -- is also of use to us.  

To make the presentation self-contained, we first remind the reader about 'lepton triality'~\cite{Ma:2010gs} 
and then discuss our model and the resulting phenomenology.

\subsection{Lepton triality in \texorpdfstring{$A_4$}{A4} models}
\label{sec:model}

\begin{table}
\centering
\ra{1.2}
\begin{tabular}{lccccccc}\toprule
 &$ \ell$ &$e_R$ &$\mu_R$ &$\tau_R$ &$\chi$&$\Phi$ &$\xi$ \\ \midrule
$ A_4$& ${\Rep{3}}$&$\MoreRep{1}{1}$&$\MoreRep{1}{3}$&$\MoreRep{1}{2}$&${\Rep{3}}$&${\Rep{3}}$&$\MoreRep{1}{1}$\\
$Z_4$&$\I$&$\I$ &$\I$&$\I$&$1$&$-1$&$-1$\\
$\SU{2}_L$&$2$&$1$ &$1$&$1$&$2$&$1$&1\\
$\U{1}_Y$&$-1/2$&$-1$ &$-1$&$-1$&$1/2$&$0$&$0$\\
\bottomrule
\end{tabular}
\caption{Particle content of the minimal model that realizes flavor symmetry breaking at the electroweak scale, which may be UV completed in the fashion of Ref.~\cite{Holthausen:2012wz}. The flavon $\chi$ contains the Higgs field and ties the electroweak to the flavor breaking scale.
\label{tab:partcontent-EWscale}}
\end{table}
\ra{1}

We here describe lepton triality~\cite{Ma:2010gs}, i.e.~the $Z_3$ subgroup typically conserved in the charged lepton sector of $A_4$ models where the Higgs transforms as a triplet ${\Rep{3}}$ under $A_4$. The discrete symmetry group $A_4$ is the smallest group containing an irreducible 3-dimensional representation; we use the basis
\begin{align}
\rho(S)=\left( \begin{array}{ccc}1&0&0\\0&-1&0\\0&0&-1\end{array}\right),\qquad \rho(T)=\left( \begin{array}{ccc}0&1&0\\0&0&1\\1&0&0\end{array}\right)
\label{eq:Ma-Basis}
\end{align}
and implement a model describing the lepton sector at the electroweak scale, following Refs.~\cite{Ma:2001fk,Ma:2006km,Kubo:2006kx,Hirsch:2009mx,Ibanez:2009du,Ma:2010gs,Adelhart-Toorop:2010fk,Adelhart-Toorop:2010uq,Bhattacharyya:2010hp,Cao:2011df,Cao:2010mp,Bhattacharyya:2012ze,*Bhattacharyya:2012pi,Holthausen:2012wz, Holthausen:2011vd}, only caring about the charged lepton sector for now.
The particle content is given in \Tabref{tab:partcontent-EWscale}. 
The necessary vacuum configuration for $\chi \equiv (\chi_1,\chi_2,\chi_3)^T \sim \Rep{3}$,
\begin{align}
\ev{\chi_i}=\left( \begin{array}{c}0\\ \frac{v}{\sqrt{6}} \end{array}\right),\qquad i=1,2,3 ,
\label{eq:vac-conf}
\end{align}
can be naturally obtained from the most general scalar potential following the discussion in Ref.~\cite{Holthausen:2011vd}. Obviously these fields break the discrete symmetry group $A_4$ down to the subgroup $\ev{T\vert T^3=E}\cong Z_3$, while simultaneously breaking the electroweak gauge group $\SU{2}_L\times \U{1}_Y$ down to the electromagnetic $\U{1}_{\mathrm{em}}$.
The normalization in Eq.~\eqref{eq:vac-conf} is chosen such that $v$ corresponds to the SM value, i.e.~$v^2\equiv\sum_i\vev{\chi_i^0}^2= 3\left(\sqrt{2}\frac{v}{\sqrt{6}}\right)^2=(\sqrt{2}G_F)^{-1}\simeq (246 \GeV)^2$.
The charged lepton sector is described by the couplings\footnote{ As there is only one $A_4$ invariant that can be formed out of these fields, we do not specify the contraction here. In ambiguous cases, we always specify the contraction.}
\begin{align}
-\mathcal{L}_e = y_e\bar \ell {\chi} e_R 
+y_\mu \bar \ell {\chi} \mu_R  
+y_\tau \bar\ell {\chi} \tau_R   +\hc
\label{eq:chargedlepton-EW}
\end{align}
Because of the unbroken $Z_3$ symmetry in the charged lepton sector it is useful to change to the basis where this symmetry is represented diagonally: 
\begin{align}
\left(\varphi, {\varphi^{\prime}}, {\varphi^{\prime \prime}} \right)^T \equiv \Omega_T^\dagger \chi\sim (1,\omega^2,\omega)\,, \qquad L\equiv\left(L_e, L_\mu, L_\tau \right)^T \equiv \Omega_T^\dagger \ell\sim (1,\omega^2,\omega)\,,
\label{eq:physical-basis-triality}
\end{align}
with a unitary matrix $\Omega_T$
\begin{align}
\label{eq:SigmaTdef}
\Omega_T\equiv\frac{1}{\sqrt{3}}
\left(
\begin{array}{ccc}
 1 & 1 & 1 \\
 1 & \omega ^2  & \omega  \\
 1 & \omega  & \omega ^2 
\end{array}
\right) \quad \text{ and } \omega \equiv e^{2\pi i/3} \,.
\end{align}
In \eqref{eq:physical-basis-triality} we have indicated the transformation properties under the unbroken subgroup $\vev{T}\cong Z_3$, under which $({e_R},{\mu_R},{\tau_R})$ transform as $(1,\omega^2,\omega)$. This has been denoted flavor triality in Ref.~\cite{Ma:2010gs} and naturally suppresses flavor changing effects, which usually severely constrain multi-Higgs doublet models. To see this, note that in this basis the vacuum configuration~\eqref{eq:vac-conf} implies that only the field $\varphi$ acquires a vacuum expectation value (VEV) $\vev{\varphi}=\left(0,v/\sqrt{2} \right)^T,$ 
while $\varphi^\prime$ and $\varphi^{\prime \prime}$ are inert (VEV-less) doublets.
In the basis of \Eqref{eq:physical-basis-triality} the Yukawa terms read
\begin{align}
\begin{split}
-\mathcal{L}_e =&\,\, {\varphi}\left(y_e \bar L_e e_R +y_\mu \bar L_\mu \mu_R+y_\tau \bar L_\tau \tau_R\right)\
+{\varphi}'\left(y_e \bar L_\tau e_R +y_\mu \bar L_e \mu_R+y_\tau \bar L_\mu \tau_R\right)\\
&+{\varphi}''\left(y_e \bar L_\mu e_R +y_\mu \bar L_\tau \mu_R+y_\tau \bar L_e \tau_R\right)\; +\hc
\end{split}
\label{eq:triality-EW}
\end{align}
and we thus see that $\varphi$ couples diagonally to leptons while $\varphi^{\prime}$ and $\varphi^{\prime \prime}$ do not. The mass matrix, defined by $\vev{\mathcal{L}_e}=\bar e_L M_e e_R $ with $e_L=\ell^{-}$, is thus given by
\begin{align}
M_e =\frac{v}{\sqrt{2}} \Omega_T\,\diag(y_e,y_\mu,y_\tau)\,.
\label{eq:triality_lepton_matrix}
\end{align}
$M_e$ is diagonal in the $Z_3$ basis of Eq.~\eqref{eq:physical-basis-triality}, which therefore corresponds to the charged-lepton mass basis for the case of unbroken triality with $y_\ell = \sqrt{2} 	m_\ell/v$. 
As it stands, the model (which was originally motivated from neutrino
 considerations) does not exhibit tree-level LFV Higgs decays, as
can be read-off of \Eqref{eq:triality-EW}. The scalars $\varphi$, $\varphi'$, and $\varphi''$ do not mix because they carry different charges under the unbroken $Z_3$ symmetry. Corrections to the VEV
alignment \eqref{eq:vac-conf} are thus needed for LFV, as will be discussed in the next
section.\footnote{The only LFV lepton decays allowed by the $Z_3$ are $\tau^\pm \to \mu^\pm \mu^\pm e^\mp$ and $\tau^\pm \to e^\pm e^\pm \mu^\mp$, others being induced exclusively by breaking of triality~\cite{Ma:2010gs}.\label{footnote-LFV}} We will show later that lepton mixing can successfully be reproduced in this model as well.

This model seems to be an excellent starting point when discussing
Higgs LFV decays: first of all, we have introduced multiple Higgses
(which are a necessity for LFV, according to
Paschos--Glashow--Weinberg~\cite{Paschos:1976ay, Glashow:1976nt}) {\it
  without} introducing additional free Yukawa couplings; the Yukawa
couplings of the additional Higgses are not free, but rather dictated
by lepton masses. Furthermore, there is a well-defined SM limit, which
is the 'lepton triality' case, giving an 'explanation' for why we have
not seen LFV processes yet.
Finally, the tau Yukawa is the only large Yukawa coupling and the model therefore predicts large LFV processes predominately in processes involving taus.

\subsection{Perturbation to the vacuum alignment}
\label{sec:VEV-pert}

The potential for the electroweak doublets $\chi \sim \Rep{3}$ is given by\footnote{See Ref.~\cite{Holthausen:2012wz} for a definition of the various Clebsch-Gordon coefficients and the notation. $r^*$ is the complex conjugate representation, i.e. $r^*=r$ except for $\MoreRep{1}{2}^*=\MoreRep{1}{3}$.}
\begin{align}
V_\chi(\chi)&=\mu^2_\chi \chi^\dagger\chi+ \hspace{-.5cm}\sum_{r=\MoreRep{1}{1,2},\MoreRep{3}{S,A}}\hspace{-.5cm}\lambda_{\chi r} (\chi^\dagger \chi)_{r}(\chi^\dagger \chi)_{r^*}+\lambda_{\chi A}\im \left[ (\chi^\dagger \chi)_{\MoreRep{3}{S}}(\chi^\dagger \chi)_{\MoreRep{3}{A}} \right],
\label{eq:chi-potential}
\end{align}
which leads to the VEV of \Eqref{eq:vac-conf} for a certain choice of
parameters (see for example Ref.~\cite{Holthausen:2011vd} and references therein).
In the following, we will always present results in the limit $\lambda_{\chi A}=0$, which simplifies the mixing in the scalar sector. We do not expect qualitative changes for small non-zero $\lambda_{\chi A}$, merely additional small mixing among the scalars.

The choice $\lambda_{\chi A}=0$ lets the potential gain another symmetry, namely the exchange of $\chi_2$ and $\chi_3$ , generated by the $Z_2$ generator    
\begin{align}
\rho(U)=\left( \begin{array}{ccc}1&0&0\\0&0&1\\0&1&0\end{array}\right) .
\label{eq:U}
\end{align}
Together with $A_4$, this leads to an $S_4$ symmetry of the potential, which protects $\lambda_{\chi A}=0$ from corrections of the other scalar couplings. However, as the Yukawa couplings do not respect this symmetry, the (technically) natural size of $\lambda_{\chi A}$ is of the order $y_\tau^4/(16 \pi^2)$.

To discuss symmetry breaking, we should also discuss how the symmetry
is implemented in the neutrino sector. Following standard literature,
we assume the existence of a scalar singlet field $\Phi\sim \Rep{3}$
(see \Tabref{tab:partcontent-EWscale}) to break the $A_4$ symmetry in the $(1,0,0)$ direction, as well as an $A_4$ singlet $\xi$ which breaks the $Z_4$. Since we are interested 
in a phenomenological analysis, we assume the following VEV hierarchy:
\begin{align}
\VEV{\Phi}\gg v \,.
\label{eq:VEV_hierarchy}
\end{align} 
The alignment then proceeds as follows:
\begin{itemize}
\item the potential for $\Phi$ is decoupled from the other scalars and $\Phi$ obtains a 
VEV $\vev{\Phi}\sim (1,0,0)$. This is a natural outcome for a 
large range of potential parameters (see e.g.\ \cite[p.\ 34]{Holthausen:2012xwa} or \cite{Ivanov:2014doa} and references therein).
\item the interaction $\lambda_{m} (\Phi \Phi)_{\MoreRep{1}{3}}
  (\chi^\dagger\chi)_{\MoreRep{1}{2}}$ is the only term communicating the $A_4$ breaking to~$\chi$. Effectively, this results in the soft-$A_4$-breaking term 
\begin{align}
\lambda_{m} (\Phi \Phi)_{\MoreRep{1}{3}} (\chi^\dagger\chi)_{\MoreRep{1}{2}} +\hc \quad \to \quad  M_S^2 \left((\chi^\dagger \chi)_{\MoreRep{1}{2}}+\hc\right)
\label{eq:MSsquared}
\end{align} 
in the scalar potential of $\chi$, that has to be added to $V_\chi
(\chi)$ in Eq.~\eqref{eq:chi-potential}. 
Let us remind the reader that the VEV of $\Phi$ points in the $(1,0,0)$ direction, so the VEV of $\Phi \Phi$ is only non-zero when coupled to a singlet, i.e.~$\langle (\Phi \Phi)_{\Rep{3}} \rangle = 0$. A trivial singlet $(\Phi \Phi)_{\MoreRep{1}{1}}$ just redefines $\mu^2_\chi$ in $V_\chi$, so the above is the only relevant coupling. 

\item the inclusion of Eq.~\eqref{eq:MSsquared} then leads to a VEV shift in $\chi$ (without back-reaction on $\Phi$) with the following structure:
\begin{align}
\vev{\chi}\sim (1+2 \epsilon, 1-\epsilon,1-\epsilon) \quad \Leftrightarrow \quad \vev{\left(\varphi, {\varphi^{\prime}}, {\varphi^{\prime \prime}} \right)^T}=\frac{v}{\sqrt{2}}(1,\epsilon, \epsilon) \,,
\label{eq:triality-breaking}
\end{align}
where $\epsilon \propto M_S^2/v^2$, defined properly below in Eq.~\eqref{eq:VEV-condition}. We thus need the soft $A_4$ breaking below the electroweak scale, which can be achieved with small $\lambda_m$ despite the hierarchy of Eq.~\eqref{eq:VEV_hierarchy}.
Note that these VEVs are in the CP-even neutral direction.
\end{itemize}
This triality-breaking VEV correction \eqref{eq:triality-breaking} with identical entries in $\chi_2$ and $\chi_3$ is a consequence of the symmetry $U$ of the potential, which is left invariant by this VEV. Its form has been observed before in alignment models with driving fields~\cite{Altarelli:2005yx} and non-trivial group extensions~\cite{Holthausen:2011vd}. 
Contrary to the philosophy employed in those references, we do not assume $\epsilon \ll 1$, and therefore rather use the parametrization\footnote{We use the standard abbreviations $c_\beta=\cos \beta$, $s_\beta=\sin \beta$ and $t_\beta=\tan \beta$.} 
\begin{align}
\vev{\left(\varphi, {\varphi^{\prime}}, {\varphi^{\prime \prime}} \right)^T}= \frac{v}{\sqrt{2}} \left({c_\beta},\frac{1}{\sqrt{2}} {s_\beta}, \frac{1}{\sqrt{2}} {s_\beta}\right) .
\label{eq:VEV-form}
\end{align}
A non-zero $\beta$ will give rise to lepton flavor violating Higgs decays as well as rare leptonic decay modes, e.g.~$\ell_i \to \ell_j \gamma$, otherwise forbidden by triality (see also footnote~\ref{footnote-LFV} on page~\pageref{footnote-LFV}).
Since the VEV structure~\eqref{eq:VEV-form} leaves invariant the generator $U$ it makes sense to define $\psi_{1,2}=\frac{1}{\sqrt{2}}\left(\varphi^{\prime }\pm{\varphi^{\prime\prime }}\right)$. 
Of these additional two Higgs doublets, only $\psi_1$ develops a non-vanishing VEV: $\vev{\psi_1}\sim\epsilon$.
Effectively, we therefore have a 2HDM-like model with an additional
VEV-less doublet $\psi_2$.\footnote{Care has to be taken when comparing our $\tan\beta$ to other two-Higgs-doublet models (2HDMs), as the replacement $\tan\beta\to 1/\tan\beta$ can easily be more appropriate depending on the fermion couplings.}

To see precisely how $M_S$ of Eq.~\eqref{eq:MSsquared} leads to the
quoted VEV configuration of Eq.~\eqref{eq:VEV-form}, we consider the
minimization conditions $\frac{\partial V}{\partial \eta}=0$, where
$\eta$ is any of the scalar fields including their neutral components 
$\varphi^0$ and $\psi_{1,2}^0$. Assuming the form \Eqref{eq:VEV-form}, they all vanish except for
\begin{align}
0=\frac{\partial V}{\partial \varphi^0}& \quad \Rightarrow \quad \mu _{\chi }^2=-\frac{1}{3} v^2 \left(\sqrt{3} \lambda _{\text{$\chi $ }1_1}+\lambda _{\text{$\chi $ }3_{1,S}}\right), \\
0=\frac{\partial V}{\partial \psi_1^0}& \quad \Rightarrow \quad M_S^2=-\frac{1}{12} v^2 {s_\beta}  \left({s_\beta} +2 \sqrt{2} {c_\beta} \right) \left(\sqrt{3} \lambda _{\text{$\chi $
   }1_2}-\lambda _{\text{$\chi $ }3_{1,S}}\right).\label{eq:VEV-condition}
\end{align}
This shows that the VEVs can be obtained from the potential once one adds a soft-breaking term (which may originate from the coupling to the neutrino-flavon $\Phi$ as in Eq.~\eqref{eq:MSsquared}).
Note the simplicity of the minimization conditions as a result of the non-Abelian symmetry of the model.
The scalar mass spectrum will lead to the conditions $\lambda _{\text{$\chi $ }3_{1,S}} <0$ (see Eq.~\eqref{eq:masses-charged}) and $\lambda _{\text{$\chi $ }1_2} > \lambda_{\text{$\chi $ }3_{1,S}}/\sqrt{3} $ (see Eq.~\eqref{eq:pseudoscalarmass}), while $M_S^2$ can take on any sign.
The sign difference between $\langle \varphi^0\rangle$ and $\langle \psi_1^0\rangle$ -- the sign of $\beta$ -- is physical and cannot be rotated away, as the Higgs fields originate from the same multiplet.  
For small $\beta \ll 1$, we find from Eq.~\eqref{eq:VEV-condition}
\begin{align}
  \epsilon = s_\beta \simeq \beta \simeq \frac{-3\sqrt{2} }{\sqrt{3} \lambda_{\text{$\chi $
   }1_2}-\lambda_{\text{$\chi $ }3_{1,S}}} \frac{M_S^2}{v^2} 
	\simeq  -\frac{\sqrt{2}\, M_S^2}{\sqrt{3} \, M^2  }
	\,,
\end{align}
as expected from the observation that $M_S\to 0$ reinstates triality. For the last equation we already inserted the scalar mass $M$, to be introduced in the next section (see Eq.~\eqref{eq:pseudoscalarmass}). Since values of interest to explain the CMS excess in $h\to\mu\tau$ lie around $|\beta|\sim 0.2$, we will actually only occasionally make use of the small-$\beta$ limit to gain analytic insights but otherwise use the full expression for $\beta$.

\subsection{Scalar masses}
\begin{figure}
\begin{center}
\includegraphics[width=0.5\textwidth]{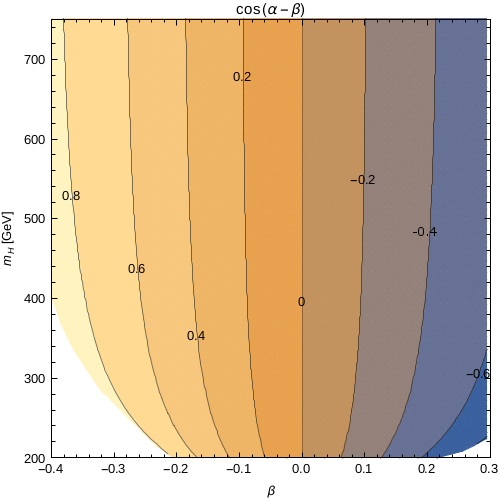}
\end{center}
\caption{$\cos (\alpha -\beta)$ as function of $m_H$ and $\beta$. In the parameter regions where the contour is white, some masses are negative/imaginary, and the VEV is not a minimum.}
\label{fig:cosab}
\end{figure}

After symmetry breaking, the nine physical scalars contained in $\chi$ arrange themselves in the following multiplets under the remnant $\U{1}_{\rm em}\times Z_3^T\times Z_2^U$ symmetry of the $\chi$'s: The first four degrees of freedom are  in the charged scalars $ H^+={c_\beta}\psi_1^{+}-{s_\beta} \varphi^+$ and $\psi_2^{ +}$, which both have the mass
 \begin{align}
m^2_{H^+} 
=-\frac{ \lambda_{\text{$\chi $ }3_{1,S}}}{2\sqrt{3}} v^2 \,.
\label{eq:masses-charged}
\end{align}
The quartic coupling $\lambda _{\text{$\chi $ }3_{1,S}}$ is thus
required to be negative for an electrically neutral vacuum, which leads
to consistency conditions on the parameters of
Eq.~\eqref{eq:chi-potential} by demanding boundedness of the
potential. 
The next two degrees of freedom are $A=\sqrt{2}({c_\beta}\;\im \psi_1^{0}-{s_\beta}\; \im \varphi^0)$ and $\sqrt{2}\re \psi_2^0$, which are degenerate with mass
\begin{align}
m_A^2=m^2_{H^+} -\frac{ \lambda _{\text{$\chi $ }3_{1,A}}}{2\sqrt{3}}  v^2\,.
\label{eq:masses-mA}
\end{align}
We also have the neutral state $\sqrt{2} \im \psi_2^0 $ with mass
\begin{align}
m^2(\sqrt{2} \im \psi_2^0)=\frac{1}{3} \left( \lambda _{\text{$\chi $ }1_2} v^2+2 m^2_{H^+}\right)\left(\frac{1}{4}(3+{c_{2\beta}}-2 \sqrt{2}{s_{2\beta}})\right)\equiv{M^2}\left(\frac{1}{4}(3+{c_{2\beta}}-2 \sqrt{2}{s_{2\beta}})\right).
\label{eq:pseudoscalarmass}
\end{align}
In the last line we defined a new mass parameter $M$ for convenience,
which corresponds to the mass of $\sqrt{2} \im \psi_2^0 $ in the triality limit $\beta\to 0$.
The final two real scalars sit in the two complex neutral scalars $\psi_1^0$ and $\varphi^0$ that acquire VEVs. The mass eigenstates are given by the neutral scalars 
\begin{equation}
\label{eq:mixVarphiNeutral}
\left(\begin{array}{c}{H}\\ {h} \end{array} \right)
=\left(\begin{array}{cc} {c_{\alpha}} &{s_{\alpha}}\\ -{s_{\alpha}} & {c_{\alpha}} \end{array}\right)\left(\begin{array}{c}\sqrt{2} \re \varphi^0\\\sqrt{2} \re \psi_1^0 \end{array} \right) ,
\end{equation}
with masses $m_h^2=(m_h^0)^2 -\Delta$ and $m_H^2=(m_H^0)^2+\Delta$. 
We can express the last remaining
potential parameter in terms of physical quantities:
\begin{align}
(m_h^0)^2&=\frac{2}{3} \left( \lambda _{\text{$\chi $ }1_1} v^2-2 m_{H^+}^2\right),\qquad  (m_H^0)^2=\frac{1}{4}{M^2} \left(2 \sqrt{2} {s_{2\beta}}-{c_{2\beta}}+5\right)
\intertext{and}
\Delta
&=\frac{{(m_h^0)}^2 s^2_{\beta} \left(4 \sqrt{2} s_{2 \beta}+7 c_{2 \beta}+9\right)}{2 \sqrt{2} s_{2 \beta}-c_{2 \beta}+5}+\mathcal{O}\left({(m_h^0)}^4/(m_H^0)^2\right).
\end{align}
Positivity of masses restricts the values of $\beta$, 
see Fig.~\ref{fig:cosab}. 
Note that the mass splitting is predicted in terms of the other scalar
masses; this non-trivial relation is due to the fact that there is a smaller number of parameters in the scalar
sector than in the general case, courtesy of the non-Abelian flavor symmetry. 
In the same vein, the mixing angle $\alpha$ is predicted in terms of scalar masses:
\begin{align}
\tan 2 \alpha &=-\frac{4 s_{\beta} \left(2 \left({{M}^2}+{(m_h^0)}^2\right) c_\beta+\sqrt{2} {{M}^2} s_\beta)\right)}{\left(3{M^2}-4(m_h^0)^2\right) c_{2 \beta}+{{M}^2} \left(2 \sqrt{2} s_{2 \beta}+1\right)} \,.
\label{eq:scalar_mixing_angle}
\end{align}
Note that the CP-even $\re \psi_2^0$ does not mix with $H$ and $h$
because it is odd under the $Z_2^U$ we obtained by setting
$\lambda_{\chi A}=0$. Since the $Z_2^U$ is broken by the Yukawa
interactions, $\re \psi_2^0$ is not stable and will mix with $h$ and
$H$ at loop level. The same comment applies to the mixing of the
charged scalars and pseudoscalars. 
We will neglect this complication, which is anyways expected to give only small modifications to our results. 

The state $h$ will play the role of the SM-like Higgs particle that
has been produced at the LHC.
The limit of $\cos (\alpha-\beta) = 0 $ is the SM limit, as in other 2HDMs~\cite{Branco:2011iw}. We can eliminate $m_h^0$ by using $(125 \GeV)^2 \simeq m_h^2=(m_h^0)^2 -\Delta $ and therefore end up with the free parameters 
$m_{H^+}$, $m_{A}$, $M$ and $\beta$. 
Note that we have $\cos (\alpha-\beta) \simeq -2 \beta$ in the limit of small $\beta$ and $m_H\gg m_h$ (see Fig.~\ref{fig:cosab}).
The parameters $m_{H^+}$ and $m_{A}$ are not particularly important for the following discussion and can be made large to evade experimental constraints (see also the discussion for the Abelian model in section~\ref{sec:LmuLtau}). Lower limits on $m_{H^+}$ typically range from $90\GeV$ (LEP) up to $\mathcal{O}(300)\GeV$ ($B$ physics)~\cite{Agashe:2014kda}, but depend strongly on the $H^+$ couplings to quarks, which are not specified in our model (see Sec.~\ref{sec:quarks}). Similar comments apply to~$m_A$.

As a numerical example, we consider $\beta = 0.2$ and $M = 400\GeV$, which leads to $\sqrt{\Delta} \simeq 51\GeV$, $m_h^0\simeq 135\GeV$ -- in order to obtain the Higgs mass $m_h = 125\GeV$ --  $m_H \simeq 460\GeV$ and the scalar mixing angle $\sin\alpha\simeq -0.98$ (and hence $\cos (\alpha-\beta)\simeq -0.4$). Keep in mind that our notation for $\alpha$ and $\beta$ is somewhat different from the standard 2HDM notation.
The state $\sqrt{2} \im \psi_2^0 $ has mass $336\GeV$, whilst the other four scalars have masses that depend on an additional coupling (Eqs.~\eqref{eq:masses-charged} and~\eqref{eq:masses-mA}). The soft-breaking parameter from Eq.~\eqref{eq:VEV-condition} is given by $M_S^2\simeq -(200\GeV)^2$.

\subsection{Lepton masses}

\begin{figure}[t]
\centering
\includegraphics[height=0.3\linewidth]{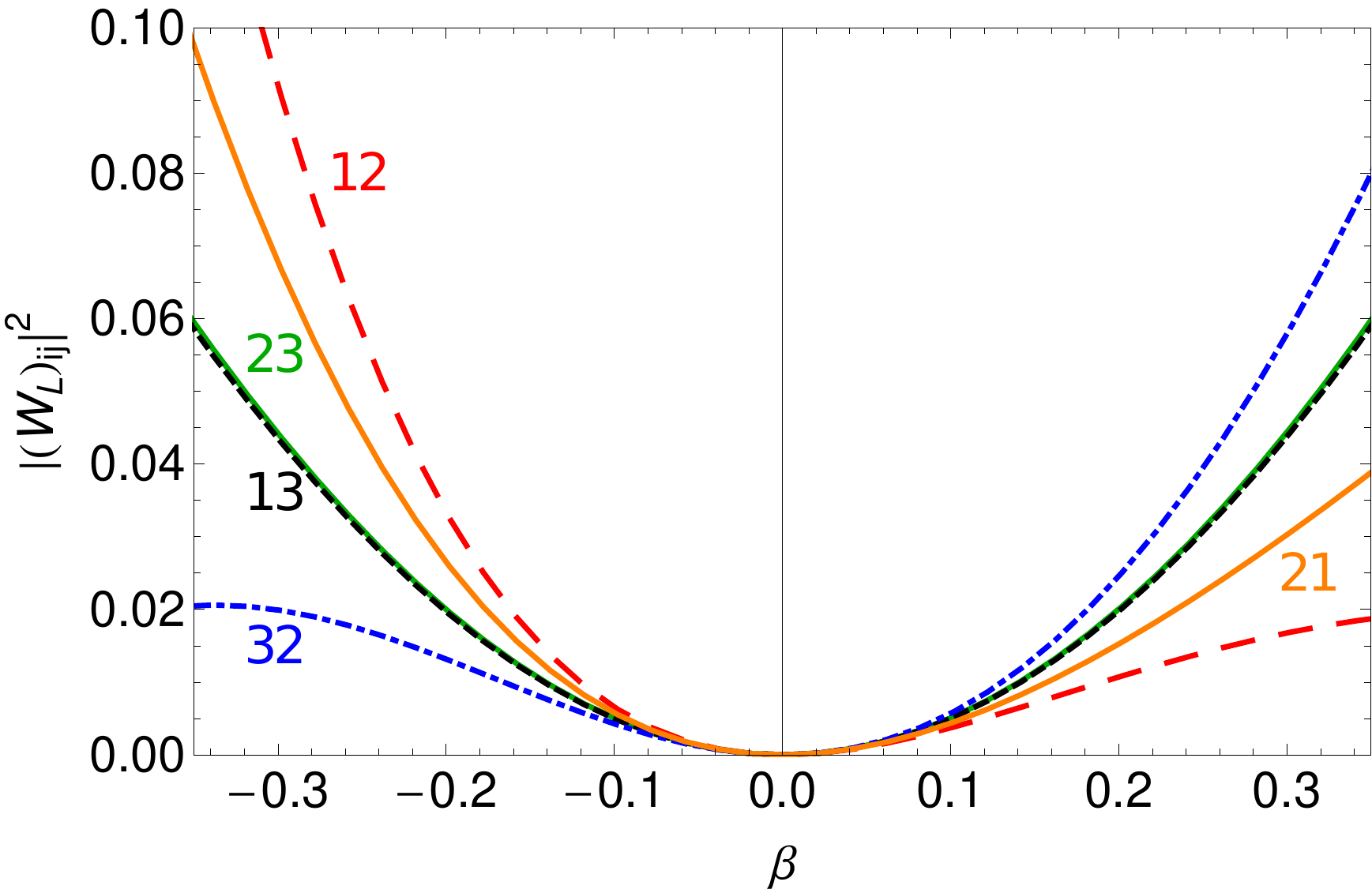}
\includegraphics[height=0.3\linewidth]{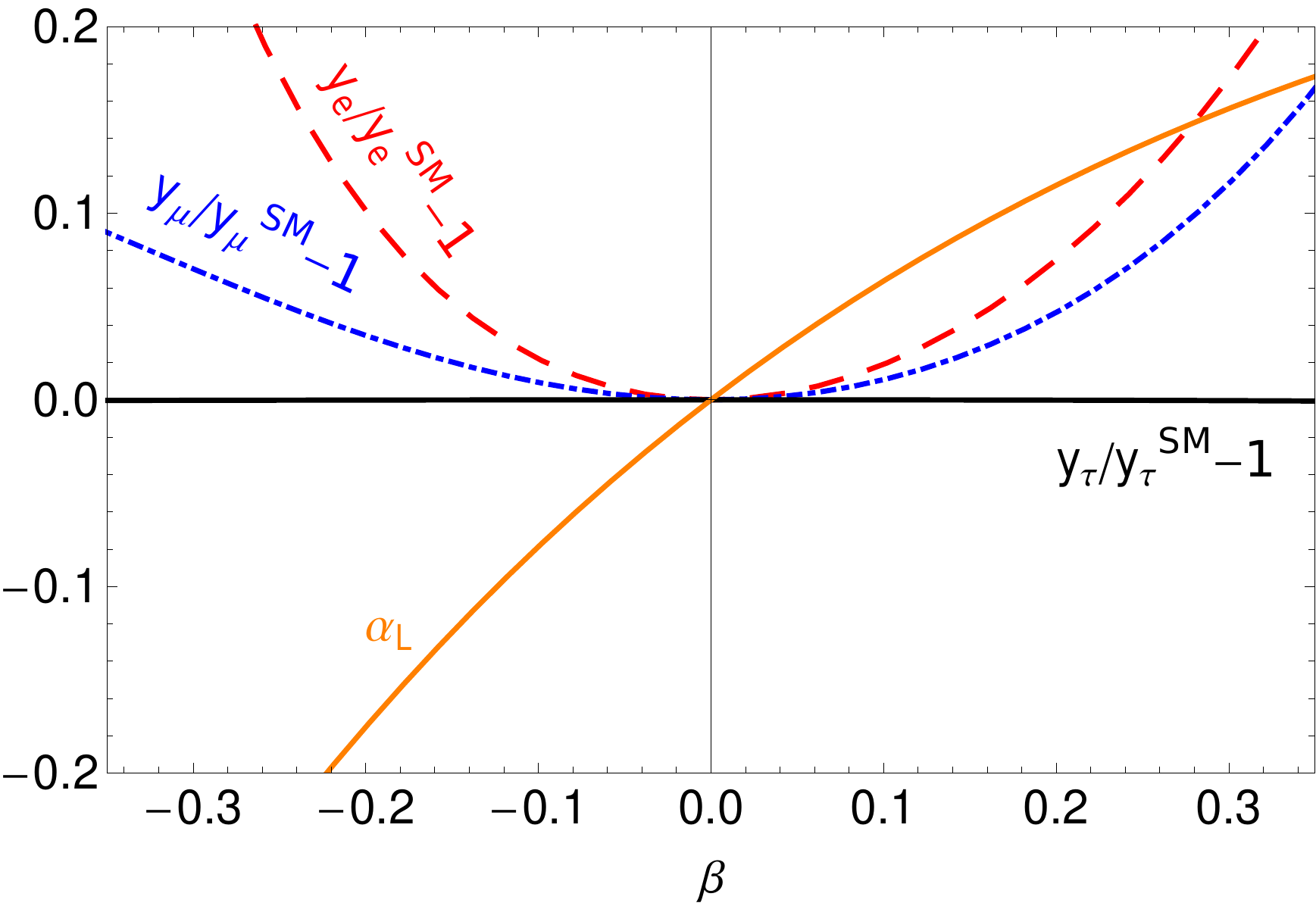}
\caption[]{Relevant couplings in the charged-lepton sector as functions of the triality-violating angle $\beta$.
Left: Off-diagonal charged-lepton mixing matrix elements $|(W_L)_{ij}|^2$, with $(W_L)_{31} \simeq (W_L)_{21}$ and $(W_L)_{13} \simeq (W_L)_{23}$.
Right: Yukawa couplings of the charged leptons relative to the SM values $y_{\alpha}^\text{SM}=\sqrt{2}m_\alpha/v$ (see Eq.~\eqref{eq:charged_lepton_mass_matrix}), as well as the angle $\alpha_L$ of $W_L$ (see Eq.~\eqref{eq:alphaL}).}
\label{fig:plotyuks}
\end{figure}

With the Lagrangian of Eq.~\eqref{eq:chargedlepton-EW} and VEV structure of Eq.~\eqref{eq:VEV-form} we find the charged-lepton mass matrix
\begin{align}
M_e = \frac{v}{\sqrt{2}} \Omega_T \left[ c_\beta \matrixx{y_e  & & \\ & y_\mu & \\ & & y_\tau} + \frac{s_\beta}{\sqrt{2}} \matrixx{ & y_\mu & y_\tau\\y_e & & y_\tau \\ y_e & y_\mu & } \right],
\label{eq:charged_lepton_mass_matrix}
\end{align}
which reduces to the matrix of Eq.~\eqref{eq:triality_lepton_matrix} in the triality limit $\beta\to 0$.
The off-diagonal mass-matrix elements all scale with $s_\beta$ and their relative magnitude is fixed by the charged lepton masses (for small $\beta$ we have the SM-like relations $y_\ell \simeq \sqrt{2} m_\ell/v$).
In particular, the $e\tau$ and $\mu\tau$ entries dominate and have the
same magnitude, which will ultimately lead to large rates for $h\to \mu\tau$, $e\tau$ of similar magnitude, discussed below.
We go to the charged-lepton mass basis $e^0_L$, $e^{0}_R$,  
\begin{equation}
e_L=V_{e_L} e_L^0\,, \qquad {e_R}=V_{{e_R}}e_R^{0}\,,
\end{equation}
where the unitary matrices satisfy
\begin{equation}
V_{e_L}^\dagger M_e V_{{e_R}}=\diag(m_e,m_\mu,m_\tau) \,.
\end{equation}
Both mixing matrices will be functions of $\beta$ and the lepton masses. A good approximation to the left-handed rotation matrix can be parametrized as follows
\begin{align}
V_{e_L} \equiv \Omega_T W_L\simeq \Omega_T R \, O_{23}(\beta)  R^T O_{12}(\alpha_L), \qquad R \equiv\left(
\begin{array}{ccc}
 -\frac{1}{\sqrt{2}} & \frac{1}{\sqrt{2}} & 0 \\
 \frac{1}{\sqrt{2}} & \frac{1}{\sqrt{2}} & 0 \\
 0 & 0 & 1 \\
\end{array}
\right),  
\label{eq:VeL-good}
\end{align}
where $\Omega_T$ is defined in \Eqref{eq:SigmaTdef}. $W_L$ describes the deviation from the triality case $\beta = 0$, which just has $V_{e_L} = \Omega_T$. Here we have expanded in small Yukawa couplings ($y_e\ll y_\mu \ll y_\tau$), but not in small values of $\beta$. The 
$O_{ij}$ denote rotations in the $ij$ plane, and we have 
\begin{align}
\tan 2 \alpha_L \simeq \frac{{s_{\beta}} \left(-3 {s_{\beta}}-7 {s_{3\beta}}+12 \sqrt{2} {c_{\beta}}+4
   \sqrt{2} {c_{3\beta}}\right)}{8 \sqrt{2} {s^3_\beta}+6 {c_{\beta}}+10 c_{3 \beta}}\,,
	\label{eq:alphaL}
\end{align}
or approximately $\alpha_L= \frac{\beta }{\sqrt{2}}-\frac{3\beta^2}{4} +\mathcal{O}\left(\beta^3\right)$, which is true to relative order in small Yukawas and to leading order only depends on $\beta$.
For small $\beta$, this simply yields
\begin{align}
W_L \simeq  \matrixx{1 & \alpha_L & \beta/\sqrt{2}\\ -\alpha_L & 1 & \beta/\sqrt{2} \\ -\beta/\sqrt{2}& -\beta/\sqrt{2} & 1} \simeq  \matrixx{1 & \beta/\sqrt{2} & \beta/\sqrt{2}\\ -\beta/\sqrt{2} & 1 & \beta/\sqrt{2} \\ -\beta/\sqrt{2}& -\beta/\sqrt{2} & 1} ,
\label{eq:VeL-stupid}
\end{align}
which gives non-negligible contributions to the
Pontecorvo--Maki--Nakagawa--Sakata (PMNS) mixing matrix for the values
required to explain the CMS excess (as we will see, values of interest are around $|\beta|\sim 0.2$). The approximation of Eq.~\eqref{eq:VeL-stupid} is pretty good for the $13$ and $23$ elements of $W_L$, but quickly breaks down for all others, see Fig.~\ref{fig:plotyuks}. This is where our definition of $\alpha_L$ kicks in. Note that our parametrization of $W_L$ from Eq.~\eqref{eq:VeL-good} obeys $(W_L)_{23} = (W_L)_{13}$, which is valid to order $m_\mu^2/m_\tau^2$ (see Fig.~\ref{fig:plotyuks}) and $(W_L)_{31} = (W_L)_{21}$, valid to order $m_e^2/m_\mu^2$. These are dictated by the flavor structure in $M_e$ with its equal $23$ and $13$ elements, etc.\ (see Eq.~\eqref{eq:charged_lepton_mass_matrix}).

The right-handed mixing angles are all suppressed by
small Yukawas and it therefore suffices to expand in first order:  
\begin{align}
V_{{e_R}} \simeq
\left(
\begin{array}{ccc}
 1 & -\sqrt{2} \frac{y_e}{y_\mu} \sin\beta & -\sqrt{2}  \frac{y_e}{y_\tau}  \sin\beta \\
 \sqrt{2} \frac{y_e}{y_\mu} \sin\beta  & 1 & -\sqrt{2} \frac{y_\mu}{y_\tau} \sin\beta  \\
 \sqrt{2} \frac{y_e}{y_\tau} \sin\beta & \sqrt{2} \frac{y_\mu}{y_\tau} \sin\beta  & 1 \\
\end{array}
\right).
\end{align}
The Yukawa couplings $y_\ell$ deviate from their SM values for $\beta\neq 0$; the relative corrections are larger for the first and second 
generation Yukawa couplings, with a behavior at small $\beta$ reading 
\begin{align}\label{eq:yuks}
y_{e}\simeq \frac{m_e}{v/\sqrt{2}}\left( 1 + 2 \beta^2\right) , &&
y_{\mu}\simeq \frac{m_\mu}{v/\sqrt{2}}\left( 1 + \beta^2\right) , &&
y_{\tau}\simeq \frac{m_\tau}{v/\sqrt{2}}\left( 1 - \frac{m_\mu^2}{m_\tau^2} \beta^2\right) .
\end{align}
The relations between the Yukawa couplings $y_\alpha$ and their SM values of
$y_{\alpha}^\text{SM}=\sqrt{2}m_\alpha/v$ are shown in
Fig.~\ref{fig:plotyuks}. 

For the neutrino sector, we have introduced a scalar field $\Phi\sim
\Rep{3}$ that breaks the group $A_4$ to the subgroup generated by $S$
of \Eqref{eq:Ma-Basis}, and therefore has a VEV in the $(1,0,0)$
direction~\cite{Holthausen:2011vd}. Its $Z_4$ charge is $-1$ in order
to couple only to neutrinos, similar to the $A_4$ singlet scalar $\xi$.
Using the particle content of Tab.~\ref{tab:partcontent-EWscale} we then obtain the leading order effective operators
\begin{align}
\begin{split}
\mathcal{L} \ &\supset \ x_a \left(\ell^T \sigma_2 \vec{\sigma} \ell\right)_{\MoreRep{1}{1}} \left(\chi^T \sigma_2 \vec{\sigma} \chi\right)_{\MoreRep{1}{1}} {\xi}
 +x_d\left(\ell^T \sigma_2 \vec{\sigma} \ell\right)_{{\Rep{3}}} \left(\chi^T \sigma_2 \vec{\sigma} \chi\right)_{\MoreRep{1}{1}} {\Phi}\\
&\quad + x_e\left(\ell^T \sigma_2 \vec{\sigma} \ell\right)_{{\Rep{3}}} \left(\chi^T \sigma_2 \vec{\sigma} \chi\right)_{{\Rep{3}}} {\xi}
+ \sum_{i=2,3} x_{b_i}\left(\ell^T \sigma_2 \vec{\sigma} \ell\right)_{\MoreRep{1}{i}} \left[\left(\chi^T \sigma_2 \vec{\sigma} \chi\right)_{{\Rep{3}}} {\Phi}\right]_{\MoreRep{1}{i}^*} \\
&\quad + \sum_{i=2,3} x_{c_i}\left(\ell^T \sigma_2 \vec{\sigma} \ell\right)_{\MoreRep{1}{i}} \left(\chi^T \sigma_2 \vec{\sigma} \chi\right)_{\MoreRep{1}{i}^*} \xi +\hc ,
\end{split}
\end{align}
where $\vev{{\Phi}}\sim (1,0,0)$ and the $x_j$ have mass dimension $-2$. 
The Majorana neutrino mass matrix is then given by
\begin{align}
M_\nu=\left(
\begin{array}{ccc}
 a+{b_2}+{b_3}& e & e \\
 e & {b_3}\omega ^2+{b_2} \omega +a & d+e \\
 e & d+e & {b_2} \omega ^2+{b_3}\omega +a \\
\end{array}
\right) ,
\label{eq:Mnu}
\end{align}
with 
\begin{align}
\begin{split}
a&=\frac{1}{3}v^2 {x_a} \abs{\vev{\xi}}\,, \qquad
d=\frac{1}{24} v^2 \left[4 \sqrt{3} {x_d} \abs{\vev{\Phi}}+6 {x_e}\abs{\vev{\xi}} {s_{\beta}} \left({s_{\beta}}-\sqrt{2} {c_{\beta}}\right)\right] ,\\
e&=\frac{1}{24} v^2 {x_e} \abs{\vev{\xi}}\left(\sqrt{2} {s_{2\beta}}+4 {c_{2\beta}}\right), \\
b_i &=\frac{1}{36} v^2 {x_{b_i}}\abs{\vev{\Phi}} \left(-2 \sqrt{2} {s_{2\beta}}+{c_{2\beta}}+3\right) + \frac{1}{12} v^2 {x_{c_i}}\abs{\vev{\xi}} s_\beta \left(s_\beta + 2\sqrt{2} c_\beta\right).
\end{split}
\end{align}
The matrix is diagonalized by going to the mass basis $\nu_L=V_\nu \nu_L^0$ with
\begin{equation}
V_\nu^T M_\nu V_\nu=\diag(m_{\nu_1},m_{\nu_2},m_{\nu_3}) \,,
\end{equation}
leading to the unitary PMNS matrix $U \equiv V_{e_L}^\dagger V_\nu = W_L^\dagger \Omega_T^\dagger V_\nu$ relevant for charged-current interactions.  

In the limit $b_2 = b_3$ the matrix $M_\nu$ becomes $\mu$--$\tau$
symmetric and hence gives a $V_\nu$ with $\theta_{13}^\nu = 0$ and
$\theta_{23}^\nu = \pi/4$  (setting further $b_2 = b_3 = d/3$ gives tri-bimaximal mixing (TBM) values in $V_\nu$, i.e.~additionally $\sin^2 \theta_{12}^\nu = 1/3$). Neglecting the triality-breaking $W_L$ would then result in $U\simeq \Omega_T^\dagger V_\nu$ with $\theta_{13} = 0$ and $\theta_{23} = \pi/4$, incompatible with current data~\cite{Gonzalez-Garcia:2014bfa}. 
Triality breaking $W_L\neq \mathbb{I}$ contributes corrections of order $\beta/\sqrt{2}$ (see Fig.~\ref{fig:plotyuks}), and thus roughly of order $\theta_{13}$ when the CMS excess is to be explained ($\beta \sim 0.2$, see below). One could thus hope to take the $\mu$--$\tau$-symmetric (or TBM) limit in $M_\nu$ as a starting point and use the $W_L$ corrections to generate a non-zero $\theta_{13}$. Unfortunately this does not work; the reason for this is the relation $(W_L)_{31} = (W_L)_{21}$ (see Fig.~\ref{fig:plotyuks}), ultimately due to the mass matrix structure in $M_e$ (Eq.~\eqref{eq:charged_lepton_mass_matrix}). This gives $U_{13} = ((W_L)_{31} - (W_L)_{21})/\sqrt{2} \sim m_e^2/m_\mu^2$, so $\theta_{13}$ is highly suppressed ($\theta_{13}\simeq 4\times 10^{-6}$ for $\beta = 0.2$). $W_L$ does hence lead only to $\beta/\sqrt{2}$ corrections to $\theta_{12}$ and $\theta_{23}$.

We thus need a $\mu$--$\tau$-asymmetric (non-TBM) structure in $V_\nu$, easily accomplished for
$b_2\neq b_3$ ($\neq d/3$). If all the $x_j$ are of similar order, this means in particular that the VEVs of $\Phi$ and $\xi$ should be non-hierarchical, $\langle \Phi\rangle \sim \langle\xi\rangle$, to get a large enough $\theta_{13}$. 
The mass matrix $M_\nu$ in Eq.~\eqref{eq:Mnu} has sufficient 
parameters to fit the present global data, so we omit a detailed discussion. The
flavor symmetry can then no longer predict specific values for mixing
angles (and/or sum-rules for neutrino masses \cite{Barry:2010yk}), 
but rather just motivate the
mixing angle hierarchy. Definite predictions arise, however, in the LFV
observables, as discussed below.

\subsection{Quark couplings}
\label{sec:quarks}

Having discussed the lepton sector of the model, which serves as a major motivation for the discrete flavor group Ansatz, we turn to the other fermions.
To extract experimental limits on the scalars, in particular the  SM-like $h$, one has to take the quark sector into account. So far, all introduced scalars carried charges under the flavor group $A_4\times Z_4$ in order to generate viable lepton mixing patterns. Having treated $h$ as the potential candidate for the $125$ GeV scalar discovered at the LHC, we have to specify its couplings to quarks and how quark masses/mixing arises in our model. This is important, because the very same scalar particle that we study below via its $h\to\mu\tau$ decay has been observed to decay/couple to third-generation quarks, forcing us to include quarks in our discussion.
While the coupling of $h$ to bottom quarks is not yet established at a statistically significant level (around $1$--$2\sigma$~\cite{Aad:2014xzb, Chatrchyan:2013zna}) and the top-quark couplings are so far only inferred indirectly (e.g.~via the loop-induced gluon production rate of $h$), we will not entertain the ludicrous idea of $h$ not coupling to quarks.
Two qualitatively different scenarios emerge~\cite{Holthausen:2012wz}: 
\begin{enumerate}
	\item Including the quarks in the flavor group and generating their masses by the VEV of~$\chi$. One possibility is to generate quark masses analogously to lepton masses, by putting $Q_L\sim \Rep{3}$ and $u^i_R, d^i_R\sim \MoreRep{1}{i}$, which gives the couplings
\begin{align}
	-\mathcal{L}_Q = y_d\bar Q_L {\chi} d_R 
	+y_s\bar Q_L {\chi} s_R  
	+y_b \bar Q_L {\chi} b_R +y_u\bar Q_L \tilde{\chi} u_R 
		+y_c\bar Q_L \tilde{\chi} c_R  
		+y_t \bar Q_L \tilde{\chi} t_R  +\hc , \nonumber
\end{align}
	in complete analogy to the charged leptons. For simplicity we
        insert the SM Yukawa couplings $y_q$ in the above formula. 
For vanishing
        $\beta$ (triality limit) one finds a trivial
        Cabibbo--Kobayashi--Maskawa (CKM) matrix
        $V_\text{CKM}=\mathbb{I}$. 
A non-zero $\beta\simeq 0.2$ introduces non-trivial mixing, but too
small to accommodate the rather large Cabibbo angle. This can already be observed from the diagonalization in Eq.~\eqref{eq:VeL-good}: the left-handed rotations $V_{u_L}$ and $V_{d_L}$ will depend to high accuracy only on $\beta$ and not on the Yukawa couplings/masses, so even though both rotations have off-diagonal entries $\mathcal{O}(\beta)$, the overlap $V_{u_L} V_{d_L}^\dagger$ remains very close to $\mathbb{I}$ (for $\beta=0.2$ the Cabibbo angle is $\simeq 6 \times 10^{-4}$). Consequently, one has
to introduce a higher-dimensional operator to generate viable CKM mixing, e.g.\ $\left(
  \overline{Q}_L \chi\right)_{\Rep{3}} \Phi\xi d_R /\Lambda^2$. This
introduces an additional parameter, which we can adjust to reproduce
the Cabibbo angle. 
As an example, we give the CKM matrix for $\beta=0.2$ and operator strength $\abs{\langle\Phi\rangle\langle\xi\rangle} /\Lambda^2=7\times 10^{-4}$,
\begin{align}
	          \vert V_\text{CKM}\vert=\left(
	          \begin{array}{ccc}
	           0.972356 & 0.233472 & 0.003771 \\
	           0.233458 & 0.972356 & 0.004559 \\
	           0.004534 & 0.003801 & 0.999982 \\
	          \end{array}
	          \right),
\end{align}
which does not appear to be completely unrealistic. Here, we have only used SM Yukawa couplings and have not fitted all parameters of the theory.
Obviously, including other operators will allow us to fit the CKM matrix to even better precision, and also to include the quark masses. Flavor-violating Higgs decays will also be induced, suppressed by small Yukawas, and more importantly heavily depending on the various possible higher-dimensional operators and on details of charge assignments.
We are confident that such an analysis can be performed and the point of this discussion is to outline ways of how this can be achieved.

          	\item Introduction of an additional scalar doublet $H$, uncharged under the flavor group, which couples to quarks in the usual manner and acquires a VEV $\langle H \rangle \neq 0$. The VEV of $\chi$ (split among its components with angle $\beta$ as in Eq.~\eqref{eq:VEV-form}) is then no longer fixed to yield $246 \GeV$, but we rather have $\langle H \rangle^2 +\sum_i \langle \chi_i^0\rangle^2 = (246 \GeV)^2$, so another angle $\beta'$ has to be introduced in order to describe the ratio $\langle H \rangle^2/\sum_i \langle \chi_i^0\rangle^2$. New scalar mixing angles $\alpha'$ arise as well, giving rise to a rather large parameter space.
						
\end{enumerate}
We conclude that while the quark sector of this model is not
completely satisfactory, options exist which can make the framework
holistic, and which can render the quark part largely decoupled from the lepton part. With our main focus on lepton flavor physics phenomenology, we leave the discussion on the quark sector as it is. Independent of the fermion couplings one can set a limit of $|\cos (\beta-\alpha)| < 0.45$ at $95\%$~C.L.~using the vector boson couplings of $h$ alone~\cite{Barger:2014qva}. This is the minimal bound employed in this paper.

\subsection{Higgs interactions}

The SM state that is carrying all the VEV, and therefore couples with
SM strength to the gauge bosons, is given by $H_{\rm SM}=\sqrt{2}{c_\beta}
\re \varphi^0  +\sqrt{2}{s_\beta} \re \psi_1^0 $. 
The coupling to gauge bosons of the state $h$ is therefore suppressed when compared to the SM,
\begin{align}
\frac{g_{hWW}}{g_{hWW}^{\rm SM}}=\frac{g_{hZZ}}{g_{hZZ}^{\rm SM}}=\sin(\beta-\alpha)\,,
\end{align}
just like in other 2HDMs. As discussed above, measurements of the Higgs--vector boson couplings give a limit $|\cos (\beta-\alpha)| < 0.45$ at $95\%$~C.L., typically strengthened depending on the underlying 2HDM couplings to fermions. We will not perform a scan of the currently allowed parameter range of our 2HDM-like model, but rather focus on the LFV aspects. We always display the employed $\cos (\beta-\alpha)$ (which is closely related to the triality-violating angle~$\beta$, but easier to access experimentally in this form) to enable cross checks with LHC results.

\begin{figure}[t]
\centering
\includegraphics[width=0.6\linewidth]{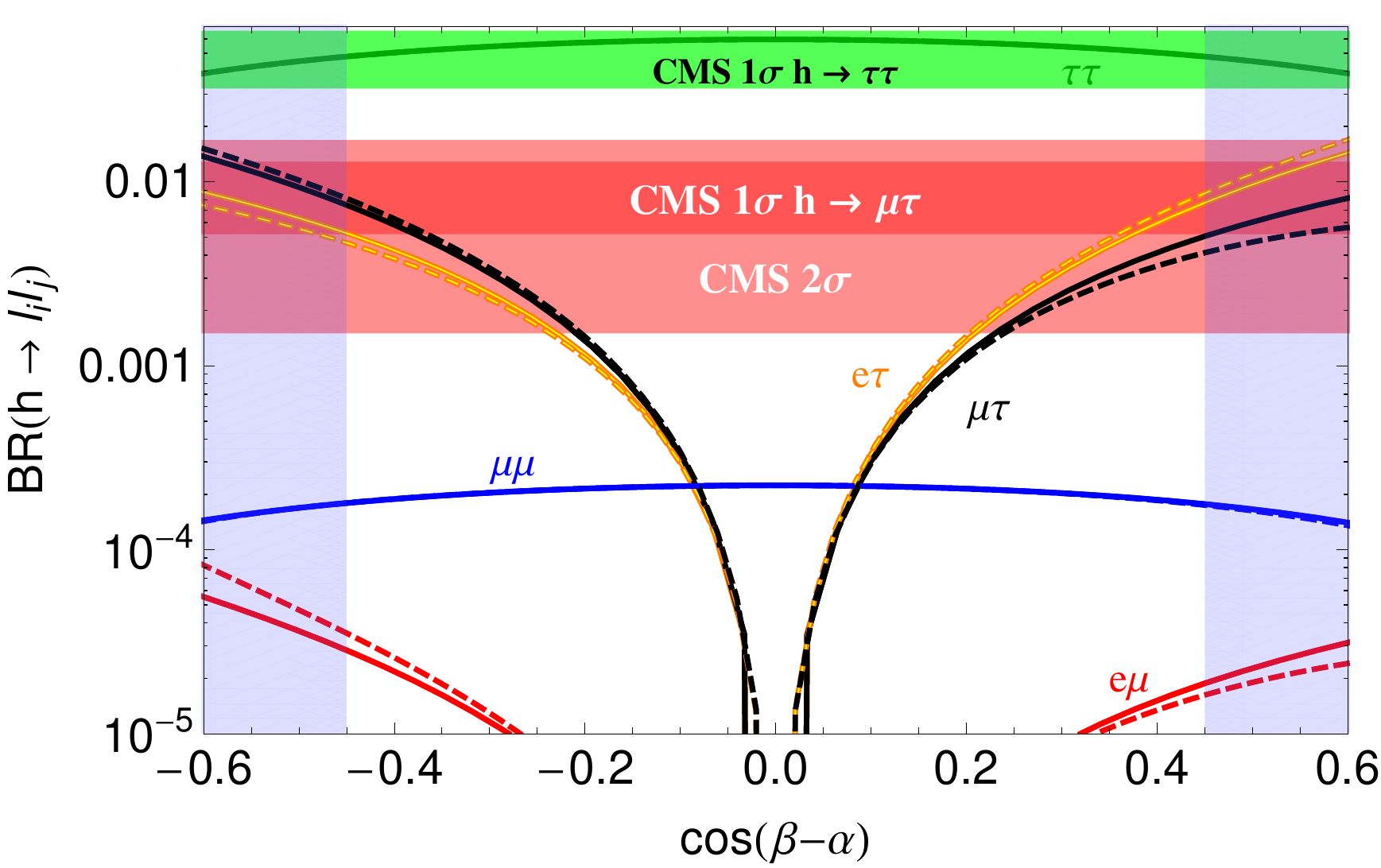}
\caption[]{The leptonic branching ratios $h\to\ell_i \ell_j$ as a
  function of $\cos (\beta-\alpha)$. The shaded horizontal areas
  denote the $1\sigma$ and $2\sigma$ ranges of CMS for $h\to\mu\tau$ (red)~\cite{CMS:2014hha} and $h\to\tau\tau$ (green, only $1\sigma$)~\cite{Chatrchyan:2014nva}. There
  is a small dependence on the heavy Higgs mass $m_H$ through the
  scalar mixing angle $\alpha_L$ (Eq.~\eqref{eq:scalar_mixing_angle}): solid
  (dashed) lines are for $m_H = \unit[200]{GeV}$
  ($\unit[800]{GeV}$). The shaded vertical areas $|c_{\beta-\alpha}|
  \geq 0.45$ are excluded by the conservative limits from
  Higgs--vector--vector measurements~\cite{Barger:2014qva}. The SM-values are recovered for 
$\cos (\beta-\alpha) = 0$.}
\label{fig:BRvsCos}
\end{figure}

The interaction Lagrangian of the lepton mass eigenstates with the Higgs mass eigenstate $h$ is given by
\begin{align}
-\mathcal{L}_{h\bar ff} =y_{\alpha \beta}{\bar e}^{0} _{L\alpha} h {e^{0}_R }_{\beta}+\hc
\end{align}
and the parameters are given to first order in
$\mathcal{O}(y_\mu/y_\tau)$ as 
(see \eqref{eq:yuks} and Fig.\ \ref{fig:plotyuks})
\begin{align}
y_{e \tau} = \frac{m_\tau}{\sqrt{2} v} (c_{\alpha_L}-s_{\alpha_L}) c_{\alpha-\beta}\,,\qquad
y_{\mu \tau} = \frac{m_\tau}{\sqrt{2} v} (c_{\alpha_L}+s_{\alpha_L}) c_{\alpha-\beta}\, ,\qquad
y_{\tau \tau} =  \frac{m_\tau}{v} s_{\beta-\alpha} \,.
\end{align}
All other couplings vanish in this approximation. 
Note again that the couplings become SM-like for $c_{\beta-\alpha} \to 0$ as in other 2HDMs. The third LFV coupling is suppressed by the muon Yukawa coupling 
\begin{align}
\begin{split}
y_{e \mu} &= \frac{y_{\mu}}{4} \left\{ {c_\alpha} \left[{c_\beta} ({c_{\alpha_L}}-{s_{\alpha_L}})+\sqrt{2} {s_{\beta}} ({c_{\alpha_L}}-{s_{\alpha_L}})+{s_{\alpha_L}}+{c_{\alpha_L}}\right]\right.\\
&\quad\left.+2   \sqrt{2} \sin \alpha  \sec \beta  \left[{s_{\alpha_L}} (3 \cos
   \beta -2) {c^2_{\beta/2}}+{c_{\alpha_L}} s^2_{\beta/2} (3 {c_\beta}+2)\right]\right\} +\mathcal{O}\left(y_{\mu}^2\right),
\end{split}
\end{align}
and hence small, but tightly constrained by the LFV decay $\mu\to e\gamma$. While not obvious in any way, $y_{e \mu}$ also vanishes in the limit $c_{\beta-\alpha} \to 0$.
With the off-diagonal $\mu\tau$ couplings of $h$ at our disposal, we can determine the parameter values necessary to explain the CMS excess (Eq.~\eqref{eq:cmsyukawas}) as
\begin{align}
|y_{\mu\tau}| = |\frac{m_\tau}{\sqrt{2} v}(c_{\alpha_L}+s_{\alpha_L})c_{\alpha-\beta}| \simeq 7\times 10^{-3} |\sin (\alpha_L + \pi/4) \cos (\alpha-\beta)|\stackrel{!}{\simeq} 3\times 10^{-3} \,.
\end{align}
Note that the chiral coupling $\overline{\mu}_L \tau_R h$ dominates the decay $h\to\mu\tau$ in this model.
The branching ratio depends only on the parameter $\beta$ (slightly on $m_H$ due to the scalar mixing angle $\alpha$, see Fig.~\ref{fig:cosab}), but we show it as a function of $\cos(\beta-\alpha)$ in Fig.~\ref{fig:BRvsCos} because this quantity is directly related to the Higgs couplings to vector bosons. We see that rather large values $|c_{\beta-\alpha}| \simeq 0.4$ ($|\beta| \simeq 0.2$) are required to describe the CMS excess. Because of the few free parameters in our flavor model, this has direct consequences for other LFV rates.
For one thing, the LFV rate $h\to e\tau$ is expected to be close to the $h\to\mu\tau$ rate,
\begin{align}
\frac{\BR (h\to \mu\tau)}{\BR (h\to e\tau)} \simeq \left( \frac{c_{\alpha_L}+s_{\alpha_L}}{c_{\alpha_L}-s_{\alpha_L}} \right)^2 = \tan^2 (\alpha_L + \pi/4) \xrightarrow{\beta = 0.2} 1.59\,.
\end{align}
A sensitivity to $h\to e\tau$ of similar order as $h\to\mu\tau$ seems feasible at the LHC~\cite{Harnik:2012pb}, even though a dedicated analysis has so far only been performed in the $\mu\tau$ channel. The above prediction will thus serve as the most important discriminator between models once this channel has been probed. The rate $h\to e\mu$ is suppressed by $y_\mu^2/y_\tau^2$ and hence unobservably small compared to the other two LFV channels.
The flavor conserving rates $h\to \mu\mu$, $\tau\tau$ are reduced in this model, but only slightly so (compared to the Abelian explanation of the CMS excess in Sec.~\ref{sec:LmuLtau}). The $h\to\tau\tau$ rate lies comfortably in the $1\sigma$ region of CMS~\cite{Chatrchyan:2014nva}: $0.78\pm 0.27$ (relative to the SM), as shown in Fig.~\ref{fig:BRvsCos}.

Not only the Higgs--vector--vector coupling limit $|\cos
(\beta-\alpha)| < 0.45$ constrains the non-Abelian CMS explanation, the induced LFV rate $\mu\to e\gamma$ further impacts our model, as we will discuss now, and actually excludes the region of interest with positive $c_{\beta-\alpha}$.

\subsection{Bounds from indirect measurements}

\noindent
The most stringent constraints are expected from the rare decays $\mu \rightarrow e \gamma $~\cite{Adam:2013mnn} and $\tau \rightarrow \mu \gamma$~\cite{Aubert:2009ag}.
Note that the decay $\tau \to \mu\gamma$ gives typically stronger limits on the scalar sector than $\tau\to 3\mu$, even though the experimental limit on the branching ratio is a factor $\sim 2$ weaker. This is because $\tau\to 3\mu$ is either suppressed by an additional muon Yukawa coupling (tree-level scalar exchange) or fine-structure coupling (off-shell photon in $\tau\to\mu\gamma\to 3\mu$)~\cite{Harnik:2012pb}. The same holds for $\mu\to e\gamma$ vs.~$\mu\to 3 e$.
The Wilson coefficients $c_L$ and $c_R$, which affect the rate for $\tau \rightarrow \mu \gamma$ as
\begin{align}
\Gamma(\tau \rightarrow \mu \gamma ) = \frac{\alpha m_\tau ^5}{64 \pi^4}\left( \abs{c_L}^2+\abs{c_R}^2 \right),
\end{align}
are given at one-loop as
\begin{align}
c_L &=\sum_{\substack{\alpha=e, \mu, \tau  \\ s=h,H,A,\re \psi_2,\im \psi_2}}  F(m_\tau, m_\alpha, m_\mu, m_s,0,Y_s) \,,\\
c_R &=\sum_{\substack{\alpha=e, \mu, \tau  \\ s=h,H,A,\re \psi_2,\im \psi_2}}  F(m_\tau, m_\alpha, m_\mu, m_s,0,Y_s^\dagger)\,,
\end{align}
with Yukawa coupling matrix $Y_s$ of scalar $s$ and the loop function $F$ given in Eq.~(A.1) of Ref.~\cite{Harnik:2012pb}. 
The corresponding equations for $\mu\rightarrow e \gamma$ can be 
obtained by obvious replacements. 
Note that these complicated expressions only depend on $\beta$, $m_A$ and ${M}$ (or equivalently, $m_H$) as free parameters.

\begin{figure}[t]
\centering
\includegraphics[width=0.6\linewidth,height=7cm]{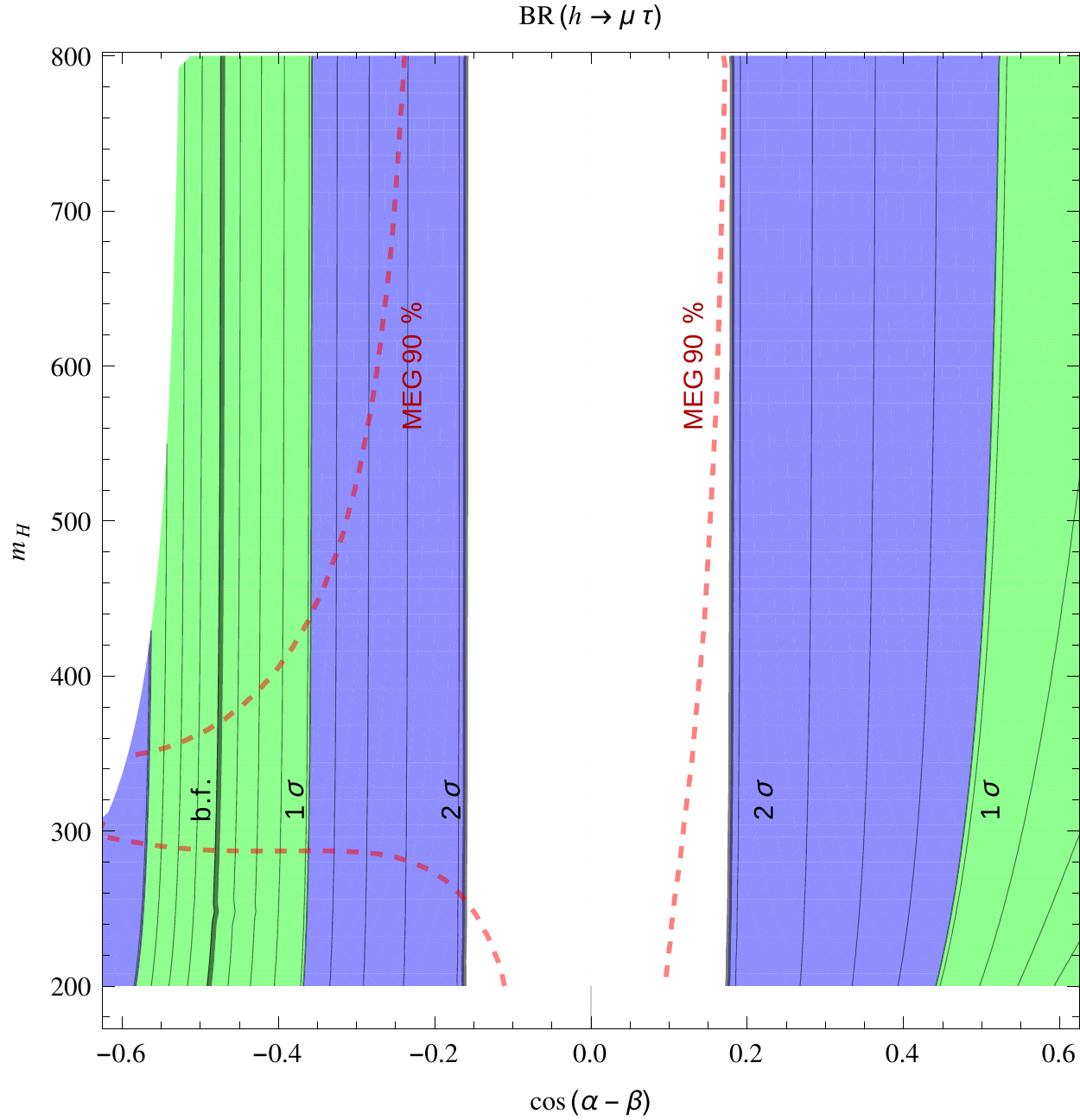}
\caption[]{Relevant parameter space of the model in order to explain the CMS excess in $h\to\mu\tau$ (colored regions, the light lines give steps in 0.001). The dashed red contour denotes the $90\%$~C.L.~bound from $\mu\rightarrow e \gamma$ (MEG~\cite{Adam:2013mnn}); the region inside is allowed. The mass of $A$ is taken to be $m_A=600 \GeV$.
 }
\label{fig:parameter-space2}
\end{figure}

\begin{figure}[tbh]
\centering
\includegraphics[width=0.85\linewidth]{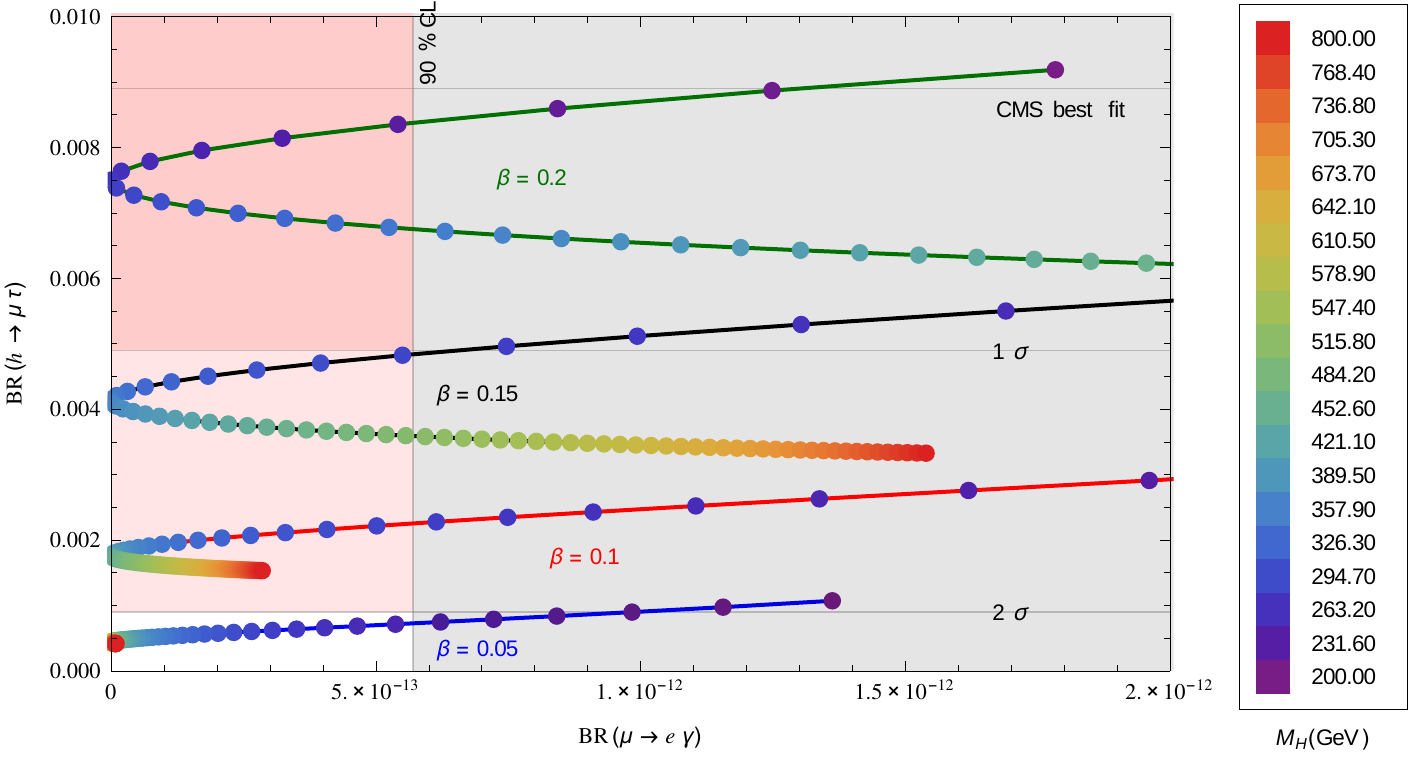}
\caption[]{Branching ratios of $h \to \mu \tau$ vs.\ $\mu \rightarrow e \gamma$. Horizontal lines are best-fit value and the $1 \sigma$ or $2\sigma$ ranges for the Higgs branching ratio, see \Eqref{eq:yeah}. The vertical line is the MEG bound on 
$\mu \rightarrow e \gamma$~\cite{Adam:2013mnn}. The various lines correspond to the different values for $\beta$ indicated in the plot; color coding is in $m_H$. $m_A$ is fixed to $m_A=600 \GeV$.
 }
\label{fig:BRplot2}
\end{figure}

\begin{figure}[tbh]
\centering
\includegraphics[width=0.85\linewidth]{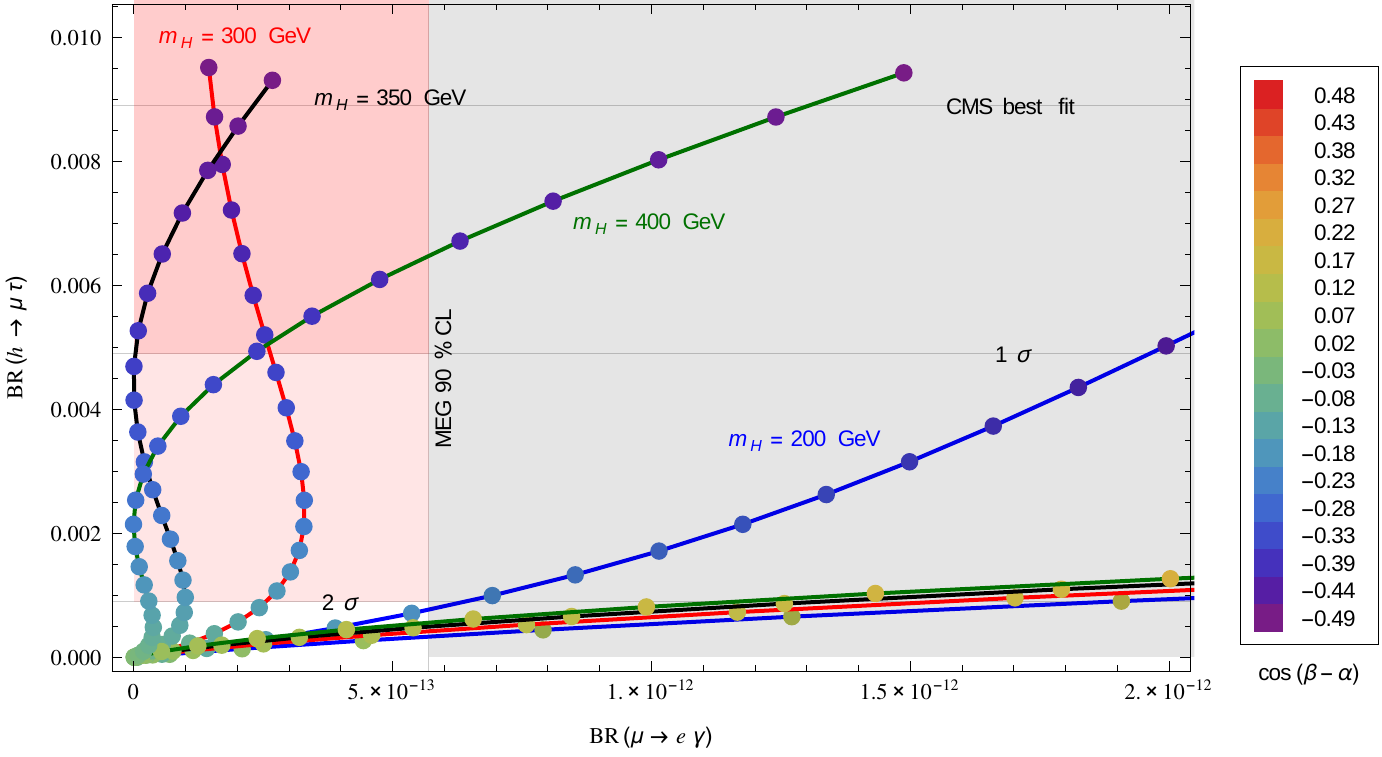}
\caption[]{Branching ratios of $h \to \mu \tau$ vs.\ $\mu \rightarrow e \gamma$. Horizontal lines are best-fit value and the $1 \sigma$ or $2\sigma$ ranges for the Higgs branching ratio, see \Eqref{eq:yeah}. The vertical line is the MEG bound on 
$\mu \rightarrow e \gamma$~\cite{Adam:2013mnn}. The various lines correspond to the different values for $m_H$ indicated in the plot; color coding is in $\cos(\beta-\alpha)$. $m_A$ is fixed to $m_A=600 \GeV$.
}
\label{fig:BRplot}
\end{figure}

We find the most constraining bound to come from $\mu \rightarrow e\gamma$ with recent MEG result~\cite{Adam:2013mnn}
\begin{align}
\BR (\mu\to e \gamma) <  5.7 \times 10^{-13} \text{ at } 90\% \text{ C.L.,}
\end{align}
see Fig.~\ref{fig:parameter-space2}. We also plot the relevant branching ratios, $\BR (h\to\mu\tau)$ and $\BR (\mu \to e\gamma)$, against each other in Fig.~\ref{fig:BRplot2} and Fig.~\ref{fig:BRplot}. 
The MEG bound is so strong that it forbids a resolution of the CMS excess for positive $\cos (\beta-\alpha)$ (negative $\beta$);
A cancellation among the scalar contributions to $\mu\to e\gamma$ occurs however for negative $\cos(\beta-\alpha)$ (positive $\beta$) for $m_H \simeq 300$--$\unit[400]{GeV}$, opening up parameter space in CMS' $1\sigma$ region for $c_{\beta-\alpha} \simeq -0.4$.

Note that two-loop contributions to the radiative lepton decays $\ell_i \to \ell_j \gamma$ can be
dominant in some parts of parameter space because the stronger scalar
coupling to top quarks or vector bosons compared to leptons can
compensate the additional loop suppression. Since this requires a specific model
for the quark couplings we do not take it into account here, but this
will pose a challenge for the way quarks are included in the model;
Ref.~\cite{Harnik:2012pb} found that an SM-like $h$ with LFV couplings
would have dominating two-loop contributions to $\mu\to e\gamma$,
ultimately resulting in $\BR (h\to e \mu) \lesssim 10^{-8}$, orders of magnitude below our prediction (see Fig.~\ref{fig:BRvsCos}). We most
likely need some fine-tuning to suppress $\mu\to
e\gamma$ in our $A_4$ model once we take quark couplings and two loops
into account. 

The constraint from $\mu\to e\gamma$ is shown in Fig.~\ref{fig:parameter-space2} on top of the relevant parameter space for $h\to\mu\tau$. For further visualization of the parameter space and constraints we directly plot $\BR (h\to\mu\tau)$ against $\BR (\mu\to e\gamma)$, fixing either $\beta$ (Fig.~\ref{fig:BRplot2}) or $m_H$ (Fig.~\ref{fig:BRplot}). We observe again the cancellation that suppresses $\mu\to e\gamma$ for certain values of $m_H$ and $\beta$.
Seeing as our model demands a large $\cos (\beta-\alpha) \sim -0.4$ to explain the CMS excess in $h\to\mu\tau$ and a rather light 'heavy' Higgs $m_H \simeq 280$--$\unit[380]{GeV}$ for sufficient one-loop cancellation of $\mu\to e\gamma$, it might be possible to see $H$ at the LHC. The specifics depend strongly on the employed quark couplings and will be left for a future publication.

\clearpage

\section{Abelian case: An \texorpdfstring{$L_\mu - L_\tau$}{Lmu-Ltau} example}
\label{sec:LmuLtau}

In the second part of this paper, we study the realization of $h\to\mu\tau$ in the framework of \emph{Abelian} flavor symmetries, specifically $U(1)_{L_\mu-L_\tau}$. Not only is this an anomaly-free global symmetry within the SM~\cite{He:1990pn, Foot:1990mn, He:1991qd}, it is also a good zeroth-order symmetry for neutrino mixing with a quasi-degenerate mass spectrum, predicting maximal atmospheric and vanishing reactor mixing angles~\cite{Binetruy:1996cs, Bell:2000vh, Choubey:2004hn}. Breaking of $L_\mu-L_\tau$ is, of course, necessary for a viable neutrino sector, and can also induce the $\Delta (L_\mu-L_\tau) = 2$ process 
$h\to \mu \tau$, as we will show below. This will also lead to the lepton-flavor-violating decays $\tau\to3\mu$ and $\tau\to \mu\gamma$~\cite{Dutta:1994dx,Heeck:2011wj}.
Since the $Z'$ of a gauged $U(1)_{L_\mu-L_\tau}$ does not couple to first generation fermions, the experimental limits are not as stringent as for other $U(1)'$ models, and it might even be possible to use (a light) $Z'$ to resolve the longstanding $3$--$4\sigma$ anomaly surrounding the muon's magnetic moment~\cite{Gninenko:2001hx, Baek:2001kca, Ma:2001md, Ma:2002df, Heeck:2011wj, Altmannshofer:2014cfa, Altmannshofer:2014pba,Araki:2014ona, Gninenko:2014pea}. An even lighter $Z'$ may induce long-range forces modifying neutrino oscillations~\cite{Heeck:2010pg}, although this is not the limit of interest here.

We work within gauged $U(1)_{L_\mu-L_\tau}$ with three right-handed neutrinos $N_{e,\mu,\tau}$, qualitatively similar to Ref.~\cite{Heeck:2011wj}.
For symmetry breaking, we introduce two scalar doublets $\Phi_{1,2}$, with $L_\mu-L_\tau$ charge~$-2$ and~$0$, respectively, as well as an SM-singlet scalar $S$ with $L_\mu-L_\tau$ charge~$+1$ (see Tab.~\ref{tab:partcontent-LmuLtau}). 
A small VEV of $\Phi_1$ -- induced by the larger VEV of $S$ that generates right-handed neutrino masses -- will break $L_\mu-L_\tau$ by two units in the charged-lepton sector and subsequently lead to the LFV decay mode $h\to \mu \tau$. A particular feature of this model is  LFV only in the $\mu\tau$ sector, evading strong constraints from, e.g., $\mu\to e\gamma$. This is opposite to the model Ref.~\cite{Heeck:2011wj}, where $\Phi_1$ was given the $L_\mu-L_\tau$ charge $+1$, leading to charged-lepton processes with $\Delta (L_\mu-L_\tau) = \pm 1$, with $\Delta (L_\mu-L_\tau) = \pm 2$ being highly suppressed. We will comment on variations of our model in Sec.~\ref{sec:variants}, which have a similar structure but different phenomenology.

\subsection{Scalar potential and Yukawa couplings}

\begin{figure}
\centering
\includegraphics[width=0.5\textwidth]{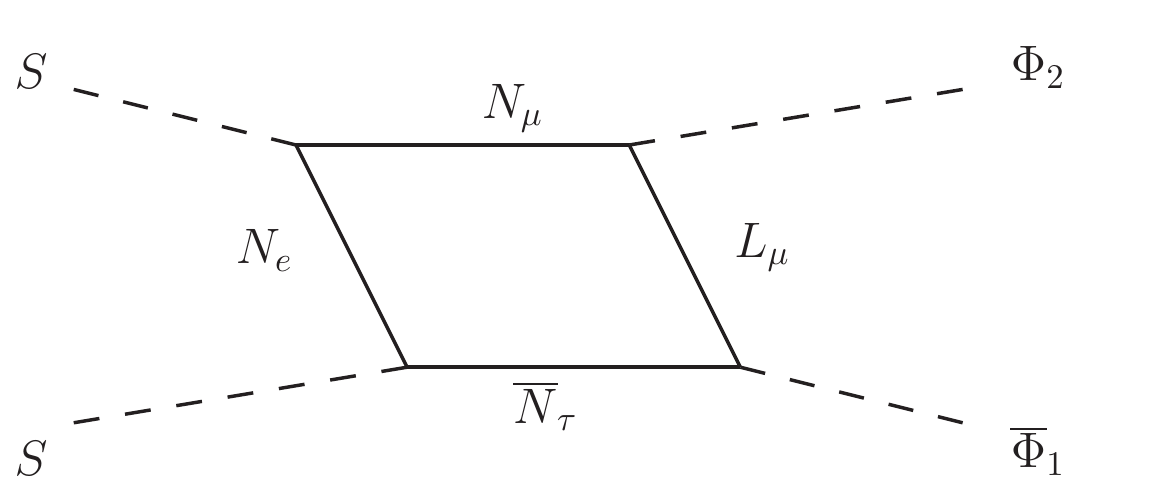}
\caption{Example of a loop contribution to the scalar coupling $S^2 \Phi_2^\dagger \Phi_1$ that generates $\langle \Phi_1 \rangle \neq 0$ and ultimately $h\to\mu\tau$.}
\label{fig:delta_operator_loop}
\end{figure}

With the particle content from Tab.~\ref{tab:partcontent-LmuLtau}, the scalar potential takes the form
\begin{align}
V(\Phi_1, \Phi_2,S) &=  m_1^2 |\Phi_1|^2 + \tfrac{\lambda_1}{2} |\Phi_1|^4- m_2^2 |\Phi_2|^2 + \tfrac{\lambda_2}{2} |\Phi_2|^4 + \lambda_3 |\Phi_1|^2 |\Phi_2|^2 + \lambda_4 |\Phi_1^\dagger \Phi_2|^2\nonumber \\
&\quad -\mu_S^2 |S|^2 + \tfrac{\lambda_S}{2}|S|^4 + \lambda_{\Phi_1 S} |\Phi_1|^2 |S|^2 + \lambda_{\Phi_2 S} |\Phi_2|^2 |S|^2\\
&\quad - \delta \ S^2 \Phi_2^\dagger \Phi_1  +\hc\nonumber
\end{align}
The scalar $S$ acquires a high-scale VEV, and for simplicity we assume it also to be heavy and have negligible mixing with the other scalars (similar to the flavon field $\Phi$ in the $A_4$ model, see Eq.~\eqref{eq:MSsquared}). In this limit, we can simply consider the effective 2HDM potential (after renaming coefficients)
\begin{align}
V(\Phi_1, \Phi_2) &\simeq  m_1^2 |\Phi_1|^2 + \tfrac{\lambda_1}{2} |\Phi_1|^4- m_2^2 |\Phi_2|^2 + \tfrac{\lambda_2}{2} |\Phi_2|^4 + \lambda_3 |\Phi_1|^2 |\Phi_2|^2 + \lambda_4 |\Phi_1^\dagger \Phi_2|^2\\
&\quad - m_3^2 \Phi_2^\dagger \Phi_1  +\hc ,
\end{align}
which is just a $U(1)$-invariant 2HDM~\cite{Branco:2011iw}, softly broken by the mass-mixing term in the last line, $m_3^2 \equiv \delta \langle S\rangle^2$, again similar to the soft-breaking term $M_S^2$ in the $A_4$ potential.\footnote{The model can also be identified with a CP-conserving 2HDM with softly broken $Z_2$ symmetry and $\lambda_5 = 0$~\cite{Branco:2011iw}.} Our choice $U(1)_{L_\mu-L_\tau}$ acts here as a very simple anomaly-free horizontal symmetry in the scalar potential (see Ref.~\cite{Ko:2012hd} for other $U(1)_H$ choices).
In Sec.~\ref{sec:neutrinomasses} we will see that $\langle S \rangle$ contributes to the right-handed neutrino masses, and is therefore expected to be close to the seesaw scale, at least $\langle S \rangle \gg v$. We work with a low-scale seesaw in mind in order to have more interesting $Z'$ phenomenology ($M_{Z'} \simeq g' \langle S \rangle$), but a high-scale seesaw is of course possible. In this case, $\delta$ might have to be chosen very small if we still want the new scalars to be at the electroweak scale. In this regard we note that $\delta \to 0$ would lead to an additional global $U(1)$ symmetry in the scalar potential, but \emph{not} in the full Lagrangian, so a small $\delta$ is not technically natural; loop contributions to this operator arise at one loop, see for example Fig.~\ref{fig:delta_operator_loop}.

\begin{table}
\centering
\ra{1.2}
\begin{tabular}{lcccccccccccc}\toprule
 &$ L_e$&$ L_\mu$&$ L_\tau$ &$e_R$ &$\mu_R$ &$\tau_R$ &$N_e$&$N_\mu$&$N_\tau$&$\Phi_1$&$\Phi_2$ &$S$ \\ \midrule
$\U{1}_{L_\mu-L_\tau}$& 0 & 1 & $-1$ & 0 & 1 & $-1$ & 0 & 1 & $-1$ & $-2$ & 0 & 1\\
$\SU{2}_L$&$2$&$2$ &$2$&$1$&$1$&$1$&$1$&$1$&$1$&2 & 2 & 1\\
$\U{1}_Y$&$-1/2$&$-1/2$&$-1/2$&$-1$ &$-1$&$-1$&$0$&$0$&$0$&$1/2$&$1/2$& 0\\
\bottomrule
\end{tabular}
\caption{Particle content of the $L_\mu-L_\tau$ model; quarks are uncharged under the new $U(1)$. $\Phi_j$ and $S$ denote the scalar bosons of the model, uncharged under the color group $\SU{3}_C$.
\label{tab:partcontent-LmuLtau}}
\end{table}
\ra{1}

With positive $m_{1,2}^2$, $\Phi_2$ acquires a VEV from its Mexican-hat potential, and the $m_3^2$ term subsequently induces a small VEV for $\Phi_1$: $\langle \Phi_1\rangle \simeq \langle \Phi_2\rangle m_3^2/m_1^2$, where we neglected the portal couplings $\lambda$. We will assume the hierarchy $\tan\beta \equiv\vev{\Phi_2}/\vev{\Phi_1}= v_2/v_1 \gg 1$ in the following, as this suffices for our purposes. Again neglecting the portal terms, the new scalars contained in $\Phi_1$, namely the heavy CP-even $H$, the CP-odd $A$, and the charged $H^+$, are then degenerate with mass $m_A^2 = m_3^2/s_\beta c_\beta \simeq m_1^2$.

More accurately, the charged scalar has mass $m_+^2 = m_A^2 - \lambda_4 v^2$, whereas the neutral CP-even scalars $h_j$ inside $\Phi_j = (\phi_j^+, (v_j+h_j-i z_j)/\sqrt{2})^T$ mix according to the symmetric mass matrix
\begin{align}
\matrixx{\lambda_1 v_1^2 + m_3^2 t_\beta & (\lambda_3+\lambda_4) v_1 v_2 -m_3^2\\\cdot & \lambda_2 v_2^2 + m_3^2/t_\beta} \equiv \matrixx{c_\alpha & -s_\alpha \\ s_\alpha & c_\alpha} \matrixx{m_h^2 & \\ & m_H^2} \matrixx{c_\alpha & s_\alpha \\ -s_\alpha & c_\alpha},
\end{align}
which leads to a light SM-like scalar $h = c_\alpha h_2 - s_\alpha h_1$ and a heavy $H= c_\alpha h_1 + s_\alpha h_2$. 
Note that the seven real parameters of the potential can be expressed in terms of the physical quantities $m_h$, $m_H$, $m_A$, $m_+$, $v$, $\tan\beta$, and $\alpha$ (see, e.g., Ref.~\cite{Kanemura:2004mg} for the interchangeable parameter sets).
The couplings of $h$ ($H$) to $ZZ$ and $WW$ are given by their respective SM values times $\sin (\beta-\alpha)$ ($\cos (\beta-\alpha)$), as usual in 2HDMs, so the value $\beta-\alpha = \pi/2$ makes $h$ SM-like and decouples $H$ in the gauge boson sector. The limit $\beta-\alpha = \pi/2$ also reduces all fermion couplings of $h$ to their SM values, but does not decouple $H$ from them.

Yukawa couplings and fermion mass terms are dictated by the quantum numbers of Tab.~\ref{tab:partcontent-LmuLtau} and take the form
\begin{align}
-\L_Y &=  \overline{L}_L Y_{\ell_1} \Phi_1 \ell_R + \overline{L}_L Y_{N_1} \tilde{\Phi}_1 N_R \nonumber \\
&\quad + \overline{L}_L Y_{\ell_2} \Phi_2 \ell_R+ \overline{L}_L Y_{N_2} \tilde{\Phi}_2 N_R + \overline{Q}_L Y_u \tilde{\Phi}_2 u_R + \overline{Q}_L Y_d \Phi_2 d_R \\
&\quad + \tfrac{1}{2} \overline{N}^c_R \mathcal{M}_N N_R + \tfrac{1}{2}\overline{N}^c_R Y_{S_1} S N_R + \tfrac{1}{2}\overline{N}^c_R Y_{S_2} \overline{S} N_R +\hc\nonumber
\end{align}
The quark Yukawa matrices $Y_{u,d}$ are arbitrary and quarks will receive their masses (and CKM mixing) just from $\Phi_2$. Consequently, there are no tree-level flavor-changing neutral currents in the quark sector, as enforced by our $U(1)$ gauge symmetry.
In the lepton sector, on the other hand, the $U(1)_{L_\mu-L_\tau}$ symmetry enforces diagonal $\Phi_2$ couplings
\begin{align}
Y_{\ell_2} = \diag (y_e, y_\mu, y_\tau)\,, \quad Y_{N_2} = \diag (y_1,y_2,y_3)\,,
\end{align}
and even more selective $\Phi_1$ couplings
\begin{align}
Y_{\ell_1} = \matrixx{ 0 & 0 & 0\\ 0 & 0 & 0\\ 0 & \xi_{\tau\mu} & 0} , \quad Y_{N_1} = \matrixx{ 0 & 0 & 0\\ 0 & 0 & \xi_{23}\\ 0 & 0 & 0} .
\label{eq:Phi1couplings}
\end{align}
It is the off-diagonal $\tau$--$\mu$ entry in $Y_{\ell_1}$ that will ultimately lead to $h\to\mu\tau$.
The right-handed neutrino Majorana mass matrix will be build from the pieces
\begin{align}
\M_N = \matrixx{M_1 & & \\ & & M_2\\ & M_2 & } , \quad Y_{S_1} = \matrixx{ &  & a_{13}\\  &  & \\ a_{13} &  & }, \quad Y_{S_2} = \matrixx{ & a_{12} &  \\ a_{12} &  & \\   &  & }.
\label{eq:yukawasN}
\end{align}
Overall we recognize our model as being basically a 2HDM of type I (albeit slightly restricted in the scalar potential through $\lambda_5 = 0$~\cite{Branco:2011iw}), plus $\Phi_1$ interactions~\eqref{eq:Phi1couplings} that exclusively modify the $\mu\tau$ lepton sector. Since only the off-diagonal charged-lepton coupling $\overline{\tau}_L \mu_R$ to scalars exists, as enforced by the gauge symmetry $U(1)_{L_\mu-L_\tau}$, our model provides a very minimal explanation for the fact that flavor violation has only been (potentially) observed in the $\mu\tau$ sector. No flavor-changing neutral currents arise in the quark sector, nor will $\Delta (L_\mu-L_\tau) = \pm 1$ processes such as $\mu\to e\gamma$ be generated at an observable rate.

\subsection{Neutrino masses and mixing}
\label{sec:neutrinomasses}

Having defined our setup, let us first take a look at neutrino masses and mixing, which serve as a major motivation for $L_\mu-L_\tau$, independent of any charged-lepton flavor violation.
The symmetric right-handed neutrino mass matrix has contributions from $L_\mu-L_\tau$ symmetric parts ($M_{1,2}$) and $\Delta (L_\mu-L_\tau) = \pm 1$ pieces induced by the VEV $\langle S \rangle$:
\begin{align}\label{eq:MR}
M_N \equiv \M_N + \langle S\rangle Y_{S_1}+ \langle S\rangle Y_{S_2} =  \matrixx{M_1 & a_{12} \langle S \rangle & a_{13} \langle S \rangle\\ \cdot & 0 & M_2\\ \cdot & \cdot  & 0} .
\end{align}
The Dirac mass matrix, coupling $N_R$ to the active left-handed neutrinos, similarly contains a diagonal $L_\mu-L_\tau$ symmetric part and one off-diagonal term generated by $\langle \Phi_1\rangle$:
\begin{align}\label{eq:mD}
m_D = \langle \Phi_2\rangle Y_{N_2}+\langle \Phi_1\rangle Y_{N_1} =  \frac{v}{\sqrt{2}} \matrixx{y_1 s_\beta & & \\ & y_2 s_\beta & \xi_{23} c_\beta \\ &  & y_3 s_\beta} .
\end{align} 
In the seesaw limit, we obtain the active-neutrino Majorana mass matrix $\overline{\nu}^c_L \M_\nu \nu_L$
\begin{align}
 \mathcal{M}_\nu \simeq -m_D^* (M_N^*)^{-1} m_D^\dagger \,.
 \label{eq:mass_matrix}
\end{align} 
Diagonalization $V_\nu^T \M_\nu V_\nu= \diag (m_{\nu_1},m_{\nu_2},m_{\nu_3})$ then gives the neutrino mixing matrix $V_\nu$, which forms the PMNS mixing matrix together with the charged-lepton contribution $V_{e_L}$ via $U = V_{e_L}^\dagger V_\nu$. As we will see in Sec.~\ref{sec:chargedleptonmasses}, $V_{e_L}$ consists only of a 23 rotation, so it influences exclusively the atmospheric mixing angle $\theta_{23}$ (in the standard parametrization for $U$).

The Abelian gauge group $U(1)_{L_\mu-L_\tau}$ obviously introduces a structure in $\M_\nu$, making it a \emph{flavor} symmetry. Without the breaking terms from $\vev{S}$ and $\vev{\Phi_1}$, $\M_\nu$ would have the same $L_\mu-L_\tau$ symmetric structure as $\M_N$ in Eq.~\eqref{eq:yukawasN}, predicting maximal atmospheric mixing and vanishing solar and reactor angles, as well as $\Delta m_{32}^2 \equiv m_{\nu_3}^2-m_{\nu_2}^2= 0$~\cite{Choubey:2004hn}. 
The $U(1)_{L_\mu-L_\tau}$ breaking terms are thus needed in order to accommodate the observed neutrino parameters, and it has been shown that small perturbations suffice to obtain viable mass splittings and angles for quasi-degenerate neutrinos~\cite{Choubey:2004hn, Heeck:2011wj}.

Let us briefly comment on our specific setup:
$M_N$ has two texture zeroes due to the $\Delta (L_\mu-L_\tau) = \pm 1$ breaking structure of $S$.
With $\xi_{23}=0$ (or $\langle \Phi_1\rangle = 0$), these two texture zeroes propagate to $\M_\nu$ as two vanishing minors, i.e.~$(\M_\nu^{-1})_{33} = 0 = (\M_\nu^{-1})_{22}$~\cite{Araki:2012ip}, as is obvious from the seesaw formula. These two vanishing minors ultimately imply correlations among the mixing angles and phases~\cite{Lavoura:2004tu, Lashin:2007dm}, and our texture is among the seven patterns that are compatible with current data~\cite{Araki:2012ip}. One of the vanishing minors survives even for $\xi_{23}\neq 0$, namely $(\M_\nu^{-1})_{22} = 0$, which leads to a weaker relation among the parameters~\cite{Lashin:2009yd}. The $\Delta (L_\mu-L_\tau) = 2$ perturbation from the Dirac sector thus helps to reduce the required fine-tuning otherwise necessary for viable two-texture-zero/two-vanishing-minor neutrino mass matrices. The corrections from the charged-lepton sector, $U = V_{e_L}^\dagger V_\nu$, only influence the atmospheric mixing angle, leaving $\theta_{12}$ and $\theta_{13}$ to be determined solely from $\M_\nu$. Since the left-handed charged-lepton mixing angle will turn out to be small, see Eq.~\eqref{eq:charged_lepton_angles}, even the atmospheric mixing angle is essentially determined by the neutrino mass matrix only, and hence expected to be close-to-maximal due to the approximate $L_\mu-L_\tau$ structure.

\begin{figure}
\centering
\includegraphics[width=0.7\textwidth]{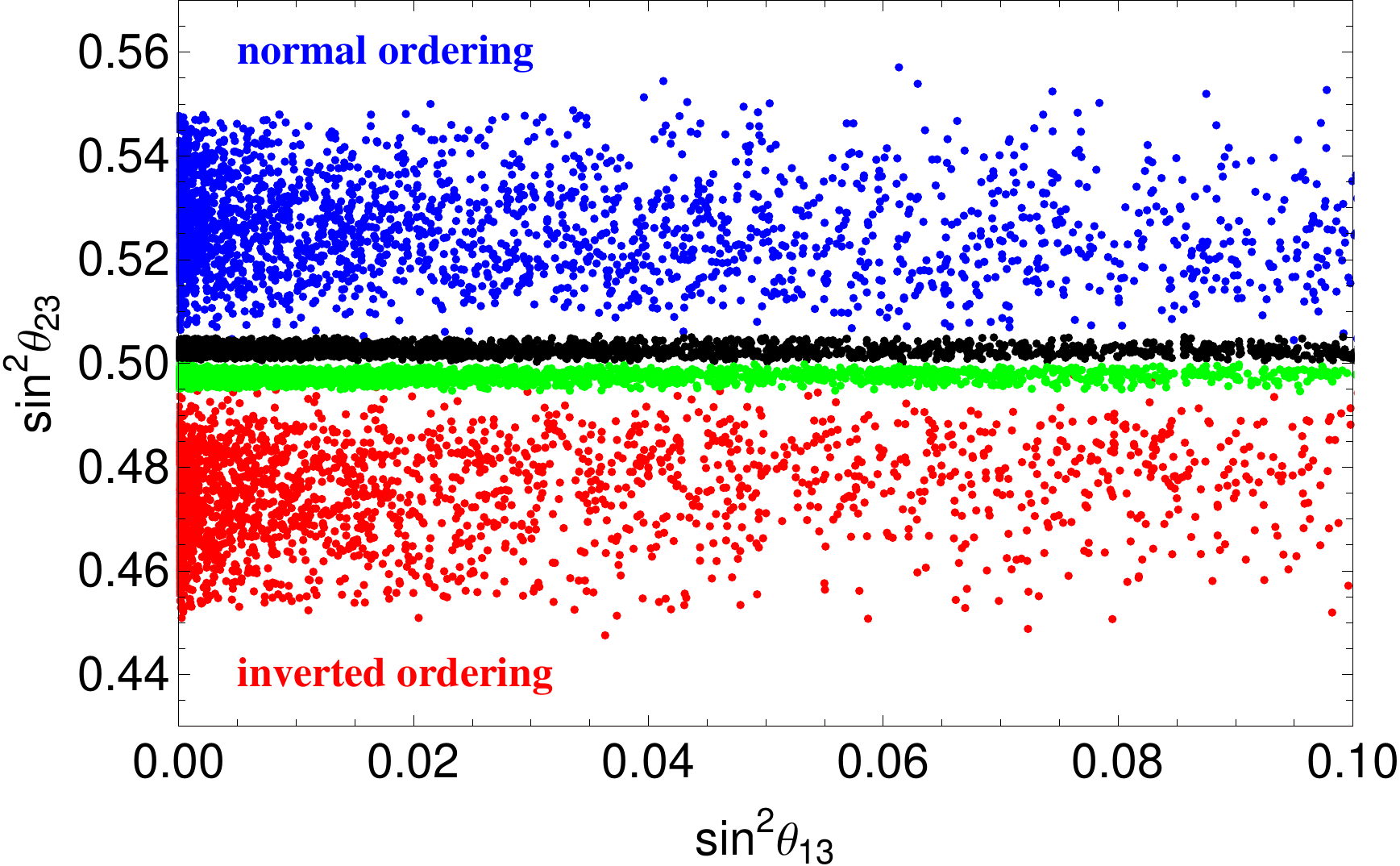}
\caption{Scatter plot of neutrino mixing angles $\theta_{23,13}$ of $\M_\nu$ (Eq.~\eqref{eq:mass_matrix}). The red (blue) points have $\epsilon = 0.1$ and inverted (normal) ordering, the green (black) points have $\epsilon = 0.01$ and inverted (normal) ordering.
The global-fit $3\sigma$ ranges are $\sin^2\theta_{23} = 0.385$--$0.644$ and $\sin^2\theta_{13} = 0.0188$--$0.0251$~\cite{Gonzalez-Garcia:2014bfa}.
}
\label{fig:scatterLmLt}
\end{figure}

We visualize the dependence of $\theta_{23}$ on the breaking scale with a scatter plot in Fig.~\ref{fig:scatterLmLt}. Here, the $L_\mu-L_\tau$ symmetric entries in $m_D$ and $M_N$ are generated with absolute values $\in [1,3]$ and random phases in order to generate the desired quasi-degenerate neutrino mass spectrum. Entries breaking $L_\mu-L_\tau$, i.e.~$(M_N)_{12,13}$ and $(m_D)_{23}$ in \eqref{eq:MR} and \eqref{eq:mD}, are taken random $\in [0,\epsilon]$ with random phases. We impose the $3\sigma$ constraints on $\sin\theta_{12}$ and $\Delta m_{21}^2/\Delta m_{31}^2$ from Ref.~\cite{Gonzalez-Garcia:2014bfa}. The units/scale of $m_D$ and $M_N$, and hence $\M_\nu$, are not fixed, but we expect a quasi-degenerate spectrum, i.e.~$m_{\nu_j}\simeq 0.1$--$\unit[1]{eV}$, with $0\nu 2\beta$ rates testable in the near future~\cite{Rodejohann:2011mu}. 
As can be seen, normal ordering (NO) and inverted ordering (IO) correspond to $s_{23}^2 >1/2$ and $s_{23}^2<1/2$, respectively, with perturbations $|s_{23}^2 - 1/2|\lesssim \epsilon/2$.\footnote{The relation of the octant of $\theta_{23}$ and the mass ordering is dictated by our breaking structure in $m_D$, leading to $(\M_\nu^{-1})_{22} = 0$. Had we chosen an $L_\mu-L_\tau$ charge $+2$ instead of $-2$ for $\Phi_1$, effectively replacing $m_D$ by its transpose, the correlation would be flipped, i.e.~$s_{23}^2 >1/2$ ($<1/2$) for inverted (normal) ordering, the vanishing minor being $(\M_\nu^{-1})_{33} = 0$. We will comment on this scenario in Sec.~\ref{sec:variants}.} Note that the global fit of Ref.~\cite{Gonzalez-Garcia:2014bfa} prefers $s_{23}^2 < 1/2$ ($>1/2$) for normal (inverted) ordering, opposite to our prediction; while this is not statistically relevant at the moment, it will become an important constraint on the breaking structure of $L_\mu-L_\tau$ in the future (note that other global fits have different preferences~\cite{Capozzi:2013csa, Forero:2014bxa}).

Ultimately, the relation $s_{23}^2 \lessgtr 1/2$ (NO/IO) is fixed by the chiral structure in $h\to \overline{\mu} P_{L,R} \tau$ in our $L_\mu-L_\tau$ model, at least if the charged-lepton contribution to $\theta_{23}$ is small. As we will see in Sec.~\ref{sec:variants}, a flip in the $L_\mu-L_\tau$ charge of $\Phi_1$ will modify the chiral structure to $h\to \overline{\mu} P_R \tau$ and give $s_{23}^2 < 1/2$ for NO, opposite to the case discussed above. Determination of the $\theta_{23}$ octant as well as the mass ordering are hence important discriminators for our model.

Note that the generic perturbations required for successful neutrino phenomenology are in the range $\langle S \rangle /M_{1,2} \simeq 10^{-1}$--$10^{-2}$ (Yukawa couplings of order one), so a lower bound on $\langle S \rangle \simeq M_{Z'}/g'$ also implicitly bounds the seesaw scale $M_{1,2}$.

As a final remark, the solar mixing angle $\theta_{12}$ vanishes in the $L_\mu-L_\tau$-symmetric limit, but can easily be large for a quasi-degenerate neutrino spectrum. In the $\mu$--$\tau$-symmetric case with $\theta_{13}=0$ and $\theta_{23}=\pi/4$, it is given in the form
\begin{align}
\tan 2\theta_{12} \propto \frac{\langle S \rangle /M_{1,2}}{1- (\M_\nu)_{11}/(\M_\nu)_{23}} \,,
\end{align}
$(\M_\nu)_{11,23}$ being the $L_\mu-L_\tau$-symmetric entries of $\M_\nu$, naturally of similar magnitude. $\theta_{12}$ is hence given by the ratio of two small numbers~\cite{Choubey:2004hn}: $U(1)_{L_\mu-L_\tau}$-breaking entries and deviations from degeneracy. The expressions for $\theta_{12}$ become more intricate for the realistic cases with $\theta_{13}\neq 0$, but remain qualitatively similar.

\subsection{Charged lepton masses}
\label{sec:chargedleptonmasses}

After spontaneous symmetry breaking of $SU(2)_L\times U(1)_Y \times U(1)_{L_\mu-L_\tau}$, the charged-lepton mass matrix takes the form
\begin{align}
\begin{split}
M_e = \frac{v}{\sqrt{2}}\matrixx{ y_e s_\beta & & \\ & y_\mu s_\beta & \\ & \xi_{\tau\mu} c_\beta & y_\tau s_\beta} 
&\equiv \matrixx{1 & & \\ & c_L & s_L\\ & -s_L & c_L} \matrixx{m_e & & \\ & m_\mu & \\ & & m_\tau}  \matrixx{1 & & \\ & c_R & -s_R\\ & s_R & c_R} \\
&\equiv V_{e_L} \diag (m_e,m_\mu,m_\tau) V_{e_R}^\dagger\,,
\end{split}
\end{align}
which defines the charged-lepton mass basis via $\ell_L^0 = V_{e_L}^\dagger \ell_L$, $\ell_R^0 = V_{e_R}^\dagger \ell_R$. We assumed all Yukawa couplings to be real here, so the two new mixing angles $\theta_{L,R}$ are given by
\begin{align}
\frac{\tan \theta_L}{\tan\theta_R} = \frac{m_\mu}{m_\tau} \ll 1  \quad \text{and}\quad
\sin \theta_R \simeq \frac{v}{m_\tau} \frac{\xi_{\tau\mu}}{\sqrt{2}} \cos\beta \,,
\label{eq:charged_lepton_angles}
\end{align}
and replace $\xi_{\tau\mu}$ as a parameter. Note that $\theta_L$ is automatically small and does not play a significant role for neutrino mixing.

\subsection{Lepton flavor violating interactions}

\begin{figure}
\centering
\includegraphics[width=0.7\textwidth]{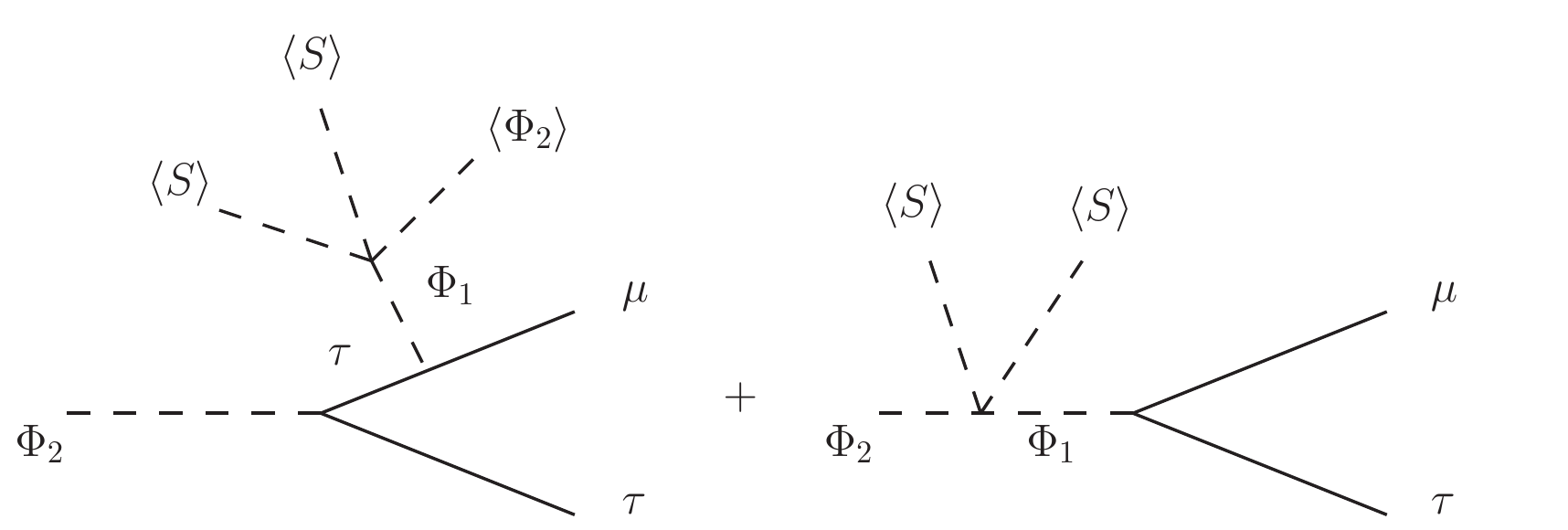}
\caption{Pictorial depiction of the $U(1)_{L_\mu-L_\tau}$-breaking structure behind $h\to \mu\tau$.}
\label{fig:vev_insertions}
\end{figure}

The couplings of the light scalar $h$ to the lepton mass eigenstates, $-\L_Y\supset \overline{\ell}_L^0 y \ell_R^0 h$, can be calculated in a straightforward manner to be
\begin{align}
y \simeq \diag ( m_e, m_\mu, m_\tau) \frac{c_\alpha}{v s_\beta} - s_R \frac{m_\tau}{v} \frac{\cos (\alpha-\beta)}{c_\beta s_\beta} \matrixx{0 & & \\ & -c_R s_L & -s_L s_R\\ & c_L c_R & c_L s_R} .
\label{eq:hcoupling}
\end{align} 
The first (diagonal) term corresponds to the standard type-I 2HDM couplings, proportional to $c_\alpha/s_\beta$, just like in the quark sector~\cite{Branco:2011iw}; the second matrix is proportional to $\xi_{\tau\mu}$ (or, equivalently, $\sin\theta_R$) and obviously induces the desired lepton flavor violation in the $\mu\tau$ sector. The $\tau\mu$ entry is dominant due to the mixing-angle hierarchy $s_L\ll s_R$, as expected from the structure of the Yukawa couplings in $Y_{\ell_1}$.
Writing $\cos (\alpha-\beta) =\cos\alpha\cos\beta + \sin\alpha \sin \beta $ allows us to picture the induced $h\to \mu\tau$ as a combination of lepton mixing (proportional to the type-I $c_\alpha/s_\beta$) and scalar mixing ($s_\alpha/c_\beta$), see Fig.~\ref{fig:vev_insertions}.

Note that the choice $\cos(\beta - \alpha) = 0$ reduces all couplings of $h$ (to vector bosons and fermions) to their SM values~\cite{Branco:2011iw}, and eliminates the LFV couplings. In order to obtain a measurable $h\to\mu\tau$ rate, we thus need to allow for $\cos (\alpha-\beta) \neq 0$ and hence generically predict (slightly) reduced $h$ rates compared to the SM. This is easily allowed in our type-I-like 2HDM even in the limit of large $\tan\beta$, as discussed below in Sec.~\ref{sec:constraints}.

The coupling of the heavy scalar to leptons ($-\L_Y\supset\overline{\ell}_L^0 y^H \ell_R^0 H$) follows from the $h$ couplings with $\alpha\to \alpha-\pi/2$:
\begin{align}
y^H \simeq \diag ( m_e, m_\mu, m_\tau) \frac{s_\alpha}{v s_\beta} - s_R \frac{m_\tau}{v} \frac{\sin (\alpha-\beta)}{c_\beta s_\beta} \matrixx{0 & & \\ & -c_R s_L & -s_L s_R\\ & c_L c_R & c_L s_R} .
\end{align} 
In particular, the $\tau\mu$ entry is given by $y^H_{\tau\mu} =t_{\alpha-\beta} y^h_{\tau\mu} $. Also note the chiral nature of these couplings, due to our non-Hermitian coupling matrix.

The charged-scalar couplings to active neutrinos take the form
\begin{align}
-\L_Y &\supset \frac{\sqrt{2}}{v t_\beta}\ \overline{\nu}_L^0 U^\dagger \diag (m_e,m_\mu,m_\tau) \ell_R^0 H^+
-\frac{1}{s_\beta}\ \overline{\nu}_L^0 V_\nu^\dagger Y_{\ell_1} V_{e_R} \ell_R^0 H^+ + \hc
\end{align}
with $\overline{\nu} \mu_R H^+$ dominating in the large $\tan\beta$ limit, as expected. The first term is again the type-I coupling, while the second one has the by now familiar matrix structure
\begin{align}
-\frac{\xi_{\tau\mu}}{s_\beta}\ \overline{\nu}_L^0 U^\dagger \matrixx{0 & & \\ & -c_R s_L & -s_L s_R\\ & c_L c_R & c_L s_R} \ell_R^0 H^+ \,.
\end{align}
Pseudoscalar interactions, following our parametrization $\Phi_j = (\phi_j^+, (v_j+h_j-i z_j)/\sqrt{2})^T$ and $A \equiv c_\beta z_2 - s_\beta z_1$, are given by
\begin{align}
-\L_Y \supset  \left(\frac{-i}{v t_\beta}\right) \overline{\ell}^0_L \diag (m_e,m_\mu,m_\tau) \ell^0_R \ A +\left(\frac{i \xi_{\tau\mu} }{\sqrt{2} s_\beta}\right) \overline{\ell}^0_L \matrixx{0 & & \\ & -c_R s_L & -s_L s_R\\ & c_L c_R & c_L s_R}\ell^0_R \ A +\hc
\end{align}
Again, the dominant term in the large $\tan\beta$ limit is $i \xi_{\tau\mu}/\sqrt{2}\ \overline{\tau} P_R \mu A$. Note the chiral nature even of the pseudoscalar couplings (only in the off-diagonal couplings since we assume $\xi_{\tau\mu}$ to be real for simplicity).

\subsection{Hints and constraints}
\label{sec:constraints}

Since we have basically a type-I 2HDM (slightly restricted via $\lambda_5=0$, which however barely changes the phenomenology~\cite{Ko:2013zsa}), we inherit the bounds on masses and mixing angles from  Ref.~\cite{Dumont:2014wha, *Dumont:2014kna}. We only have to worry about the new interactions we introduced in the $\mu\tau$ sector.

Without going into any details, let alone a scan of the huge 2HDM parameter space, we simply take $|\cos(\beta-\alpha)| \lesssim 0.4$ for $\tan\beta\gtrsim 3$, following the recent scan of the type-I 2HDM with LHC bounds from Ref.~\cite{Dumont:2014wha, *Dumont:2014kna}. This means in particular $A$, $H$, and $H^+$ masses below roughly $\unit[800]{GeV}$~\cite{Dumont:2014wha, *Dumont:2014kna}, otherwise $c_{\beta-\alpha}$ is highly suppressed (decoupling regime) and makes our job in explaining $h\to\mu\tau$ slightly more difficult. The main difference of our 2HDM to type-I is the absence of the $\lambda_5$ term in the potential, fixing the two otherwise free mass parameters $m_{A}$ and $m_3$ via $m_A^2 = m_3^2/s_\beta c_\beta$, which has little impact on the valid $c_{\beta-\alpha}$--$t_\beta$ values~\cite{Ko:2013zsa}. Furthermore, our model features additional fermion couplings of $H$, $A$, and $H^+$, predominantly in a LFV manner to $\mu$ and $\tau$ fermions. This should in principle strengthen the bound on $H^+$ compared to the type-I 2HDM, seeing as $H^+\to \mu \nu$ is potentially enhanced; since the fermionic decay modes under investigation at colliders are however only $H^+\to \tau \nu $ or quarks, there are no additional constraints.

This leaves the additional non-collider constraints from e.g.~$\tau \to \mu \gamma$ and $(g-2)_\mu$ to impose on our model, which we will discuss below, as well as the decay $h\to\mu\tau$ that motivates the our study.

With the $\overline{\tau}_L \mu_R h$ coupling at our disposal, we can explain the CMS excess~\cite{CMS:2014hha} in $h\to\mu\tau$ with Yukawa couplings (see Eq.~\eqref{eq:cmsyukawas})
\begin{align}
|y_{\tau\mu}| = \frac{m_\tau}{v}\left|\frac{\cos (\alpha-\beta)}{c_\beta s_\beta}c_R c_L s_R \right|\simeq 7 \times 10^{-3}\ \left|\frac{\cos (\alpha-\beta)}{s_\beta c_\beta} c_R s_R \right|\stackrel{!}{\simeq} 3\times 10^{-3} \,.
\label{eq:y_taumu}
\end{align}
Working in the limit $c_\beta \sim s_R \ll 1$, this simply fixes the parameter combination $|\xi_{\tau\mu} c_{\alpha-\beta}| \simeq 4\times 10^{-3}$; deviations of $h$'s couplings with respect to the SM, parametrized by $\sin (\alpha-\beta)\neq 1$, can hence be easily made unobservably small, even for perturbatively small Yukawa coupling~$\xi_{\tau\mu}$. The scalar $h$ is then very much SM-like, and will not lead to additional (LFV) processes in conflict with observation, e.g.~$\tau\to\mu\gamma$ or $(g-2)_\mu$, following the work of Ref.~\cite{Harnik:2012pb}. For example, the current bound from $\tau\to\mu\gamma$ translates into $|y_{\tau\mu}| < 0.016$ at $90\%$~C.L.~for a sufficiently SM-like $h$. Since all non-$h$ rates can be suppressed by choosing $m_H$, $m_A$, and $m_{H^+}$ large enough, we clearly have a large allowed parameter space at our disposal.

\begin{figure}
\centering
\includegraphics[width=0.7\textwidth]{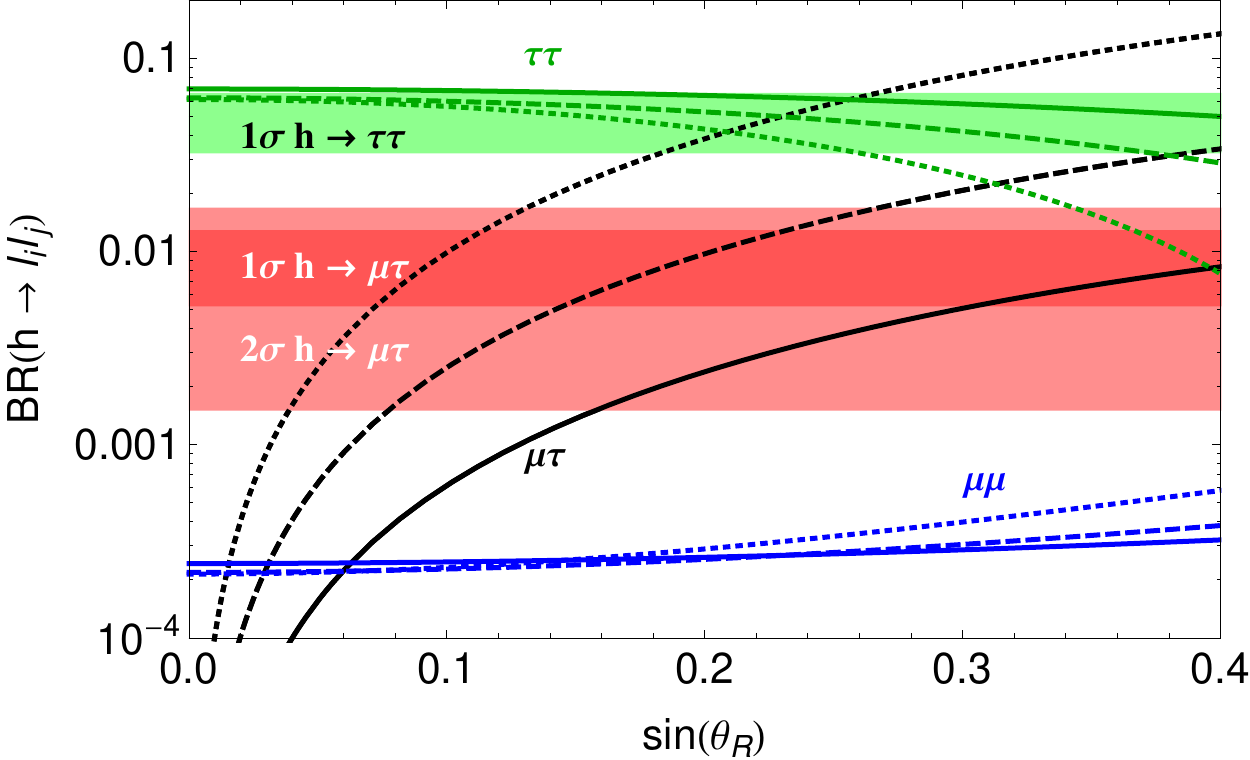}
\caption{The branching ratios $h\to \mu\tau$ (black), $h\to\tau\tau$ (green), and $h\to\mu\mu$ (blue) as a function of $\sin\theta_R \simeq \tfrac{v}{m_\tau} \xi_{\tau\mu} \cos\beta/\sqrt{2}$. Solid lines are for $\tan\beta = 3$, $\cos(\alpha-\beta)= -0.3$, dashed lines for $\tan\beta = 10$, $\cos (\alpha-\beta) = -0.2$, dotted lines for $\tan\beta = 20$, $\cos (\alpha-\beta) = -0.2$.
The colored region show the 1$\sigma$ and 2$\sigma$ ranges for the CMS hint of $h\to\mu\tau$~\cite{CMS:2014hha} (red) and the 1$\sigma$ range for CMS $h\to\tau\tau$~\cite{Chatrchyan:2014nva} (green).}
\label{fig:HiggsBranching}
\end{figure}

Let us however briefly discuss possible effects of our model away from the decoupling limit mentioned above. From the matrix structure of the $h$ couplings in Eq.~\eqref{eq:hcoupling} we see that a large $\theta_R$ will induce changes in the $\mu\mu$ and $\tau\tau$ couplings of $h$, and hence to potentially observable modified rates for $h\to\mu\mu$ and $h\to\tau\tau$. We plot the three branching ratios of interest in Fig.~\ref{fig:HiggsBranching} for some sample values of $\alpha$ and $\beta$. The $h\to \mu\mu$ branching ratio, even when enhanced in our model, is currently not experimentally accessible~\cite{Khachatryan:2014aep}. The di-tauon rate on the other hand has been observed by CMS~\cite{Chatrchyan:2014nva} and ATLAS~\cite{atlastautau}, with rates (relative to the SM) of $0.78\pm 0.27$ and $1.42^{+0.44}_{-0.38}$, respectively. In our case, the modified $\tau$ rate is
\begin{align}
\frac{\BR (h\to\tau \tau)}{\BR (h\to\tau \tau)|_\mathrm{SM}} \simeq \left( \frac{c_\alpha}{s_\beta} + y_{\tau\mu}^h \frac{v}{m_\tau} t_R\right)^2 \simeq \left( 1 \pm 0.4\ |t_R|\right)^2 ,
\label{eq:htautau}
\end{align}
inserting $|y_{\tau\mu}^h|\simeq 3\times 10^{-3}$ in the last step and assuming $c_\alpha/s_\beta\simeq 1$. The rate is enhanced (reduced) for $c_{\alpha-\beta}<0$ ($>0$).
As expected, a large $\theta_R$ can strongly modify the rate $h\to\tau\tau$; we could fit the CMS rate nicely with $t_R \simeq 1/4$, which obviously worsens the agreement with ATLAS, or take $t_R\simeq 1/2$ to match ATLAS' enhanced $h\to\tau\tau$ rate, worsening agreement with CMS. Future improvement in the accuracy of the di-tauon rate can hence provide important information for our model, complementary to the $h\to\mu\tau$ rate.

We stress again that, as mentioned above, we can in any case work in the limit $\theta_R\ll 1$, while still explaining the $h\to\mu\tau$ rate, thus rendering even the $h\to\tau\tau$ and $h\to\mu\mu$ rates effectively SM-like. Nevertheless, the generic predictions of our $L_\mu-L_\tau$ explanation of the LFV excess $h\to\mu\tau$ are modified di-tauon and di-muon rates, together with in general not-too-small $c_{\alpha-\beta}$, thus suppressing the $h$ couplings to gauge bosons, as in any other 2HDM.

\begin{figure}
\centering
\includegraphics[width=0.8\textwidth]{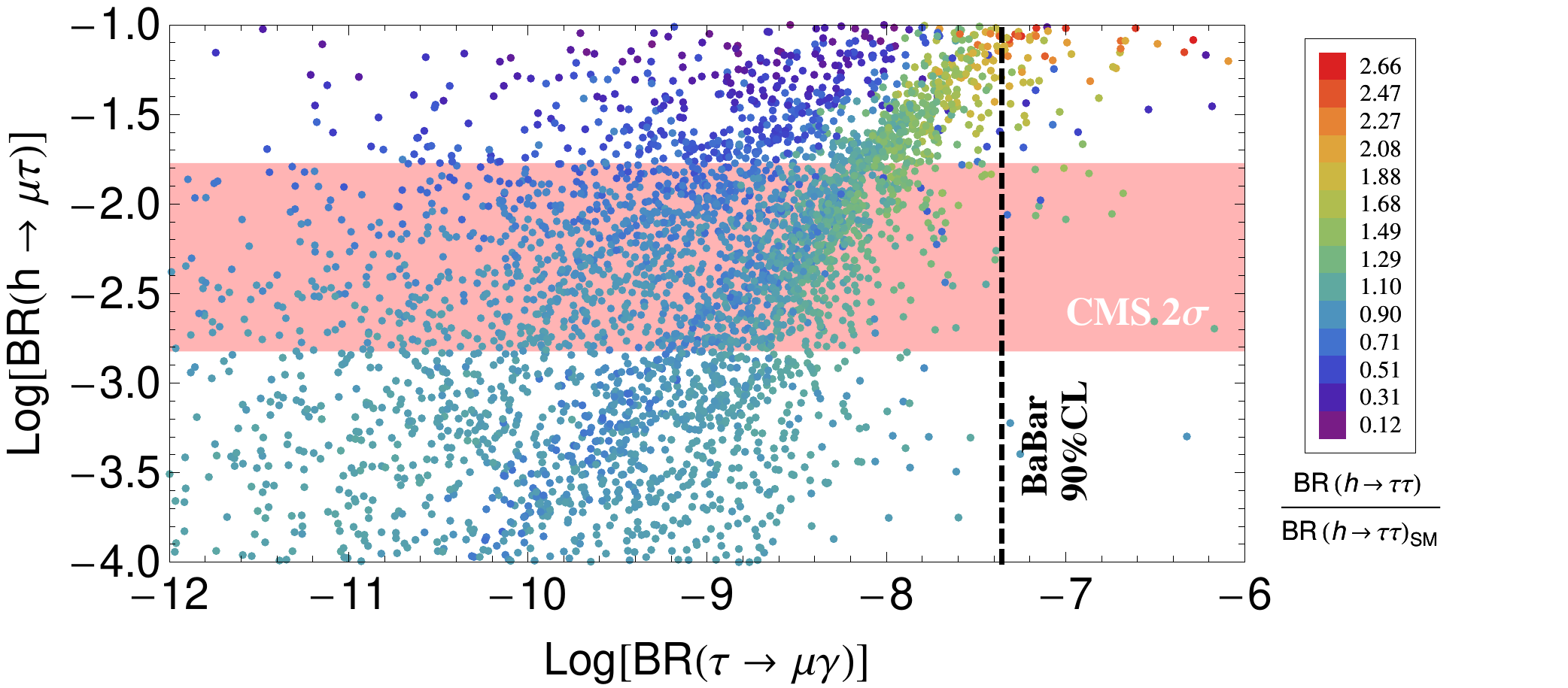}
\caption{The LFV branching ratios $h\to \mu\tau$ and $\tau\to\mu\gamma$ for the random values of Eq.~\eqref{eq:2HDMscatter}.
The color coding of the points refers to the rate $h\to\tau\tau$ relative to the SM prediction.
Also shown are the $2\sigma$ ranges for the CMS hint of $h\to\mu\tau$~\cite{CMS:2014hha} (light red area) and the current $90\,\%$~C.L.~upper limit on $\BR (\tau\to\mu\gamma)$ from BaBar~\cite{Aubert:2009ag} (vertical black dashed line). }
\label{fig:TauToMuGamma}
\end{figure}

On to other LFV processes:
An SM-like $h$ with $y_{\tau\mu}\simeq 3\times 10^{-3}$ does not lead to $\tau\to\mu\gamma$ rates in conflict with current constraints, as shown in Ref.~\cite{Harnik:2012pb}.
One might still expect additional LFV processes induced by the other scalars, seeing as they couple more dominantly to $\mu\tau$ the more SM-like $h$ becomes. This is not necessarily the case, though, because additional suppression factors arise. For $\tau\to\mu\gamma$, not only the coupling $y_{\tau\mu}^\eta$ is required for $\eta \in \{H,A\}$ to run in the loop, but also $y_{\tau\tau}^\eta$ (one loop) or $y_{tt,WW}^\eta$ (two loop). 
Since these couplings are suppressed by $s_\alpha/s_\beta$, $\cot \beta$, or $c_{\alpha-\beta}$, the rates are typically small. 
Using the formulae from Ref.~\cite{Harnik:2012pb} for the one-loop contribution of $h$, $H$, and $A$ to $\tau\to\mu\gamma$ as well as the dominant two-loop diagrams with top-quark and $W$-boson loops, we can find a weak correlation between $h\to\mu\tau$ and $\tau\to\mu\gamma$, see Fig.~\ref{fig:TauToMuGamma}. This is not surprising, as all LFV scales with $\xi_{\tau\mu}$, the only LFV coupling in our model (outside the neutrino sector). Since $\tau\to\mu\gamma$ is additionally suppressed by the heavy masses of $A$ and $H$, the rates are typically below the current sensitivity.
In Fig.~\ref{fig:TauToMuGamma}, we randomly selected values
\begin{align}
m_A, m_H \in [150,700]\, \unit{GeV}, \quad \tan\beta \in [3,60]\,, \quad |\cos(\alpha-\beta)|<0.4\,, \quad \sin\theta_R \in [0,0.5]\,.
\label{eq:2HDMscatter}
\end{align}
Note that we did not impose any bounds, but chose values that seem compatible with Ref.~\cite{Dumont:2014wha, *Dumont:2014kna}. Not surprisingly, we find that the two LFV rates are somewhat correlated, and also that $h\to\tau\tau$ is generically modified for large LFV.
Improvements of $\tau\to\mu\gamma$ searches down to branching ratios of $\mathcal{O}(10^{-9})$ appear feasible~\cite{Aushev:2010bq} and will have a major impact on the allowed parameter space.

Note that the decay $\tau \to \mu\gamma$ gives stronger limits on the scalar sector than $\tau\to 3\mu$, even though the experimental limit on the branching ratio is a factor $\sim 2$ weaker. This is because $\tau\to 3\mu$ is either suppressed by an additional muon Yukawa coupling (tree-level scalar exchange) or fine-structure coupling (off-shell photon in $\tau\to\mu\gamma\to 3\mu$)~\cite{Harnik:2012pb}.
Finally, the scalar contributions to the muon's magnetic moment $(g-2)_\mu$ are insignificant for the parameter values chosen above.

\subsection{Gauge boson sector}

The gauge boson $Z'$ of $U(1)_{L_\mu-L_\tau}$ couples only to muonic and tauonic leptons (neglecting kinetic mixing), and is thus subject to quite different constraints than other popular $Z'$ models. It has recently been shown in Ref.~\cite{Altmannshofer:2014cfa} that trident production $\nu_\mu N \to \nu_\mu N \mu^+ \mu^-$ provides the strongest limit for a heavy $Z'$, $M_{Z'}/g' \gtrsim\unit[550]{GeV}$ at $95\,\%$ C.L.\ from CCFR~\cite{Mishra:1991bv}, making in particular a resolution of the muon's anomalous magnetic moment $a_\mu$ impossible with a heavy $Z'$. A lighter $Z'$ below $\unit[400]{MeV}$ might still do the trick~\cite{Altmannshofer:2014pba}, but is not considered here.\footnote{A resolution of $a_\mu$ and $h\to\mu\tau$ with such a light $Z'$ would allow for the two-body decay $\tau\to\mu Z'$. Old limits from ARGUS~\cite{Albrecht:1995ht} then already imply a tiny $\theta_R < \mathcal{O}(10^{-5})$ and a correspondingly large $\tan\beta \sim 1/\theta_R$ to keep $y^h_{\tau\mu}$ in Eq.~\eqref{eq:y_taumu} large. Updated searches, e.g.~at Belle, could easily improve this limit.} 
Note that $M_{Z'}/g' \simeq \langle S \rangle$, so the trident bound effectively also acts as a lower bound on the seesaw scale (Sec.~\ref{sec:neutrinomasses}).
The trident bound can be improved in future collider measurements, as pointed out recently in Ref.~\cite{delAguila:2014soa}.

The flavor-dependent $Z'$ couplings in combination with the charged-lepton diagonalization lead to the gauge couplings
\begin{align}
 g' j^\alpha_{L_\mu-L_\tau} Z'_\alpha = g'\sum_{i= L,R} \overline{\ell}^0_i \matrixx{0 & & \\ & \cos 2\theta_i & \sin 2\theta_i\\ & \sin 2\theta_i & -\cos 2\theta_i} \gamma^\alpha \ell_i^0 \ Z'_\alpha \,,
\end{align}
which not only contain an axial-vector component due to $\theta_L\neq\theta_R$, but more importantly yield LFV couplings, e.g.~$g' \sin 2\theta_R \overline{\mu}_R\gamma^\alpha \tau_R Z'_\alpha$.
These couplings induce a tree-level decay $\tau \to 3\mu$ mediated by $Z'$, with a rate proportional to $\sin^2 2\theta_i \cos^2 2\theta_j (g'/M_{Z'})^4$ for $M_{Z'}\gg m_\tau$. The experimental limit is $\BR (\tau \to 3\mu) < 2.1\times 10^{-8}$ at $90\,\%$~C.L.~\cite{Hayasaka:2010np}, resulting in the strong constraint~\cite{Chiang:2013aha}
\begin{align}
 \left(\frac{\sin 4\theta_j}{5\times 10^{-3}}\right) \left(\frac{\unit[550]{GeV}}{M_{Z'}/g'}\right)^{2} < 1 \,, \quad j=L,R\,,
\end{align}
beating out $\tau\to\mu\gamma$~\cite{Chiang:2013aha}.
Obviously lepton flavor violation induced by the $Z'$ can be made small by increasing $M_{Z'}/g'$ even for large $\theta_R$, so this bound in no way poses a problem for $h\to\mu\tau$.
Also note that the VEV of $\Phi_1$ automatically induces a $Z$--$Z'$ mixing angle
\begin{align}
 \tan 2\theta_{Z Z'}  \simeq \frac{2 g_1 g' v^2 \cos^2\beta}{M_Z^2 - M_{Z'}^2} \simeq -\frac{10^{-3}}{g'} \left(\frac{\unit{TeV}}{M_{Z'}/g'}\right)^2 \left(\frac{10}{\tan\beta}\right)^2 ,
\end{align}
that leads to lepton non-universal $Z$ couplings.
Limits on $g' \theta_{Z Z'}$ are typically around $10^{-2}$--$10^{-3}$~\cite{Heeck:2011wj} and can be easily satisfied by increasing $M_{Z'}\propto \langle S \rangle$. There is furthermore no theoretical reason to forbid a kinetic-mixing angle between our $Z'$ and the hypercharge gauge boson, which will in any way generated radiatively.
We will not consider this any further. Note that the gauge coupling should satisfy $g' \lesssim 0.5$ in order to avoid a Landau pole below the Planck scale if $L_\mu-L_\tau$ is broken around TeV.

\subsection{Other charge assignments}
\label{sec:variants}

Having focused on the specific $L_\mu-L_\tau$ charge assignments of Tab.~\ref{tab:partcontent-LmuLtau}, let us make some comments about closely related models. First, consider a sign-flip of the $\Phi_1$ scalar doublet's $L_\mu-L_\tau$ charge, i.e.~$+2$ instead of $-2$. The effective 2HDM potential looks the same, the soft-breaking parameter $m_3^2$ being generated by the coupling $\delta \, S^2 \Phi_1^\dagger \Phi_2$. A difference arises in the Yukawa couplings of $\Phi_1$, though, as one effectively replaces $Y_{\ell_1}$ and $Y_{N_1}$ from Eq.~\eqref{eq:Phi1couplings} by their transposed matrices. Because of this, it is now the $\overline{\mu}_L\tau_R$ entry that is dominant, not $\overline{\tau}_L\mu_R$, so the chiral structure behind $h\to\mu\tau$ changes. 
For the neutrino mixing, this changes the vanishing minor from $(\M_\nu^{-1})_{22}$ to $(\M_\nu^{-1})_{33}$ and leads to the octant--ordering relation $s_{23}^2 >1/2$ (IO), $ <1/2$ (NO) for the 23 mixing angle in $V_\nu$. This is the (not yet statistically relevant) preferred correlation in the global fit of Ref.~\cite{Gonzalez-Garcia:2014bfa}, 
assuming the charged lepton contribution to the PMNS matrix is small.\footnote{Note that other global fits have different preferences~\cite{Capozzi:2013csa, Forero:2014bxa}.}

In the charged-lepton mass matrix diagonalization, the change of quantum numbers switches $\theta_L \leftrightarrow \theta_R$, so $\theta_L$ is now the dominant charged-lepton mixing angle. Besides this renaming, there are no changes in the LFV phenomenology. However, the fact that $\theta_L$ can be large has potentially a huge impact on the neutrino mixing sector, because $\theta_L$ is added to the generically close-to-maximal mixing angle $\theta_{23}^\nu$ to form the physical atmospheric mixing angle (which is measured to be close to maximal: $\theta_{23} = 38$--$53^\circ$ at $3\sigma$~\cite{Gonzalez-Garcia:2014bfa}). So, in addition to strongly modified $h\to\tau\tau$ rates (Eq.~\eqref{eq:htautau}), one expects also large deviations from maximal $23$ mixing in the region of parameter space where $\theta_L$ is large. This is however not a hard prediction, as $h\to\mu\tau$ can still be explained with $\theta_L\ll 1$ in this scenario.

If we give the doublet $\Phi_1$ a $L_\mu-L_\tau$ charge $\pm 1$ instead of $\pm 2$, we recover the model from Ref.~\cite{Heeck:2011wj}. Here, the $\Delta (L_\mu-L_\tau) = 1$ Yukawa couplings will generate LFV processes like $\mu\to e X$ or $h\to e\mu, e\tau$ (see Ref.~\cite{Harnik:2012pb} for details), mediated by the scalars and $Z'$. Stronger constraints apply, of course, due to the couplings to electrons and muons, and there are no indications in the data that could be taken as a hint.

Analogous considerations for the anomaly-free symmetries $L_e - L_\mu$ and $L_e- L_\tau$ -- which suffer from stronger bounds due to the coupling to electrons and can not be regarded as good symmetries to describe leptonic mixing -- can be found in Ref.~\cite{Dutta:1994dx}.


\section{Conclusion}
\label{sec:conclusion}

Using a recently found hint for non-standard Higgs decays $h \to \mu \tau$ as an exemplary 
benchmark value, we have analyzed the possibility to explain lepton flavor 
violating Higgs decays in 
flavor symmetry models. These models usually focus on explaining lepton mixing, but 
can also have phenomenology at the electroweak scale, when they are broken at that scale. 
Two very different approaches were followed here: a continuous Abelian and a discrete 
non-Abelian case. Both models enlarge the Higgs sector.  

The non-Abelian group $A_4$ is the most often applied flavor group, as it is the 
smallest discrete group with a 3-dimensional irreducible representation. We introduce, 
as almost always in such models, an $A_4$-triplet containing the three lepton doublets, but also 
assume an $A_4$-triplet of scalar weak doublets. The model can easily reproduce neutrino mixing 
parameters in agreement with current data. It predicts a branching ratio for 
$h\to e\tau$ with similar magnitude as the one of $h\to\mu\tau$, where this decay enjoys a chiral 
structure mainly of $\overline{\mu} P_{R} \tau$. The breaking of the symmetry 
introduces also breaking of 'triality', which usually suppresses lepton flavor violation in such 
models; hence, the decay $\mu \to e\gamma$ is induced and can pose important constraints.

The Abelian flavor symmetry $U(1)_{L_\mu-L_\tau}$ as a particularly attractive 
symmetry for quasi-degenerate neutrinos 
with close-to-maximal atmospheric mixing can also be taken as a horizontal symmetry in a 2HDM, 
allowing for flavor violation in selected lepton modes only. These are specified by the 
$L_\mu-L_\tau$ quantum numbers of the non-SM-like scalar doublet, leading to lepton Yukawa 
couplings that perturb the overall type-I 2HDM structure. We have shown how this can lead to 
LFV exclusively in the $\mu\tau$ sector in order to explain the tentative $h\to\mu\tau$ 
signal at CMS, with natural suppression/absence of other flavor violating modes typically 
occurring in other models. The model non-trivially correlates the Higgs branching ratio for 
$h\to\mu\tau$ with the decay $\tau \to \mu \gamma$, and generally predicts modified  
$h\to\mu\mu$ and $\tau\tau$ rates, as well as overall type-I 2HDM-like suppression 
of Higgs couplings to gauge bosons, on a smaller scale than the $A_4$ model. 
Additionally, the chiral structure of 
$h\to\overline{\mu} P_{R,L} \tau$ influences the correlation of mass ordering and the octant of $\theta_{23}$ in the neutrino sector. \\

In summary, the option to explain non-standard Higgs phenomenology in low scale flavor symmetry 
models is an interesting testing ground for those models, and possible in a 
variety of different approaches, predicting a large number of other phenomenological
consequences.

\section*{Acknowledgements}
This work of WR is supported by the Max Planck Society in the project MANITOP. 
The work of JH is funded in part by IISN and by Belgian Science Policy
(IAP VII/37). 
YS is supported by JSPS Postdoctoral Fellowships for Research Abroad, No.~20130600.
\\\emph{Note added}:
recently, CMS has released its final analysis of the $h\to\mu\tau$ search as a preprint~\cite{Khachatryan:2015kon}, resulting in slightly changed values -- $\BR (h\to\mu\tau) = \left( 0.84_{-0.37}^{+0.39} \right)\%$ -- with little impact on our study.

\bibliography{finalbib}

\begin{thebibliography}{90}
\expandafter\ifx\csname natexlab\endcsname\relax\def\natexlab#1{#1}\fi
\expandafter\ifx\csname bibnamefont\endcsname\relax
  \def\bibnamefont#1{#1}\fi
\expandafter\ifx\csname bibfnamefont\endcsname\relax
  \def\bibfnamefont#1{#1}\fi
\expandafter\ifx\csname citenamefont\endcsname\relax
  \def\citenamefont#1{#1}\fi
\expandafter\ifx\csname url\endcsname\relax
  \def\url#1{\texttt{#1}}\fi
\expandafter\ifx\csname urlprefix\endcsname\relax\def\urlprefix{URL }\fi
\providecommand{\bibinfo}[2]{#2}
\providecommand{\eprint}[2][]{\url{#2}}

\bibitem[{\citenamefont{Aad et~al.}(2012)}]{Aad:2012tfa}
\bibinfo{author}{\bibfnamefont{G.}~\bibnamefont{Aad}} \bibnamefont{et~al.}
  (\bibinfo{collaboration}{ATLAS Collaboration}), \bibinfo{journal}{Phys.Lett.}
  \textbf{\bibinfo{volume}{B716}}, \bibinfo{pages}{1} (\bibinfo{year}{2012}),
  \eprint{1207.7214}.

\bibitem[{\citenamefont{Chatrchyan et~al.}(2012)}]{Chatrchyan:2012ufa}
\bibinfo{author}{\bibfnamefont{S.}~\bibnamefont{Chatrchyan}}
  \bibnamefont{et~al.} (\bibinfo{collaboration}{CMS Collaboration}),
  \bibinfo{journal}{Phys.Lett.} \textbf{\bibinfo{volume}{B716}},
  \bibinfo{pages}{30} (\bibinfo{year}{2012}), \eprint{1207.7235}.

\bibitem[{\citenamefont{Bhattacharyya et~al.}(2011)\citenamefont{Bhattacharyya,
  Leser, and P{\"a}s}}]{Bhattacharyya:2010hp}
\bibinfo{author}{\bibfnamefont{G.}~\bibnamefont{Bhattacharyya}},
  \bibinfo{author}{\bibfnamefont{P.}~\bibnamefont{Leser}}, \bibnamefont{and}
  \bibinfo{author}{\bibfnamefont{H.}~\bibnamefont{P{\"a}s}},
  \bibinfo{journal}{Phys.Rev.} \textbf{\bibinfo{volume}{D83}},
  \bibinfo{pages}{011701} (\bibinfo{year}{2011}), \eprint{1006.5597}.

\bibitem[{\citenamefont{Frampton et~al.}(2010)\citenamefont{Frampton, Ho,
  Kephart, and Matsuzaki}}]{Frampton:2010uw}
\bibinfo{author}{\bibfnamefont{P.~H.} \bibnamefont{Frampton}},
  \bibinfo{author}{\bibfnamefont{C.~M.} \bibnamefont{Ho}},
  \bibinfo{author}{\bibfnamefont{T.~W.} \bibnamefont{Kephart}},
  \bibnamefont{and}
  \bibinfo{author}{\bibfnamefont{S.}~\bibnamefont{Matsuzaki}},
  \bibinfo{journal}{Phys.Rev.} \textbf{\bibinfo{volume}{D82}},
  \bibinfo{pages}{113007} (\bibinfo{year}{2010}), \eprint{1009.0307}.

\bibitem[{\citenamefont{Cao et~al.}(2011{\natexlab{a}})\citenamefont{Cao,
  Damanik, Ma, and Wegman}}]{Cao:2011df}
\bibinfo{author}{\bibfnamefont{Q.-H.} \bibnamefont{Cao}},
  \bibinfo{author}{\bibfnamefont{A.}~\bibnamefont{Damanik}},
  \bibinfo{author}{\bibfnamefont{E.}~\bibnamefont{Ma}}, \bibnamefont{and}
  \bibinfo{author}{\bibfnamefont{D.}~\bibnamefont{Wegman}},
  \bibinfo{journal}{Phys.Rev.} \textbf{\bibinfo{volume}{D83}},
  \bibinfo{pages}{093012} (\bibinfo{year}{2011}{\natexlab{a}}),
  \eprint{1103.0008}.

\bibitem[{\citenamefont{Ho and Tandean}(2013)}]{Ho:2013hia}
\bibinfo{author}{\bibfnamefont{S.-Y.} \bibnamefont{Ho}} \bibnamefont{and}
  \bibinfo{author}{\bibfnamefont{J.}~\bibnamefont{Tandean}},
  \bibinfo{journal}{Phys.Rev.} \textbf{\bibinfo{volume}{D87}},
  \bibinfo{pages}{095015} (\bibinfo{year}{2013}), \eprint{1303.5700}.

\bibitem[{\citenamefont{Blankenburg et~al.}(2012)\citenamefont{Blankenburg,
  Ellis, and Isidori}}]{Blankenburg:2012ex}
\bibinfo{author}{\bibfnamefont{G.}~\bibnamefont{Blankenburg}},
  \bibinfo{author}{\bibfnamefont{J.}~\bibnamefont{Ellis}}, \bibnamefont{and}
  \bibinfo{author}{\bibfnamefont{G.}~\bibnamefont{Isidori}},
  \bibinfo{journal}{Phys.Lett.} \textbf{\bibinfo{volume}{B712}},
  \bibinfo{pages}{386} (\bibinfo{year}{2012}), \eprint{1202.5704}.

\bibitem[{\citenamefont{Harnik et~al.}(2013)\citenamefont{Harnik, Kopp, and
  Zupan}}]{Harnik:2012pb}
\bibinfo{author}{\bibfnamefont{R.}~\bibnamefont{Harnik}},
  \bibinfo{author}{\bibfnamefont{J.}~\bibnamefont{Kopp}}, \bibnamefont{and}
  \bibinfo{author}{\bibfnamefont{J.}~\bibnamefont{Zupan}},
  \bibinfo{journal}{JHEP} \textbf{\bibinfo{volume}{1303}}, \bibinfo{pages}{026}
  (\bibinfo{year}{2013}), \eprint{1209.1397}.

\bibitem[{\citenamefont{CMS}(2014)}]{CMS:2014hha}
\bibinfo{author}{\bibnamefont{CMS}} (\bibinfo{collaboration}{CMS
  Collaboration}) (\bibinfo{year}{2014}), \bibinfo{note}{{CMS-PAS-HIG-14-005}}.

\bibitem[{\citenamefont{Dery et~al.}(2014)\citenamefont{Dery, Efrati, Nir,
  Soreq, and Susi{\v c}}}]{Dery:2014kxa}
\bibinfo{author}{\bibfnamefont{A.}~\bibnamefont{Dery}},
  \bibinfo{author}{\bibfnamefont{A.}~\bibnamefont{Efrati}},
  \bibinfo{author}{\bibfnamefont{Y.}~\bibnamefont{Nir}},
  \bibinfo{author}{\bibfnamefont{Y.}~\bibnamefont{Soreq}}, \bibnamefont{and}
  \bibinfo{author}{\bibfnamefont{V.}~\bibnamefont{Susi{\v c}}},
  \bibinfo{journal}{Phys.Rev.} \textbf{\bibinfo{volume}{D90}},
  \bibinfo{pages}{115022} (\bibinfo{year}{2014}), \eprint{1408.1371}.

\bibitem[{\citenamefont{Campos et~al.}(2014)\citenamefont{Campos,
  Hern{\'a}ndez, P{\"a}s, and Schumacher}}]{Campos:2014zaa}
\bibinfo{author}{\bibfnamefont{M.~D.} \bibnamefont{Campos}},
  \bibinfo{author}{\bibfnamefont{A.~E.~C.} \bibnamefont{Hern{\'a}ndez}},
  \bibinfo{author}{\bibfnamefont{H.}~\bibnamefont{P{\"a}s}}, \bibnamefont{and}
  \bibinfo{author}{\bibfnamefont{E.}~\bibnamefont{Schumacher}}
  (\bibinfo{year}{2014}), \eprint{1408.1652}.

\bibitem[{\citenamefont{Celis et~al.}(2014)\citenamefont{Celis, Cirigliano, and
  Passemar}}]{Celis:2014roa}
\bibinfo{author}{\bibfnamefont{A.}~\bibnamefont{Celis}},
  \bibinfo{author}{\bibfnamefont{V.}~\bibnamefont{Cirigliano}},
  \bibnamefont{and} \bibinfo{author}{\bibfnamefont{E.}~\bibnamefont{Passemar}}
  (\bibinfo{year}{2014}), \eprint{1409.4439}.

\bibitem[{\citenamefont{Sierra and Vicente}(2014)}]{Sierra:2014nqa}
\bibinfo{author}{\bibfnamefont{D.~A.} \bibnamefont{Sierra}} \bibnamefont{and}
  \bibinfo{author}{\bibfnamefont{A.}~\bibnamefont{Vicente}},
  \bibinfo{journal}{Phys.Rev.} \textbf{\bibinfo{volume}{D90}},
  \bibinfo{pages}{115004} (\bibinfo{year}{2014}), \eprint{1409.7690}.

\bibitem[{\citenamefont{Lee and Tandean}(2014)}]{Lee:2014rba}
\bibinfo{author}{\bibfnamefont{C.-J.} \bibnamefont{Lee}} \bibnamefont{and}
  \bibinfo{author}{\bibfnamefont{J.}~\bibnamefont{Tandean}}
  (\bibinfo{year}{2014}), \eprint{1410.6803}.

\bibitem[{\citenamefont{Ma}(2010)}]{Ma:2010gs}
\bibinfo{author}{\bibfnamefont{E.}~\bibnamefont{Ma}},
  \bibinfo{journal}{Phys.Rev.} \textbf{\bibinfo{volume}{D82}},
  \bibinfo{pages}{037301} (\bibinfo{year}{2010}), \eprint{1006.3524}.

\bibitem[{\citenamefont{Ishimori et~al.}(2010)\citenamefont{Ishimori,
  Kobayashi, Ohki, Shimizu, Okada et~al.}}]{Ishimori:2010au}
\bibinfo{author}{\bibfnamefont{H.}~\bibnamefont{Ishimori}},
  \bibinfo{author}{\bibfnamefont{T.}~\bibnamefont{Kobayashi}},
  \bibinfo{author}{\bibfnamefont{H.}~\bibnamefont{Ohki}},
  \bibinfo{author}{\bibfnamefont{Y.}~\bibnamefont{Shimizu}},
  \bibinfo{author}{\bibfnamefont{H.}~\bibnamefont{Okada}},
  \bibnamefont{et~al.}, \bibinfo{journal}{Prog.Theor.Phys.Suppl.}
  \textbf{\bibinfo{volume}{183}}, \bibinfo{pages}{1} (\bibinfo{year}{2010}),
  \eprint{1003.3552}.

\bibitem[{\citenamefont{King and Luhn}(2013)}]{King:2013eh}
\bibinfo{author}{\bibfnamefont{S.~F.} \bibnamefont{King}} \bibnamefont{and}
  \bibinfo{author}{\bibfnamefont{C.}~\bibnamefont{Luhn}},
  \bibinfo{journal}{Rept.Prog.Phys.} \textbf{\bibinfo{volume}{76}},
  \bibinfo{pages}{056201} (\bibinfo{year}{2013}), \eprint{1301.1340}.

\bibitem[{\citenamefont{Ma and Rajasekaran}(2001)}]{Ma:2001fk}
\bibinfo{author}{\bibfnamefont{E.}~\bibnamefont{Ma}} \bibnamefont{and}
  \bibinfo{author}{\bibfnamefont{G.}~\bibnamefont{Rajasekaran}},
  \bibinfo{journal}{Phys.Rev.} \textbf{\bibinfo{volume}{D64}},
  \bibinfo{pages}{113012} (\bibinfo{year}{2001}), \eprint{hep-ph/0106291}.

\bibitem[{\citenamefont{Babu et~al.}(2003)\citenamefont{Babu, Ma, and
  Valle}}]{Babu:2002dz}
\bibinfo{author}{\bibfnamefont{K.}~\bibnamefont{Babu}},
  \bibinfo{author}{\bibfnamefont{E.}~\bibnamefont{Ma}}, \bibnamefont{and}
  \bibinfo{author}{\bibfnamefont{J.}~\bibnamefont{Valle}},
  \bibinfo{journal}{Phys.Lett.} \textbf{\bibinfo{volume}{B552}},
  \bibinfo{pages}{207} (\bibinfo{year}{2003}), \eprint{hep-ph/0206292}.

\bibitem[{\citenamefont{Ma}(2004)}]{Ma:2004zv}
\bibinfo{author}{\bibfnamefont{E.}~\bibnamefont{Ma}},
  \bibinfo{journal}{Phys.Rev.} \textbf{\bibinfo{volume}{D70}},
  \bibinfo{pages}{031901} (\bibinfo{year}{2004}), \eprint{hep-ph/0404199}.

\bibitem[{\citenamefont{Altarelli and Feruglio}(2005)}]{Altarelli:2005yp}
\bibinfo{author}{\bibfnamefont{G.}~\bibnamefont{Altarelli}} \bibnamefont{and}
  \bibinfo{author}{\bibfnamefont{F.}~\bibnamefont{Feruglio}},
  \bibinfo{journal}{Nucl.Phys.} \textbf{\bibinfo{volume}{B720}},
  \bibinfo{pages}{64} (\bibinfo{year}{2005}), \eprint{hep-ph/0504165}.

\bibitem[{\citenamefont{Babu and He}(2005)}]{Babu:2005se}
\bibinfo{author}{\bibfnamefont{K.}~\bibnamefont{Babu}} \bibnamefont{and}
  \bibinfo{author}{\bibfnamefont{X.-G.} \bibnamefont{He}}
  (\bibinfo{year}{2005}), \eprint{hep-ph/0507217}.

\bibitem[{\citenamefont{Altarelli and Feruglio}(2006)}]{Altarelli:2005yx}
\bibinfo{author}{\bibfnamefont{G.}~\bibnamefont{Altarelli}} \bibnamefont{and}
  \bibinfo{author}{\bibfnamefont{F.}~\bibnamefont{Feruglio}},
  \bibinfo{journal}{Nucl.Phys.} \textbf{\bibinfo{volume}{B741}},
  \bibinfo{pages}{215} (\bibinfo{year}{2006}), \eprint{hep-ph/0512103}.

\bibitem[{\citenamefont{He et~al.}(2006)\citenamefont{He, Keum, and
  Volkas}}]{He:2006dk}
\bibinfo{author}{\bibfnamefont{X.-G.} \bibnamefont{He}},
  \bibinfo{author}{\bibfnamefont{Y.-Y.} \bibnamefont{Keum}}, \bibnamefont{and}
  \bibinfo{author}{\bibfnamefont{R.~R.} \bibnamefont{Volkas}},
  \bibinfo{journal}{JHEP} \textbf{\bibinfo{volume}{0604}}, \bibinfo{pages}{039}
  (\bibinfo{year}{2006}), \eprint{hep-ph/0601001}.

\bibitem[{\citenamefont{Lam}(2007)}]{Lam:2007qc}
\bibinfo{author}{\bibfnamefont{C.}~\bibnamefont{Lam}},
  \bibinfo{journal}{Phys.Lett.} \textbf{\bibinfo{volume}{B656}},
  \bibinfo{pages}{193} (\bibinfo{year}{2007}), \eprint{0708.3665}.

\bibitem[{\citenamefont{Lam}(2008{\natexlab{a}})}]{Lam:2008rs}
\bibinfo{author}{\bibfnamefont{C.}~\bibnamefont{Lam}},
  \bibinfo{journal}{Phys.Rev.Lett.} \textbf{\bibinfo{volume}{101}},
  \bibinfo{pages}{121602} (\bibinfo{year}{2008}{\natexlab{a}}),
  \eprint{0804.2622}.

\bibitem[{\citenamefont{Lam}(2008{\natexlab{b}})}]{Lam:2008sh}
\bibinfo{author}{\bibfnamefont{C.}~\bibnamefont{Lam}},
  \bibinfo{journal}{Phys.Rev.} \textbf{\bibinfo{volume}{D78}},
  \bibinfo{pages}{073015} (\bibinfo{year}{2008}{\natexlab{b}}),
  \eprint{0809.1185}.

\bibitem[{\citenamefont{Ma}(2006)}]{Ma:2006km}
\bibinfo{author}{\bibfnamefont{E.}~\bibnamefont{Ma}},
  \bibinfo{journal}{Phys.Rev.} \textbf{\bibinfo{volume}{D73}},
  \bibinfo{pages}{077301} (\bibinfo{year}{2006}), \eprint{hep-ph/0601225}.

\bibitem[{\citenamefont{Kubo et~al.}(2006)\citenamefont{Kubo, Ma, and
  Suematsu}}]{Kubo:2006kx}
\bibinfo{author}{\bibfnamefont{J.}~\bibnamefont{Kubo}},
  \bibinfo{author}{\bibfnamefont{E.}~\bibnamefont{Ma}}, \bibnamefont{and}
  \bibinfo{author}{\bibfnamefont{D.}~\bibnamefont{Suematsu}},
  \bibinfo{journal}{Phys.Lett.} \textbf{\bibinfo{volume}{B642}},
  \bibinfo{pages}{18} (\bibinfo{year}{2006}), \eprint{hep-ph/0604114}.

\bibitem[{\citenamefont{Hirsch et~al.}(2009)\citenamefont{Hirsch, Morisi, and
  Valle}}]{Hirsch:2009mx}
\bibinfo{author}{\bibfnamefont{M.}~\bibnamefont{Hirsch}},
  \bibinfo{author}{\bibfnamefont{S.}~\bibnamefont{Morisi}}, \bibnamefont{and}
  \bibinfo{author}{\bibfnamefont{J.}~\bibnamefont{Valle}},
  \bibinfo{journal}{Phys.Lett.} \textbf{\bibinfo{volume}{B679}},
  \bibinfo{pages}{454} (\bibinfo{year}{2009}), \eprint{0905.3056}.

\bibitem[{\citenamefont{Ibanez et~al.}(2009)\citenamefont{Ibanez, Morisi, and
  Valle}}]{Ibanez:2009du}
\bibinfo{author}{\bibfnamefont{D.}~\bibnamefont{Ibanez}},
  \bibinfo{author}{\bibfnamefont{S.}~\bibnamefont{Morisi}}, \bibnamefont{and}
  \bibinfo{author}{\bibfnamefont{J.}~\bibnamefont{Valle}},
  \bibinfo{journal}{Phys.Rev.} \textbf{\bibinfo{volume}{D80}},
  \bibinfo{pages}{053015} (\bibinfo{year}{2009}), \eprint{0907.3109}.

\bibitem[{\citenamefont{de~Adelhart~Toorop
  et~al.}(2011{\natexlab{a}})\citenamefont{de~Adelhart~Toorop, Bazzocchi,
  Merlo, and Paris}}]{Adelhart-Toorop:2010fk}
\bibinfo{author}{\bibfnamefont{R.}~\bibnamefont{de~Adelhart~Toorop}},
  \bibinfo{author}{\bibfnamefont{F.}~\bibnamefont{Bazzocchi}},
  \bibinfo{author}{\bibfnamefont{L.}~\bibnamefont{Merlo}}, \bibnamefont{and}
  \bibinfo{author}{\bibfnamefont{A.}~\bibnamefont{Paris}},
  \bibinfo{journal}{JHEP} \textbf{\bibinfo{volume}{1103}}, \bibinfo{pages}{035}
  (\bibinfo{year}{2011}{\natexlab{a}}), \eprint{1012.1791}.

\bibitem[{\citenamefont{de~Adelhart~Toorop
  et~al.}(2011{\natexlab{b}})\citenamefont{de~Adelhart~Toorop, Bazzocchi,
  Merlo, and Paris}}]{Adelhart-Toorop:2010uq}
\bibinfo{author}{\bibfnamefont{R.}~\bibnamefont{de~Adelhart~Toorop}},
  \bibinfo{author}{\bibfnamefont{F.}~\bibnamefont{Bazzocchi}},
  \bibinfo{author}{\bibfnamefont{L.}~\bibnamefont{Merlo}}, \bibnamefont{and}
  \bibinfo{author}{\bibfnamefont{A.}~\bibnamefont{Paris}},
  \bibinfo{journal}{JHEP} \textbf{\bibinfo{volume}{1103}}, \bibinfo{pages}{040}
  (\bibinfo{year}{2011}{\natexlab{b}}), \eprint{1012.2091}.

\bibitem[{\citenamefont{Cao et~al.}(2011{\natexlab{b}})\citenamefont{Cao,
  Khalil, Ma, and Okada}}]{Cao:2010mp}
\bibinfo{author}{\bibfnamefont{Q.-H.} \bibnamefont{Cao}},
  \bibinfo{author}{\bibfnamefont{S.}~\bibnamefont{Khalil}},
  \bibinfo{author}{\bibfnamefont{E.}~\bibnamefont{Ma}}, \bibnamefont{and}
  \bibinfo{author}{\bibfnamefont{H.}~\bibnamefont{Okada}},
  \bibinfo{journal}{Phys.Rev.Lett.} \textbf{\bibinfo{volume}{106}},
  \bibinfo{pages}{131801} (\bibinfo{year}{2011}{\natexlab{b}}),
  \eprint{1009.5415}.

\bibitem[{\citenamefont{Bhattacharyya
  et~al.}(2012{\natexlab{a}})\citenamefont{Bhattacharyya, Leser, and
  P{\"a}s}}]{Bhattacharyya:2012ze}
\bibinfo{author}{\bibfnamefont{G.}~\bibnamefont{Bhattacharyya}},
  \bibinfo{author}{\bibfnamefont{P.}~\bibnamefont{Leser}}, \bibnamefont{and}
  \bibinfo{author}{\bibfnamefont{H.}~\bibnamefont{P{\"a}s}},
  \bibinfo{journal}{Phys.Rev.} \textbf{\bibinfo{volume}{D86}},
  \bibinfo{pages}{036009} (\bibinfo{year}{2012}{\natexlab{a}}),
  \eprint{1206.4202}.

\bibitem[{\citenamefont{Bhattacharyya
  et~al.}(2012{\natexlab{b}})\citenamefont{Bhattacharyya,
  de~Medeiros~Varzielas, and Leser}}]{Bhattacharyya:2012pi}
\bibinfo{author}{\bibfnamefont{G.}~\bibnamefont{Bhattacharyya}},
  \bibinfo{author}{\bibfnamefont{I.}~\bibnamefont{de~Medeiros~Varzielas}},
  \bibnamefont{and} \bibinfo{author}{\bibfnamefont{P.}~\bibnamefont{Leser}},
  \bibinfo{journal}{Phys.Rev.Lett.} \textbf{\bibinfo{volume}{109}},
  \bibinfo{pages}{241603} (\bibinfo{year}{2012}{\natexlab{b}}),
  \eprint{1210.0545}.

\bibitem[{\citenamefont{Holthausen et~al.}(2013)\citenamefont{Holthausen,
  Lindner, and Schmidt}}]{Holthausen:2012wz}
\bibinfo{author}{\bibfnamefont{M.}~\bibnamefont{Holthausen}},
  \bibinfo{author}{\bibfnamefont{M.}~\bibnamefont{Lindner}}, \bibnamefont{and}
  \bibinfo{author}{\bibfnamefont{M.~A.} \bibnamefont{Schmidt}},
  \bibinfo{journal}{Phys.Rev.} \textbf{\bibinfo{volume}{D87}},
  \bibinfo{pages}{033006} (\bibinfo{year}{2013}), \eprint{1211.5143}.

\bibitem[{\citenamefont{Holthausen and Schmidt}(2012)}]{Holthausen:2011vd}
\bibinfo{author}{\bibfnamefont{M.}~\bibnamefont{Holthausen}} \bibnamefont{and}
  \bibinfo{author}{\bibfnamefont{M.~A.} \bibnamefont{Schmidt}},
  \bibinfo{journal}{JHEP} \textbf{\bibinfo{volume}{1201}}, \bibinfo{pages}{126}
  (\bibinfo{year}{2012}), \eprint{1111.1730}.

\bibitem[{\citenamefont{Paschos}(1977)}]{Paschos:1976ay}
\bibinfo{author}{\bibfnamefont{E.}~\bibnamefont{Paschos}},
  \bibinfo{journal}{Phys.Rev.} \textbf{\bibinfo{volume}{D15}},
  \bibinfo{pages}{1966} (\bibinfo{year}{1977}).

\bibitem[{\citenamefont{Glashow and Weinberg}(1977)}]{Glashow:1976nt}
\bibinfo{author}{\bibfnamefont{S.~L.} \bibnamefont{Glashow}} \bibnamefont{and}
  \bibinfo{author}{\bibfnamefont{S.}~\bibnamefont{Weinberg}},
  \bibinfo{journal}{Phys.Rev.} \textbf{\bibinfo{volume}{D15}},
  \bibinfo{pages}{1958} (\bibinfo{year}{1977}).

\bibitem[{\citenamefont{Holthausen}(2012)}]{Holthausen:2012xwa}
\bibinfo{author}{\bibfnamefont{M.}~\bibnamefont{Holthausen}}, Ph.D. thesis,
  \bibinfo{school}{University of Heidelberg} (\bibinfo{year}{2012}).

\bibitem[{\citenamefont{Ivanov and Nishi}(2015)}]{Ivanov:2014doa}
\bibinfo{author}{\bibfnamefont{I.}~\bibnamefont{Ivanov}} \bibnamefont{and}
  \bibinfo{author}{\bibfnamefont{C.}~\bibnamefont{Nishi}},
  \bibinfo{journal}{JHEP} \textbf{\bibinfo{volume}{1501}}, \bibinfo{pages}{021}
  (\bibinfo{year}{2015}), \eprint{1410.6139}.

\bibitem[{\citenamefont{Branco et~al.}(2012)\citenamefont{Branco, Ferreira,
  Lavoura, Rebelo, Sher et~al.}}]{Branco:2011iw}
\bibinfo{author}{\bibfnamefont{G.}~\bibnamefont{Branco}},
  \bibinfo{author}{\bibfnamefont{P.}~\bibnamefont{Ferreira}},
  \bibinfo{author}{\bibfnamefont{L.}~\bibnamefont{Lavoura}},
  \bibinfo{author}{\bibfnamefont{M.}~\bibnamefont{Rebelo}},
  \bibinfo{author}{\bibfnamefont{M.}~\bibnamefont{Sher}}, \bibnamefont{et~al.},
  \bibinfo{journal}{Phys.Rept.} \textbf{\bibinfo{volume}{516}},
  \bibinfo{pages}{1} (\bibinfo{year}{2012}), \eprint{1106.0034}.

\bibitem[{\citenamefont{Olive et~al.}(2014)}]{Agashe:2014kda}
\bibinfo{author}{\bibfnamefont{K.}~\bibnamefont{Olive}} \bibnamefont{et~al.}
  (\bibinfo{collaboration}{Particle Data Group}), \bibinfo{journal}{Chin.Phys.}
  \textbf{\bibinfo{volume}{C38}}, \bibinfo{pages}{090001}
  (\bibinfo{year}{2014}).

\bibitem[{\citenamefont{Gonzalez-Garcia
  et~al.}(2014)\citenamefont{Gonzalez-Garcia, Maltoni, and
  Schwetz}}]{Gonzalez-Garcia:2014bfa}
\bibinfo{author}{\bibfnamefont{M.}~\bibnamefont{Gonzalez-Garcia}},
  \bibinfo{author}{\bibfnamefont{M.}~\bibnamefont{Maltoni}}, \bibnamefont{and}
  \bibinfo{author}{\bibfnamefont{T.}~\bibnamefont{Schwetz}},
  \bibinfo{journal}{JHEP} \textbf{\bibinfo{volume}{1411}}, \bibinfo{pages}{052}
  (\bibinfo{year}{2014}), \eprint{1409.5439}.

\bibitem[{\citenamefont{Barry and Rodejohann}(2011)}]{Barry:2010yk}
\bibinfo{author}{\bibfnamefont{J.}~\bibnamefont{Barry}} \bibnamefont{and}
  \bibinfo{author}{\bibfnamefont{W.}~\bibnamefont{Rodejohann}},
  \bibinfo{journal}{Nucl.Phys.} \textbf{\bibinfo{volume}{B842}},
  \bibinfo{pages}{33} (\bibinfo{year}{2011}), \eprint{1007.5217}.

\bibitem[{\citenamefont{Aad et~al.}(2015)}]{Aad:2014xzb}
\bibinfo{author}{\bibfnamefont{G.}~\bibnamefont{Aad}} \bibnamefont{et~al.}
  (\bibinfo{collaboration}{ATLAS Collaboration}), \bibinfo{journal}{JHEP}
  \textbf{\bibinfo{volume}{1501}}, \bibinfo{pages}{069} (\bibinfo{year}{2015}),
  \eprint{1409.6212}.

\bibitem[{\citenamefont{Chatrchyan
  et~al.}(2014{\natexlab{a}})}]{Chatrchyan:2013zna}
\bibinfo{author}{\bibfnamefont{S.}~\bibnamefont{Chatrchyan}}
  \bibnamefont{et~al.} (\bibinfo{collaboration}{CMS Collaboration}),
  \bibinfo{journal}{Phys.Rev.} \textbf{\bibinfo{volume}{D89}},
  \bibinfo{pages}{012003} (\bibinfo{year}{2014}{\natexlab{a}}),
  \eprint{1310.3687}.

\bibitem[{\citenamefont{Barger et~al.}(2014)\citenamefont{Barger, Everett,
  Jackson, Peterson, and Shaughnessy}}]{Barger:2014qva}
\bibinfo{author}{\bibfnamefont{V.}~\bibnamefont{Barger}},
  \bibinfo{author}{\bibfnamefont{L.~L.} \bibnamefont{Everett}},
  \bibinfo{author}{\bibfnamefont{C.~B.} \bibnamefont{Jackson}},
  \bibinfo{author}{\bibfnamefont{A.~D.} \bibnamefont{Peterson}},
  \bibnamefont{and}
  \bibinfo{author}{\bibfnamefont{G.}~\bibnamefont{Shaughnessy}},
  \bibinfo{journal}{Phys.Rev.} \textbf{\bibinfo{volume}{D90}},
  \bibinfo{pages}{095006} (\bibinfo{year}{2014}), \eprint{1408.2525}.

\bibitem[{\citenamefont{Chatrchyan
  et~al.}(2014{\natexlab{b}})}]{Chatrchyan:2014nva}
\bibinfo{author}{\bibfnamefont{S.}~\bibnamefont{Chatrchyan}}
  \bibnamefont{et~al.} (\bibinfo{collaboration}{CMS Collaboration}),
  \bibinfo{journal}{JHEP} \textbf{\bibinfo{volume}{1405}}, \bibinfo{pages}{104}
  (\bibinfo{year}{2014}{\natexlab{b}}), \eprint{1401.5041}.

\bibitem[{\citenamefont{Adam et~al.}(2013)}]{Adam:2013mnn}
\bibinfo{author}{\bibfnamefont{J.}~\bibnamefont{Adam}} \bibnamefont{et~al.}
  (\bibinfo{collaboration}{MEG Collaboration}),
  \bibinfo{journal}{Phys.Rev.Lett.} \textbf{\bibinfo{volume}{110}},
  \bibinfo{pages}{201801} (\bibinfo{year}{2013}), \eprint{1303.0754}.

\bibitem[{\citenamefont{Aubert et~al.}(2010)}]{Aubert:2009ag}
\bibinfo{author}{\bibfnamefont{B.}~\bibnamefont{Aubert}} \bibnamefont{et~al.}
  (\bibinfo{collaboration}{BaBar Collaboration}),
  \bibinfo{journal}{Phys.Rev.Lett.} \textbf{\bibinfo{volume}{104}},
  \bibinfo{pages}{021802} (\bibinfo{year}{2010}), \eprint{0908.2381}.

\bibitem[{\citenamefont{He et~al.}(1991{\natexlab{a}})\citenamefont{He, Joshi,
  Lew, and Volkas}}]{He:1990pn}
\bibinfo{author}{\bibfnamefont{X.}~\bibnamefont{He}},
  \bibinfo{author}{\bibfnamefont{G.~C.} \bibnamefont{Joshi}},
  \bibinfo{author}{\bibfnamefont{H.}~\bibnamefont{Lew}}, \bibnamefont{and}
  \bibinfo{author}{\bibfnamefont{R.}~\bibnamefont{Volkas}},
  \bibinfo{journal}{Phys.Rev.} \textbf{\bibinfo{volume}{D43}},
  \bibinfo{pages}{22} (\bibinfo{year}{1991}{\natexlab{a}}).

\bibitem[{\citenamefont{Foot}(1991)}]{Foot:1990mn}
\bibinfo{author}{\bibfnamefont{R.}~\bibnamefont{Foot}},
  \bibinfo{journal}{Mod.Phys.Lett.} \textbf{\bibinfo{volume}{A6}},
  \bibinfo{pages}{527} (\bibinfo{year}{1991}).

\bibitem[{\citenamefont{He et~al.}(1991{\natexlab{b}})\citenamefont{He, Joshi,
  Lew, and Volkas}}]{He:1991qd}
\bibinfo{author}{\bibfnamefont{X.-G.} \bibnamefont{He}},
  \bibinfo{author}{\bibfnamefont{G.~C.} \bibnamefont{Joshi}},
  \bibinfo{author}{\bibfnamefont{H.}~\bibnamefont{Lew}}, \bibnamefont{and}
  \bibinfo{author}{\bibfnamefont{R.}~\bibnamefont{Volkas}},
  \bibinfo{journal}{Phys.Rev.} \textbf{\bibinfo{volume}{D44}},
  \bibinfo{pages}{2118} (\bibinfo{year}{1991}{\natexlab{b}}).

\bibitem[{\citenamefont{Binetruy et~al.}(1997)\citenamefont{Binetruy, Lavignac,
  Petcov, and Ramond}}]{Binetruy:1996cs}
\bibinfo{author}{\bibfnamefont{P.}~\bibnamefont{Binetruy}},
  \bibinfo{author}{\bibfnamefont{S.}~\bibnamefont{Lavignac}},
  \bibinfo{author}{\bibfnamefont{S.~T.} \bibnamefont{Petcov}},
  \bibnamefont{and} \bibinfo{author}{\bibfnamefont{P.}~\bibnamefont{Ramond}},
  \bibinfo{journal}{Nucl.Phys.} \textbf{\bibinfo{volume}{B496}},
  \bibinfo{pages}{3} (\bibinfo{year}{1997}), \eprint{hep-ph/9610481}.

\bibitem[{\citenamefont{Bell and Volkas}(2001)}]{Bell:2000vh}
\bibinfo{author}{\bibfnamefont{N.~F.} \bibnamefont{Bell}} \bibnamefont{and}
  \bibinfo{author}{\bibfnamefont{R.~R.} \bibnamefont{Volkas}},
  \bibinfo{journal}{Phys.Rev.} \textbf{\bibinfo{volume}{D63}},
  \bibinfo{pages}{013006} (\bibinfo{year}{2001}), \eprint{hep-ph/0008177}.

\bibitem[{\citenamefont{Choubey and Rodejohann}(2005)}]{Choubey:2004hn}
\bibinfo{author}{\bibfnamefont{S.}~\bibnamefont{Choubey}} \bibnamefont{and}
  \bibinfo{author}{\bibfnamefont{W.}~\bibnamefont{Rodejohann}},
  \bibinfo{journal}{Eur.Phys.J.} \textbf{\bibinfo{volume}{C40}},
  \bibinfo{pages}{259} (\bibinfo{year}{2005}), \eprint{hep-ph/0411190}.

\bibitem[{\citenamefont{Dutta et~al.}(1994)\citenamefont{Dutta, Joshipura, and
  Vijaykumar}}]{Dutta:1994dx}
\bibinfo{author}{\bibfnamefont{G.}~\bibnamefont{Dutta}},
  \bibinfo{author}{\bibfnamefont{A.~S.} \bibnamefont{Joshipura}},
  \bibnamefont{and}
  \bibinfo{author}{\bibfnamefont{K.}~\bibnamefont{Vijaykumar}},
  \bibinfo{journal}{Phys.Rev.} \textbf{\bibinfo{volume}{D50}},
  \bibinfo{pages}{2109} (\bibinfo{year}{1994}), \eprint{hep-ph/9405292}.

\bibitem[{\citenamefont{Heeck and
  Rodejohann}(2011{\natexlab{a}})}]{Heeck:2011wj}
\bibinfo{author}{\bibfnamefont{J.}~\bibnamefont{Heeck}} \bibnamefont{and}
  \bibinfo{author}{\bibfnamefont{W.}~\bibnamefont{Rodejohann}},
  \bibinfo{journal}{Phys.Rev.} \textbf{\bibinfo{volume}{D84}},
  \bibinfo{pages}{075007} (\bibinfo{year}{2011}{\natexlab{a}}),
  \eprint{1107.5238}.

\bibitem[{\citenamefont{Gninenko and Krasnikov}(2001)}]{Gninenko:2001hx}
\bibinfo{author}{\bibfnamefont{S.}~\bibnamefont{Gninenko}} \bibnamefont{and}
  \bibinfo{author}{\bibfnamefont{N.}~\bibnamefont{Krasnikov}},
  \bibinfo{journal}{Phys.Lett.} \textbf{\bibinfo{volume}{B513}},
  \bibinfo{pages}{119} (\bibinfo{year}{2001}), \eprint{hep-ph/0102222}.

\bibitem[{\citenamefont{Baek et~al.}(2001)\citenamefont{Baek, Deshpande, He,
  and Ko}}]{Baek:2001kca}
\bibinfo{author}{\bibfnamefont{S.}~\bibnamefont{Baek}},
  \bibinfo{author}{\bibfnamefont{N.}~\bibnamefont{Deshpande}},
  \bibinfo{author}{\bibfnamefont{X.}~\bibnamefont{He}}, \bibnamefont{and}
  \bibinfo{author}{\bibfnamefont{P.}~\bibnamefont{Ko}},
  \bibinfo{journal}{Phys.Rev.} \textbf{\bibinfo{volume}{D64}},
  \bibinfo{pages}{055006} (\bibinfo{year}{2001}), \eprint{hep-ph/0104141}.

\bibitem[{\citenamefont{Ma et~al.}(2002)\citenamefont{Ma, Roy, and
  Roy}}]{Ma:2001md}
\bibinfo{author}{\bibfnamefont{E.}~\bibnamefont{Ma}},
  \bibinfo{author}{\bibfnamefont{D.}~\bibnamefont{Roy}}, \bibnamefont{and}
  \bibinfo{author}{\bibfnamefont{S.}~\bibnamefont{Roy}},
  \bibinfo{journal}{Phys.Lett.} \textbf{\bibinfo{volume}{B525}},
  \bibinfo{pages}{101} (\bibinfo{year}{2002}), \eprint{hep-ph/0110146}.

\bibitem[{\citenamefont{Ma and Roy}(2002)}]{Ma:2002df}
\bibinfo{author}{\bibfnamefont{E.}~\bibnamefont{Ma}} \bibnamefont{and}
  \bibinfo{author}{\bibfnamefont{D.~P.} \bibnamefont{Roy}},
  \bibinfo{journal}{Phys.Rev.} \textbf{\bibinfo{volume}{D65}},
  \bibinfo{pages}{075021} (\bibinfo{year}{2002}).

\bibitem[{\citenamefont{Altmannshofer
  et~al.}(2014{\natexlab{a}})\citenamefont{Altmannshofer, Gori, Pospelov, and
  Yavin}}]{Altmannshofer:2014cfa}
\bibinfo{author}{\bibfnamefont{W.}~\bibnamefont{Altmannshofer}},
  \bibinfo{author}{\bibfnamefont{S.}~\bibnamefont{Gori}},
  \bibinfo{author}{\bibfnamefont{M.}~\bibnamefont{Pospelov}}, \bibnamefont{and}
  \bibinfo{author}{\bibfnamefont{I.}~\bibnamefont{Yavin}},
  \bibinfo{journal}{Phys.Rev.} \textbf{\bibinfo{volume}{D89}},
  \bibinfo{pages}{095033} (\bibinfo{year}{2014}{\natexlab{a}}),
  \eprint{1403.1269}.

\bibitem[{\citenamefont{Altmannshofer
  et~al.}(2014{\natexlab{b}})\citenamefont{Altmannshofer, Gori, Pospelov, and
  Yavin}}]{Altmannshofer:2014pba}
\bibinfo{author}{\bibfnamefont{W.}~\bibnamefont{Altmannshofer}},
  \bibinfo{author}{\bibfnamefont{S.}~\bibnamefont{Gori}},
  \bibinfo{author}{\bibfnamefont{M.}~\bibnamefont{Pospelov}}, \bibnamefont{and}
  \bibinfo{author}{\bibfnamefont{I.}~\bibnamefont{Yavin}},
  \bibinfo{journal}{Phys.Rev.Lett.} \textbf{\bibinfo{volume}{113}},
  \bibinfo{pages}{091801} (\bibinfo{year}{2014}{\natexlab{b}}),
  \eprint{1406.2332}.

\bibitem[{\citenamefont{Araki et~al.}(2015)\citenamefont{Araki, Kaneko,
  Konishi, Ota, Sato et~al.}}]{Araki:2014ona}
\bibinfo{author}{\bibfnamefont{T.}~\bibnamefont{Araki}},
  \bibinfo{author}{\bibfnamefont{F.}~\bibnamefont{Kaneko}},
  \bibinfo{author}{\bibfnamefont{Y.}~\bibnamefont{Konishi}},
  \bibinfo{author}{\bibfnamefont{T.}~\bibnamefont{Ota}},
  \bibinfo{author}{\bibfnamefont{J.}~\bibnamefont{Sato}}, \bibnamefont{et~al.},
  \bibinfo{journal}{Phys.Rev.} \textbf{\bibinfo{volume}{D91}},
  \bibinfo{pages}{037301} (\bibinfo{year}{2015}), \eprint{1409.4180}.

\bibitem[{\citenamefont{Gninenko et~al.}(2014)\citenamefont{Gninenko,
  Krasnikov, and Matveev}}]{Gninenko:2014pea}
\bibinfo{author}{\bibfnamefont{S.}~\bibnamefont{Gninenko}},
  \bibinfo{author}{\bibfnamefont{N.}~\bibnamefont{Krasnikov}},
  \bibnamefont{and} \bibinfo{author}{\bibfnamefont{V.}~\bibnamefont{Matveev}}
  (\bibinfo{year}{2014}), \eprint{1412.1400}.

\bibitem[{\citenamefont{Heeck and
  Rodejohann}(2011{\natexlab{b}})}]{Heeck:2010pg}
\bibinfo{author}{\bibfnamefont{J.}~\bibnamefont{Heeck}} \bibnamefont{and}
  \bibinfo{author}{\bibfnamefont{W.}~\bibnamefont{Rodejohann}},
  \bibinfo{journal}{J.Phys.} \textbf{\bibinfo{volume}{G38}},
  \bibinfo{pages}{085005} (\bibinfo{year}{2011}{\natexlab{b}}),
  \eprint{1007.2655}.

\bibitem[{\citenamefont{Ko et~al.}(2012)\citenamefont{Ko, Omura, and
  Yu}}]{Ko:2012hd}
\bibinfo{author}{\bibfnamefont{P.}~\bibnamefont{Ko}},
  \bibinfo{author}{\bibfnamefont{Y.}~\bibnamefont{Omura}}, \bibnamefont{and}
  \bibinfo{author}{\bibfnamefont{C.}~\bibnamefont{Yu}},
  \bibinfo{journal}{Phys.Lett.} \textbf{\bibinfo{volume}{B717}},
  \bibinfo{pages}{202} (\bibinfo{year}{2012}), \eprint{1204.4588}.

\bibitem[{\citenamefont{Kanemura et~al.}(2004)\citenamefont{Kanemura, Okada,
  Senaha, and Yuan}}]{Kanemura:2004mg}
\bibinfo{author}{\bibfnamefont{S.}~\bibnamefont{Kanemura}},
  \bibinfo{author}{\bibfnamefont{Y.}~\bibnamefont{Okada}},
  \bibinfo{author}{\bibfnamefont{E.}~\bibnamefont{Senaha}}, \bibnamefont{and}
  \bibinfo{author}{\bibfnamefont{C.-P.} \bibnamefont{Yuan}},
  \bibinfo{journal}{Phys.Rev.} \textbf{\bibinfo{volume}{D70}},
  \bibinfo{pages}{115002} (\bibinfo{year}{2004}), \eprint{hep-ph/0408364}.

\bibitem[{\citenamefont{Araki et~al.}(2012)\citenamefont{Araki, Heeck, and
  Kubo}}]{Araki:2012ip}
\bibinfo{author}{\bibfnamefont{T.}~\bibnamefont{Araki}},
  \bibinfo{author}{\bibfnamefont{J.}~\bibnamefont{Heeck}}, \bibnamefont{and}
  \bibinfo{author}{\bibfnamefont{J.}~\bibnamefont{Kubo}},
  \bibinfo{journal}{JHEP} \textbf{\bibinfo{volume}{1207}}, \bibinfo{pages}{083}
  (\bibinfo{year}{2012}), \eprint{1203.4951}.

\bibitem[{\citenamefont{Lavoura}(2005)}]{Lavoura:2004tu}
\bibinfo{author}{\bibfnamefont{L.}~\bibnamefont{Lavoura}},
  \bibinfo{journal}{Phys.Lett.} \textbf{\bibinfo{volume}{B609}},
  \bibinfo{pages}{317} (\bibinfo{year}{2005}), \eprint{hep-ph/0411232}.

\bibitem[{\citenamefont{Lashin and Chamoun}(2008)}]{Lashin:2007dm}
\bibinfo{author}{\bibfnamefont{E.}~\bibnamefont{Lashin}} \bibnamefont{and}
  \bibinfo{author}{\bibfnamefont{N.}~\bibnamefont{Chamoun}},
  \bibinfo{journal}{Phys.Rev.} \textbf{\bibinfo{volume}{D78}},
  \bibinfo{pages}{073002} (\bibinfo{year}{2008}), \eprint{0708.2423}.

\bibitem[{\citenamefont{Lashin and Chamoun}(2009)}]{Lashin:2009yd}
\bibinfo{author}{\bibfnamefont{E.}~\bibnamefont{Lashin}} \bibnamefont{and}
  \bibinfo{author}{\bibfnamefont{N.}~\bibnamefont{Chamoun}},
  \bibinfo{journal}{Phys.Rev.} \textbf{\bibinfo{volume}{D80}},
  \bibinfo{pages}{093004} (\bibinfo{year}{2009}), \eprint{0909.2669}.

\bibitem[{\citenamefont{Rodejohann}(2011)}]{Rodejohann:2011mu}
\bibinfo{author}{\bibfnamefont{W.}~\bibnamefont{Rodejohann}},
  \bibinfo{journal}{Int.J.Mod.Phys.} \textbf{\bibinfo{volume}{E20}},
  \bibinfo{pages}{1833} (\bibinfo{year}{2011}), \eprint{1106.1334}.

\bibitem[{\citenamefont{Capozzi et~al.}(2014)\citenamefont{Capozzi, Fogli,
  Lisi, Marrone, Montanino et~al.}}]{Capozzi:2013csa}
\bibinfo{author}{\bibfnamefont{F.}~\bibnamefont{Capozzi}},
  \bibinfo{author}{\bibfnamefont{G.}~\bibnamefont{Fogli}},
  \bibinfo{author}{\bibfnamefont{E.}~\bibnamefont{Lisi}},
  \bibinfo{author}{\bibfnamefont{A.}~\bibnamefont{Marrone}},
  \bibinfo{author}{\bibfnamefont{D.}~\bibnamefont{Montanino}},
  \bibnamefont{et~al.}, \bibinfo{journal}{Phys.Rev.}
  \textbf{\bibinfo{volume}{D89}}, \bibinfo{pages}{093018}
  (\bibinfo{year}{2014}), \eprint{1312.2878}.

\bibitem[{\citenamefont{Forero et~al.}(2014)\citenamefont{Forero, Tortola, and
  Valle}}]{Forero:2014bxa}
\bibinfo{author}{\bibfnamefont{D.}~\bibnamefont{Forero}},
  \bibinfo{author}{\bibfnamefont{M.}~\bibnamefont{Tortola}}, \bibnamefont{and}
  \bibinfo{author}{\bibfnamefont{J.}~\bibnamefont{Valle}},
  \bibinfo{journal}{Phys.Rev.} \textbf{\bibinfo{volume}{D90}},
  \bibinfo{pages}{093006} (\bibinfo{year}{2014}), \eprint{1405.7540}.

\bibitem[{\citenamefont{Ko et~al.}(2014)\citenamefont{Ko, Omura, and
  Yu}}]{Ko:2013zsa}
\bibinfo{author}{\bibfnamefont{P.}~\bibnamefont{Ko}},
  \bibinfo{author}{\bibfnamefont{Y.}~\bibnamefont{Omura}}, \bibnamefont{and}
  \bibinfo{author}{\bibfnamefont{C.}~\bibnamefont{Yu}}, \bibinfo{journal}{JHEP}
  \textbf{\bibinfo{volume}{1401}}, \bibinfo{pages}{016} (\bibinfo{year}{2014}),
  \eprint{1309.7156}.

\bibitem[{\citenamefont{Dumont et~al.}(2014{\natexlab{a}})\citenamefont{Dumont,
  Gunion, Jiang, and Kraml}}]{Dumont:2014wha}
\bibinfo{author}{\bibfnamefont{B.}~\bibnamefont{Dumont}},
  \bibinfo{author}{\bibfnamefont{J.~F.} \bibnamefont{Gunion}},
  \bibinfo{author}{\bibfnamefont{Y.}~\bibnamefont{Jiang}}, \bibnamefont{and}
  \bibinfo{author}{\bibfnamefont{S.}~\bibnamefont{Kraml}},
  \bibinfo{journal}{Phys.Rev.} \textbf{\bibinfo{volume}{D90}},
  \bibinfo{pages}{035021} (\bibinfo{year}{2014}{\natexlab{a}}),
  \eprint{1405.3584}.

\bibitem[{\citenamefont{Dumont et~al.}(2014{\natexlab{b}})\citenamefont{Dumont,
  Gunion, Jiang, and Kraml}}]{Dumont:2014kna}
\bibinfo{author}{\bibfnamefont{B.}~\bibnamefont{Dumont}},
  \bibinfo{author}{\bibfnamefont{J.~F.} \bibnamefont{Gunion}},
  \bibinfo{author}{\bibfnamefont{Y.}~\bibnamefont{Jiang}}, \bibnamefont{and}
  \bibinfo{author}{\bibfnamefont{S.}~\bibnamefont{Kraml}}
  (\bibinfo{year}{2014}{\natexlab{b}}), \eprint{1409.4088}.

\bibitem[{\citenamefont{Khachatryan
  et~al.}(2015{\natexlab{a}})}]{Khachatryan:2014aep}
\bibinfo{author}{\bibfnamefont{V.}~\bibnamefont{Khachatryan}}
  \bibnamefont{et~al.} (\bibinfo{collaboration}{CMS}),
  \bibinfo{journal}{Phys.Lett.} \textbf{\bibinfo{volume}{B744}},
  \bibinfo{pages}{184} (\bibinfo{year}{2015}{\natexlab{a}}),
  \eprint{1410.6679}.

\bibitem[{\citenamefont{{ATLAS}}(2014)}]{atlastautau}
\bibinfo{author}{\bibnamefont{{ATLAS}}} (\bibinfo{collaboration}{ATLAS
  Collaboration}) (\bibinfo{year}{2014}), \bibinfo{note}{{ATLAS-CONF-2014-061,
  ATLAS-COM-CONF-2014-080}}.

\bibitem[{\citenamefont{Aushev et~al.}(2010)\citenamefont{Aushev, Bartel,
  Bondar, Brodzicka, Browder et~al.}}]{Aushev:2010bq}
\bibinfo{author}{\bibfnamefont{T.}~\bibnamefont{Aushev}},
  \bibinfo{author}{\bibfnamefont{W.}~\bibnamefont{Bartel}},
  \bibinfo{author}{\bibfnamefont{A.}~\bibnamefont{Bondar}},
  \bibinfo{author}{\bibfnamefont{J.}~\bibnamefont{Brodzicka}},
  \bibinfo{author}{\bibfnamefont{T.}~\bibnamefont{Browder}},
  \bibnamefont{et~al.} (\bibinfo{year}{2010}), \eprint{1002.5012}.

\bibitem[{\citenamefont{Mishra et~al.}(1991)}]{Mishra:1991bv}
\bibinfo{author}{\bibfnamefont{S.}~\bibnamefont{Mishra}} \bibnamefont{et~al.}
  (\bibinfo{collaboration}{CCFR Collaboration}),
  \bibinfo{journal}{Phys.Rev.Lett.} \textbf{\bibinfo{volume}{66}},
  \bibinfo{pages}{3117} (\bibinfo{year}{1991}).

\bibitem[{\citenamefont{Albrecht et~al.}(1995)}]{Albrecht:1995ht}
\bibinfo{author}{\bibfnamefont{H.}~\bibnamefont{Albrecht}} \bibnamefont{et~al.}
  (\bibinfo{collaboration}{ARGUS Collaboration}), \bibinfo{journal}{Z.Phys.}
  \textbf{\bibinfo{volume}{C68}}, \bibinfo{pages}{25} (\bibinfo{year}{1995}).

\bibitem[{\citenamefont{del Aguila et~al.}(2015)\citenamefont{del Aguila,
  Chala, Santiago, and Yamamoto}}]{delAguila:2014soa}
\bibinfo{author}{\bibfnamefont{F.}~\bibnamefont{del Aguila}},
  \bibinfo{author}{\bibfnamefont{M.}~\bibnamefont{Chala}},
  \bibinfo{author}{\bibfnamefont{J.}~\bibnamefont{Santiago}}, \bibnamefont{and}
  \bibinfo{author}{\bibfnamefont{Y.}~\bibnamefont{Yamamoto}},
  \bibinfo{journal}{JHEP} \textbf{\bibinfo{volume}{1503}}, \bibinfo{pages}{059}
  (\bibinfo{year}{2015}), \eprint{1411.7394}.

\bibitem[{\citenamefont{Hayasaka et~al.}(2010)\citenamefont{Hayasaka, Inami,
  Miyazaki, Arinstein, Aulchenko et~al.}}]{Hayasaka:2010np}
\bibinfo{author}{\bibfnamefont{K.}~\bibnamefont{Hayasaka}},
  \bibinfo{author}{\bibfnamefont{K.}~\bibnamefont{Inami}},
  \bibinfo{author}{\bibfnamefont{Y.}~\bibnamefont{Miyazaki}},
  \bibinfo{author}{\bibfnamefont{K.}~\bibnamefont{Arinstein}},
  \bibinfo{author}{\bibfnamefont{V.}~\bibnamefont{Aulchenko}},
  \bibnamefont{et~al.}, \bibinfo{journal}{Phys.Lett.}
  \textbf{\bibinfo{volume}{B687}}, \bibinfo{pages}{139} (\bibinfo{year}{2010}),
  \eprint{1001.3221}.

\bibitem[{\citenamefont{Chiang et~al.}(2013)\citenamefont{Chiang, Nomura, and
  Tandean}}]{Chiang:2013aha}
\bibinfo{author}{\bibfnamefont{C.-W.} \bibnamefont{Chiang}},
  \bibinfo{author}{\bibfnamefont{T.}~\bibnamefont{Nomura}}, \bibnamefont{and}
  \bibinfo{author}{\bibfnamefont{J.}~\bibnamefont{Tandean}},
  \bibinfo{journal}{Phys.Rev.} \textbf{\bibinfo{volume}{D87}},
  \bibinfo{pages}{075020} (\bibinfo{year}{2013}), \eprint{1302.2894}.

\bibitem[{\citenamefont{Khachatryan
  et~al.}(2015{\natexlab{b}})}]{Khachatryan:2015kon}
\bibinfo{author}{\bibfnamefont{V.}~\bibnamefont{Khachatryan}}
  \bibnamefont{et~al.} (\bibinfo{collaboration}{CMS})
  (\bibinfo{year}{2015}{\natexlab{b}}), \eprint{1502.07400}.

\end{thebibliography}

\end{document}